%% file: paper.tex
\documentclass[twocolumn, prd, nofootinbib]{revtex4-2}

\newcommand\ForInternalReference[1]{}
\newcommand\SkipForEarlyCirculation[1]{}

\newcommand\SkipPP[1]{}
\usepackage{graphicx}
\usepackage{dcolumn}
\usepackage{bm}
\usepackage{xspace}
\usepackage{url}
\usepackage{amsmath}
\usepackage{listings}
\usepackage{multirow}
\usepackage{amssymb}
\usepackage{color}
\usepackage{tikz}
\usepackage[mathlines]{lineno}
\usepackage{longtable}

\linenumbers
\usetikzlibrary{shapes.geometric, arrows, bayesnet}
\usepackage[dvipsnames]{xcolor}
\usepackage[colorlinks=true, linkcolor=BlueViolet, citecolor=BlueViolet, urlcolor=BlueViolet]{hyperref}
\usepackage{orcidlink}
\newcommand\optional[1]{}

\newcommand\unit[1]{{\rm #1}}

\tikzstyle{startstop} = [circle, rounded corners, minimum width=1cm, minimum height=1cm,text centered, draw=black, fill=red!30]
\tikzstyle{io} = [trapezium, trapezium left angle=70, trapezium right angle=110, minimum width=2cm, minimum height=1cm, text centered, draw=black, fill=blue!30]
\tikzstyle{process} = [rectangle, minimum width=2cm, minimum height=1cm, text centered, draw=black, fill=orange!30]
\tikzstyle{decision} = [diamond, minimum width=2cm, minimum height=1cm, text centered, draw=black, fill=green!30]
\tikzstyle{arrow} = [thick,->,>=stealth]
\definecolor{amber}{rgb}{1.0, 0.75, 0.0}
\definecolor{orange}{rgb}{1.0, 0.5, 0.0}
\definecolor{amaranth}{rgb}{0.9, 0.17, 0.31}

\graphicspath{{./figures/}}

\def\ltsima{$\; \buildrel < \over \sim \;$}
\def\simlt{\lower.5ex\hbox{\ltsima}}
\def\gtsima{$\; \buildrel > \over \sim \;$}
\def\simgt{\lower.5ex\hbox{\gtsima}}

\input{symbols.tex}

\begin{document}
\nolinenumbers
\renewcommand{\arraystretch}{1.5}
\title{Assessing the imprint of eccentricity in GW signatures  using two independent waveform models }

\author{Natalie Malagon\orcidlink{0000-0002-5825-7795}} 
\affiliation{\RIT}
\author{Richard O'Shaughnessy\orcidlink{0000-0001-5832-8517}} 
\affiliation{\RIT}

\begin{abstract}
The gravitational-wave signal from merging compact binaries encodes information about their orbital and intrinsic properties. Over the last few years, state-of-the-art waveform models have begun to incorporate the effects of orbital eccentricity into their estimated signal. Over a similar period, many groups have applied these waveforms to characterize whether the imprint of eccentricity is present and, if so, measure this time-evolving property (at a suitably-defined reference point). In this work, we present a comprehensive analysis of 162 confident sources identified in the O3 and O4a observing runs of the International Gravitational Wave Network (LIGO-Virgo-KAGRA). Using the RIFT parameter inference engine, we employ two independently implemented waveform models (SEOBNRv5EHM and TEOBResumS-Dali) which account for orbital eccentricity and the effects of aligned compact object spins. Using these two waveforms, we find consistent conclusions that disfavor the eccentric hypothesis. Unlike previous work, among binary black hole candidates, we find potential evidence for eccentricity in three events: GW200129, GW231001, and GW231123. For the latter two events, the evidence for eccentricity is ambiguous, with different degrees of support from different waveforms. Consistent with previous work, we find conclusions obtained about GW200129 can be sensitive to analysis settings, as expected, given the nonstationary noise present.


%

\end{abstract}

\maketitle

\section{Introduction}

Since the  discovery of GW150914\_095045 by the Advanced LIGO \cite{2015CQGra..32g4001L} and Virgo \cite{2015CQGra..32b4001A} detectors, now joined by KAGRA \cite{2021PTEP.2021eA101A}, the first four observing runs (referred to as O1, O2, O3, and O4) of the LIGO-Virgo-KAGRA (LVK) network by the LVK Collaboration have identified the characteristic gravitational-wave signature of more than $O(\simeq 200)$ coalescing compact binaries \cite{LIGO-O3-O3a-catalog, LIGO-O3-O3a_final-catalog, LIGO-O3-O3b-catalog, LIGO-O4a-cbc-catalog_results}. Independent analyses of the public data have also contributed to the candidate event census; see, e.g., \cite{2023ApJ...946...59N,2020PhRvD.101h3030V,2023arXiv231206631W} and references therein.
For each candidate event, the gravitational wave (GW) strain data can be compared to approximate models for the emitted gravitational radiation from a coalescing compact binary, to deduce the distribution of plausible source binary parameters consistent with the data; see, eg., \cite{LIGO-O4a-cbc-catalog_methods,gwastro-RIFT-Update} for details.

While until recently these inferences were restricted to quasi-circular binaries, because few models allowed for
non-quasi-circular binaries with any spin, now multiple waveform modeling implementations account for orbital eccentricity, particularly in the case where both compact object spins are aligned with the orbital angular momentum, including SEOBNRE \cite{2020PhRvD.101d4049L}, SEOBNRv5EHM \cite{2025PhRvD.112d4038G}, 
TEOBResumS-Dali \cite{
2020PhRvD.101j1501C,%
2024PhRvD.110h4001N,%
2025PhRvD.111f4050N},
IMRPhenomTEHM \cite{
2025arXiv250313062D}, and the inspiral-only pyEFPE model \cite{
2025PhRvD.111h4052M}.
Using these and earlier models, several groups have investigated large subsets of the total GW catalog, identifying several candidate events with possible indications of orbital eccentricity (e.g., GW190521, GW190701, GW190929, GW191109, GW200105, GW200129, and GW200208\_222617) \cite{2019MNRAS.490.5210R,%
2025arXiv251207688S,%
2022NatAs...6..344G,%
2020ApJ...903L...5R,%
2022ApJ...940..171R,
2024ApJ...972...65I,%
2025PhRvD.112j4045G,%
2025ApJ...995...47P,%
2025arXiv250812460J,%
2025arXiv250800179K,%
2025arXiv250315393M,%
2025PhRvD.112f3052R,%
2025arXiv250722862M%
}.

Unambiguous signatures of eccentricity could provide complementary insight to better differentiate between different formation channels \cite{2010CQGra..27k4007M,2021ApJ...921L..43Z,2021ApJ...921L..31R,2025ApJ...994L..47S}.
There are two formation channels for stellar-mass binary black holes: isolated binary evolution and dynamic assembly \cite{2022PhR...955....1M}. The isolated binary evolution scenario entails the binary to undergo binary stellar evolution with no external interactions until the stellar binary collapses into black holes \cite{2016Natur.534..512B}. Another possible mechanism is dynamic assembly which involves independently evolved black holes which become gravitationally bound in dense stellar environments, such as globular clusters and galactic centers, and dynamically interact with other black holes \cite{2016ApJ...830L..18B, 2019ApJ...871...91Z}. Isolated compact binaries are expected to have quasi-circular orbits when they merge due to the damping of orbital eccentricity, while dynamically assembled systems may retain measurable eccentricities at the merger \cite{2022PhR...955....1M}. 

The distinctions in eccentric gravitational-wave signals serve as a potential probe for better understanding compact binary formation and evolution. However, previous experience with novel physics and parameter extremes in gravitational-wave models suggests that substantial differences may remain, complicating our ability to reliably discern source parameters with eccentricity included; see, e.g., the first analyses of GW190521 \cite{LIGO-O3-GW190521-implications} and GW231123 \cite{LIGO-O4-GW231123}. 
In addition, there is an existing degeneracy between eccentricity and spin-precession, where both physical effects can mimic each other, which complicates the ability to make confident claims for eccentricity \cite{10.1093/mnras/stad031, 2026arXiv260102260T}.
Comparative analyses incorporating both eccentric waveform models and precessing waveform models, without neglecting the relativistic anomaly parameter, contribute towards mitigating potential systematic biases in parameter estimation studies \cite{2025PhRvD.112l3004P, 2023PhRvD.108l4063R}.
At the population level, neglecting eccentricity can also bias the recovery of population parameters when the underlying population contains systems with substantial eccentricities \cite{PhysRevD.110.063009}.

In this work, we perform parameter inference on 162 sources in the O3 and O4a observing runs of the LVK network, including all previously identified as confident candidates with false alarm rates (FAR) of less than one per year as well as most other detection candidates highlighted in these two runs. Unlike previous investigations, our sample is comprehensive: we analyze additional sources above this threshold. To assess the impact of eccentricity, we perform inferences using two state-of-the-art waveform models incorporating eccentricity (SEOBNRv5EHM and TEOBResumS-Dali). We compare these inferences to each other and to previously reported results. To mitigate the high computational cost associated with evaluating these waveforms \cite{2022ApJ...940..171R}, we employ the RIFT parameter inference pipeline, whose architecture provides generic methods to mitigate this cost \cite{Lange_2018,gwastro-PENR-RIFT-GPU,gwastro-RIFT-Update,gwastro-RIFT_FinerNet}.

The remainder of this paper is organized as follows. In Section \ref{sec:methods}, we describe the parameter estimation method, \texttt{asimov} infrastructure, and analysis settings for events that are re-analyzed in this work. In Section \ref{sec:results} we survey our parameter inferences, illustrating both qualitatively and quantitatively that these two waveform models largely draw similar conclusions and briefly comment on the extent to which our eccentric inferences resemble previously-reported quasi-circular posteriors. We also identify and discuss events with at least modest indications of eccentricity or otherwise noteworthy features in Section \ref{sec:sub:ecc}, as well as events with large waveform systematics in Section \ref{sec:sub:systematics}. In addition, in Section \ref{sec:gw200129}, we examine the impact of different analysis settings on GW200129\_065458. In Section \ref{sec:discuss} we discuss how our results compare with previously published work that attempt to characterize eccentricity for these same events with other analysis frameworks, sometimes using one of the waveform models adopted here. We conclude with a summary of our findings in Section \ref{sec:conclude}. In Appendix \ref{app:A}, we supply additional plots of events with eccentric features and two events for which we find no evidence of eccentricity including a previously identified eccentric candidate, GW190701\_203306.

\section{Methods}
\label{sec:methods}

\subsection{Parameter inference with RIFT and asimov}

Coalescing compact binaries, such as binary black holes, with quasi-circular orbits can be characterized by 15 parameters with an additional parameter used to describe tidal deformability for binary neutron stars (BNSs) or neutron star-black holes (NSBHs). The eight intrinsic parameters ($\lambda$) refer to the physical properties of the individual components within each binary including the component masses, spin magnitudes, spin tilts, and azimuthal angles. The seven extrinsic parameters ($\theta$) refer to the spacetime location and orientation of the binary system: right ascension, declination, luminosity distance, inclination, polarization, orbital phase, and coalescence time. Two additional intrinsic parameters, eccentricity and mean anomaly, are inferred to describe an eccentric aligned-spin binary system.

RIFT comprises a two-step iterative process that compares gravitational-wave observations $d$ to predicted gravitational-wave signals $h(\bm{\lambda}, \bm\theta)$. In the first step, for a large number of $\bm{\lambda}$ values, RIFT computes a marginal likelihood
\begin{equation}
 {\cal L}{({\bm \lambda})}\equiv\int  {\cal L}_{\rm full}(\bm{\lambda},\bm\theta )p(\bm\theta )d\bm\theta
\end{equation}
where ${\cal L}_{\rm full}(\bm{\lambda},\bm{\theta})$ is the likelihood of the gravitational-wave signal in the multi-detector network; see \cite{Pankow_2015, Lange_2018} for a further detailed description. 

During the second step, RIFT approximates ${\cal L}(\bm{\lambda})$ based on the accumulated marginal likelihood evaluations $(\bm{\lambda},{\cal L})$. This approximation is used to construct the detector-frame posterior distribution
\begin{equation}
\label{eq:post}
p_{\rm post}(\bm{\lambda})=\frac{{\cal L}(\bm{\lambda} )p(\bm{\lambda})}{\int d\bm{\lambda} {\cal L}(\bm{\lambda} ) p(\bm{\lambda} )}.
\end{equation}
where the prior $p(\bm{\lambda})$ denotes the prior distribution on the intrinsic parameters.
After intrinsic inference is complete, extrinsic parameters $\theta$ are generated for each intrinsic sample via Monte Carlo.

\subsection{Event data and analysis settings}

We identified 162 events from GWTC-2 \cite{LIGO-O3-O3a-catalog}, GWTC-2.1 \cite{LIGO-O3-O3a_final-catalog}, GWTC-3 \cite{LIGO-O3-O3b-catalog}, and GWTC-4 \cite{LIGO-O4a-cbc-catalog_results} for analysis.  We primarily targeted the same sample used for global population inference for GWTC-3 and GWTC-4: a false alarm rate (FAR) of less than one per year. That said, given tentative indications of eccentricity among candidates not included in that sample, we also analyze 11 additional O3 events: GW190426$\_$190642, GW190514$\_$065416, GW190916$\_$200658, GW190926$\_$050336, GW191113$\_$071753, GW191126$\_$115259, GW191204$\_$110529, GW191219$\_$163120, GW200210$\_$092254, GW200220$\_$124850, and GW200306$\_$093714. Table \ref{tab:events} provides a comprehensive list of the 162 events studied here, along
with our inferences of key properties. 

To coordinate our workflows, we use the \texttt{asimov} engine \cite{asimov-paper,gwastro-mergers-rift_asimov_O3-Fernando2024}, which provides suitable analysis settings (e.g., analysis time interval, frequency ranges, sampling rates, interferometer
lists, and frame file versions).  As a result, our analyses benefit from all glitch mitigation performed on these events.
We perform inference on either the publicly-released data from GWOSC (Gravitational Wave Open Science Center) \cite{2021SoftX..1300658A}, or (to facilitate use of settings adopted in prior work) bit-equivalent internal data. 
Within this framework, we generate independent noise power spectral density (PSD) estimates for each event candidate using Bayeswave \cite{2015CQGra..32m5012C,2015PhRvD..91h4034L,2019PhRvD.100j4004C}.
Unless otherwise noted, we adopt the same event time, mass, mass ratio, and spin magnitude priors adopted in previous analyses, as disseminated via the \texttt{asimov} database. 
In this work, for brevity, we do not marginalize over calibration uncertainty, given its minimal impact.

We modify the default \texttt{asimov} settings in five physically pertinent ways. First, we require the spins to be aligned with the orbital angular
momentum, adopting the ``z prior''  for each aligned spin consistent with previous work \cite{gwastro-RIFT-Update}.
Second, we have ubiquitously employed sampling rates no less than 4096 Hz, as recommended given RIFT's internal discrete time sampling, rather than adopting previously-adopted sampling rates.
Third, we include eccentricity (and mean anomaly) as parameters, using uniform priors in both parameters over $e\in
[0,e_{\rm max}]$ 
and $\ell \in [0,2\pi]$ respectively, where $e_{\rm max}=0.9$ for events appearing in GWTC-3 and $e_{\rm max}=0.5$ for
events appearing in GWTC-4.  In practice, after performing an analysis over the extended region $e\in   [0,0.9]$, we truncate our posteriors to $e\in[0,0.5]$ to better reflect the domain of validity of the two waveform models used. Fourth, for each event in GWTC-4, we utilize a starting frequency of 13.33 Hz, while for older  events (including very high-mass or low-mass events) the starting frequency reflects previous analysis choices.  Unless otherwise noted, we do not standardize our eccentricity definition at any reference frequency between events (or waveform models). Additionally, without loss of generality for these nonprecessing models, we set the starting frequencies and references frequencies equal for each event.
Table \ref{tab:events} has columns providing the starting frequency and sampling rate adopted for each event.
Our analysis settings also correctly handle windowing for the short-duration events for which this factor has previously impacted results \cite{2025CQGra..42w5023T}.  
Finally, for O3 events, we adopt a ``Euclidean'' distance prior $p(d_L)\propto d_L^2$, as is widely adopted in early GWTC data releases.  As a result, our inferred source-frame masses are expected to differ slightly
from previously-published results in part insofar as they adopt different priors. For O4a events, following GWTC-4, we adopt a uniform merger rate in comoving volume and time. 

Since RIFT provides likelihoods, not merely posterior samples, we could efficiently reassess each of our analyses with
different priors, notably including a log-uniform prior on eccentricity or ``pinhole'' prior \cite{2026arXiv260220110R} which better reflects astrophysical context.
In the data release accompanying our publication, we provide these likelihoods and posterior samples to facilitate this and other downstream reanalyses \cite{malagon_2026_20386916}.

A recent reanalysis of GW200105\_162426 has already performed a systematic investigation of that source with multiple waveforms using RIFT, while adopting superior settings to the default choices adopted here \cite{2025arXiv250812460J}, including higher sampling rates and more conservative internal analysis settings (e.g. grid sizes).  In this work, we therefore do not provide an equally comprehensive analysis of this candidate, focusing instead on the remainder of the population.

\input{event_table_reruns_log.tex}


\subsection{Waveform models and eccentricity convention}
Parameter inferences are performed using SEOBNRv5EHM
\cite{2025PhRvD.112d4038G} and TEOBResumS-Dali
\cite{
2020PhRvD.101j1501C,%
2024PhRvD.110h4001N,%
2025PhRvD.111f4050N}.
SEOBNRv5EHM and TEOBResumS-Dali are aligned-spin, eccentric inspiral-merger-ringdown (IMR) waveform models based on the effective-one-body (EOB) formalism \cite{PhysRevD.59.084006, PhysRevD.62.064015}.
Each model is distinctly calibrated to quasi-circular numerical relativity (NR) simulations and validated against eccentric NR simulations.
SEOBNRv5EHM extends the quasi-circular SEOBNRv5HM model in the SEOBNR family to include eccentric and hyperbolic orbits \cite{PhysRevD.108.124037, PhysRevD.108.124035, 2025PhRvD.112d4038G}.
This model contains the dominant $(\ell,|m|)= (2,2)$ waveform mode and higher-order modes (2,1), (3,3), (3,2), (4, 4), and (4, 3). 
TEOBResumS-Dali is the eccentric configuration of the TEOBResumS \cite{PhysRevD.98.104052} family of EOB waveform models.
This waveform family incorporates higher-order modes up to $\ell= |m| = 5$, excluding the $m=0$ modes.
Both waveform models define eccentricity through a Keplerian parametrization quoted at a reference frequency, resulting in a gauge-dependent definition.
With SEOBNRv5EHM, the user can specify a starting frequency, expressed as a dimensionless orbit-averaged orbital frequency, and the reference frequency is set equal.
For TEOBResumS-Dali, the reference frequency can be defined at periastron, apastron, or the mean of the two.
SEOBNRv5EHM describes the binary's position in the orbit with the relativistic anomaly, while TEOBResumS-Dali utilizes the mean anomaly.

In both cases, we employ all available higher-order modes supplied by these models. In our parameter inferences, given the substantial dynamic range in mass for our binary sources, for simplicity our headline numbers and plots show eccentricity evaluated at the starting frequency for each event, using the internal conventions provided with each code.

\section{Results}
\label{sec:results}

\begin{figure*}
    \centering
    \includegraphics[width=\textwidth]{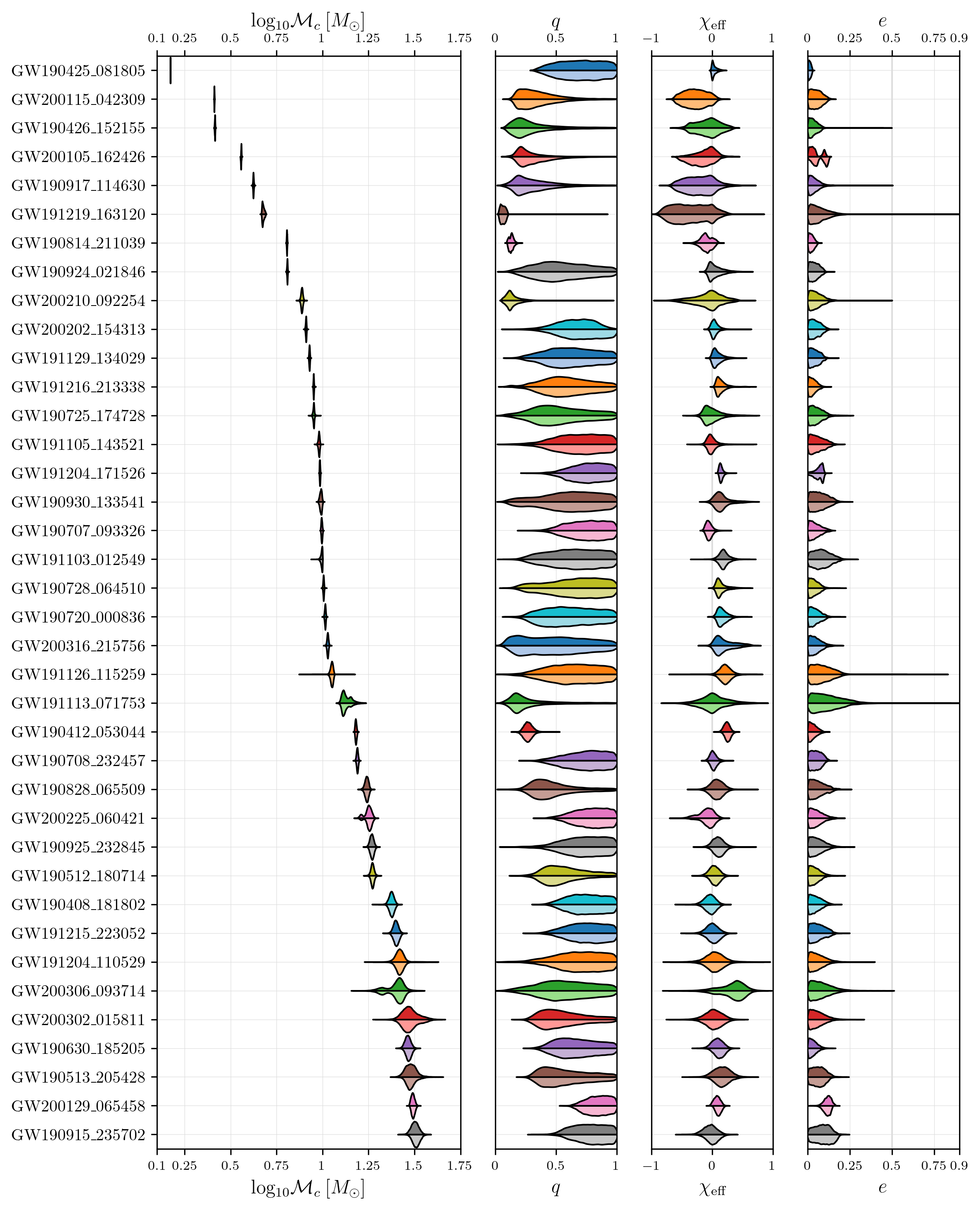}
    \label{fig:violin_A}
\end{figure*}

\begin{figure*}
    \centering
    \includegraphics[width=\textwidth]{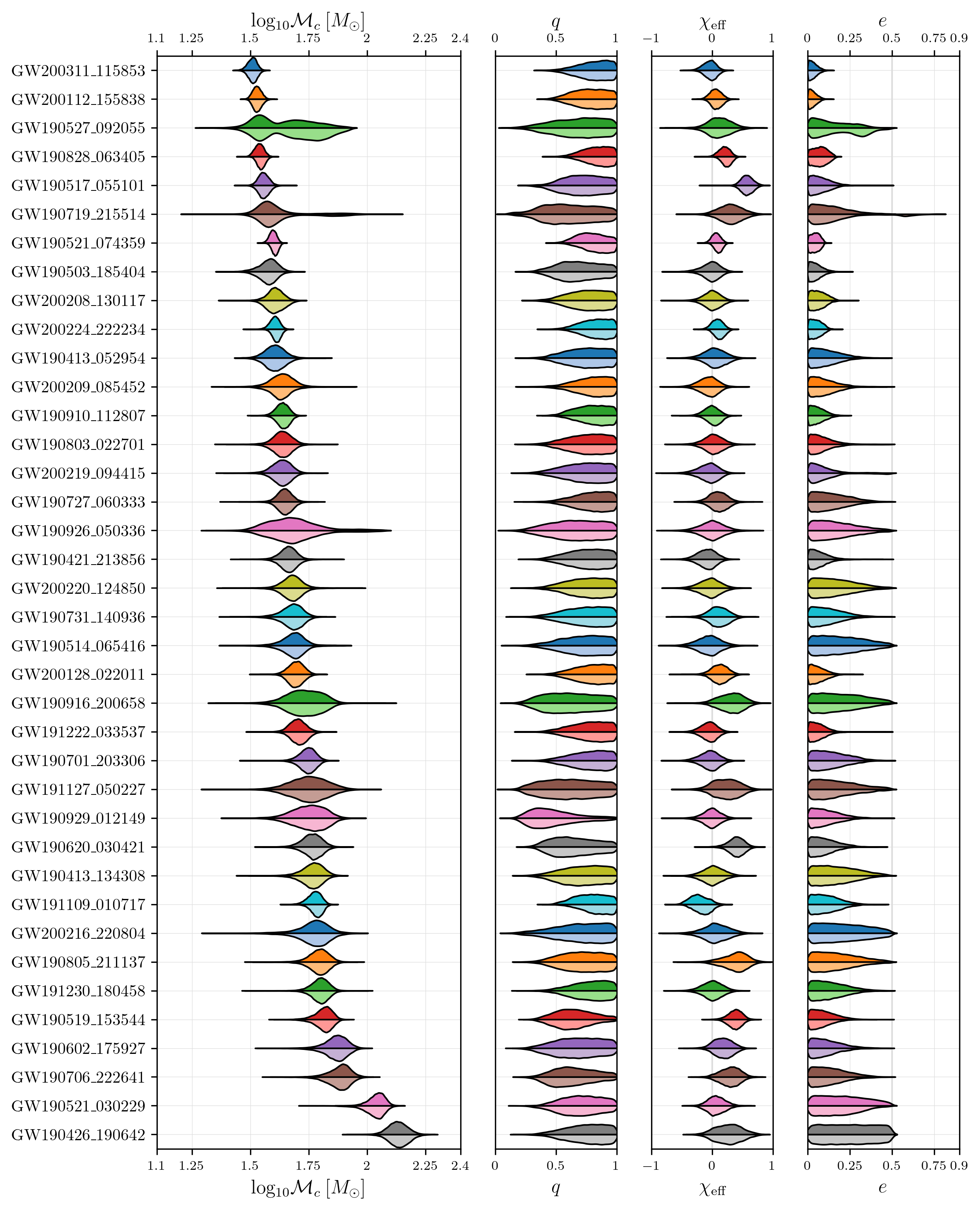} 
    \caption{The marginal probability distributions for the log-10 chirp mass $\mathcal{M}_c$ in the detector frame, mass ratio $q$, effective spin $\chi_\text{eff}$, and eccentricity $e$ of the events from O3 analyzed in this work. The upper half of each violin plot reflects the results obtained from the SEOBNRv5EHM analysis while the lower half shows the TEOBResumS-Dali result.}
    \label{fig:violin_B}
\end{figure*}

\begin{figure*}
    \centering
    \includegraphics[width=\textwidth]{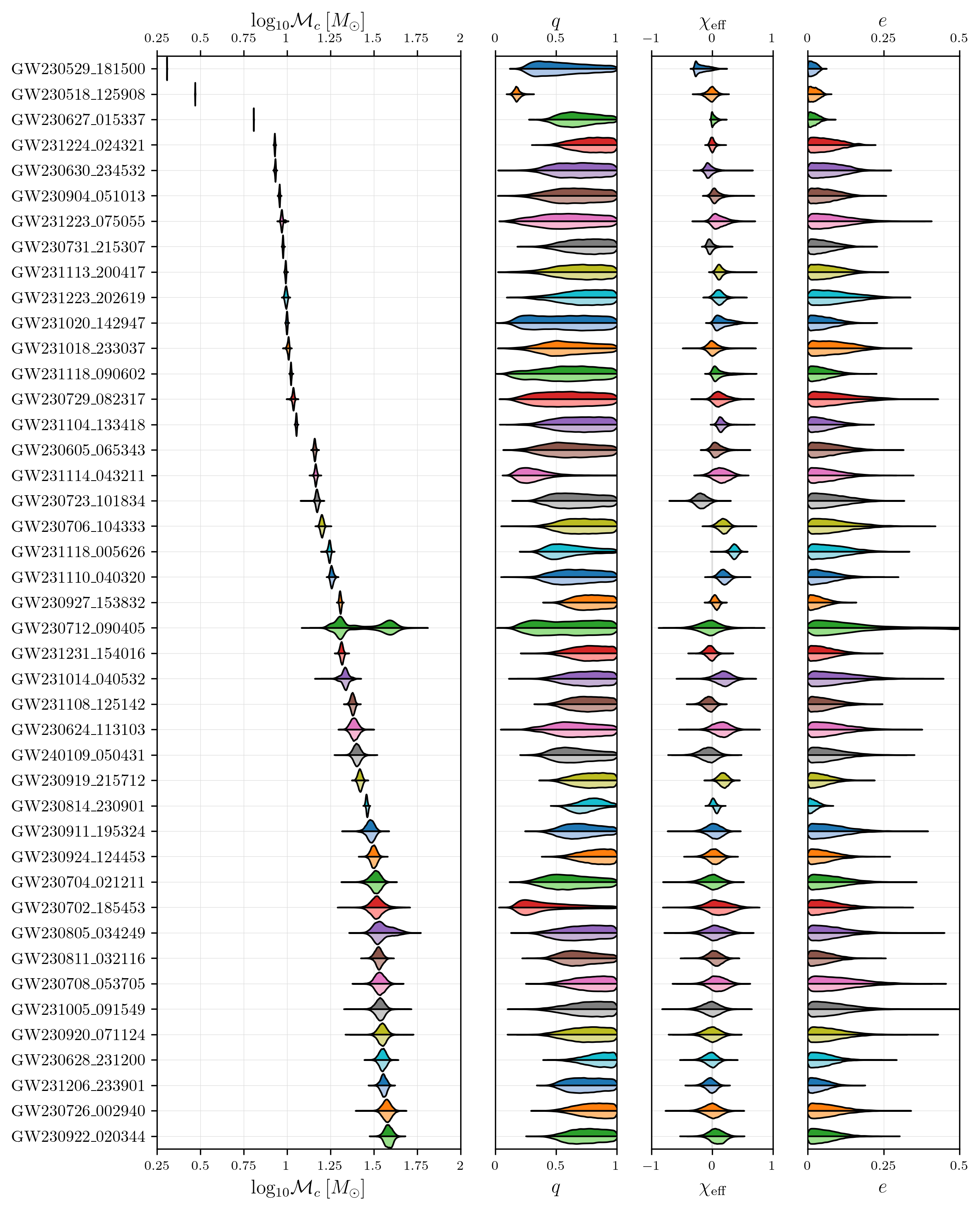} 
    \label{fig:violin_C}
\end{figure*}
\begin{figure*}
    \centering
    \includegraphics[width=\textwidth]{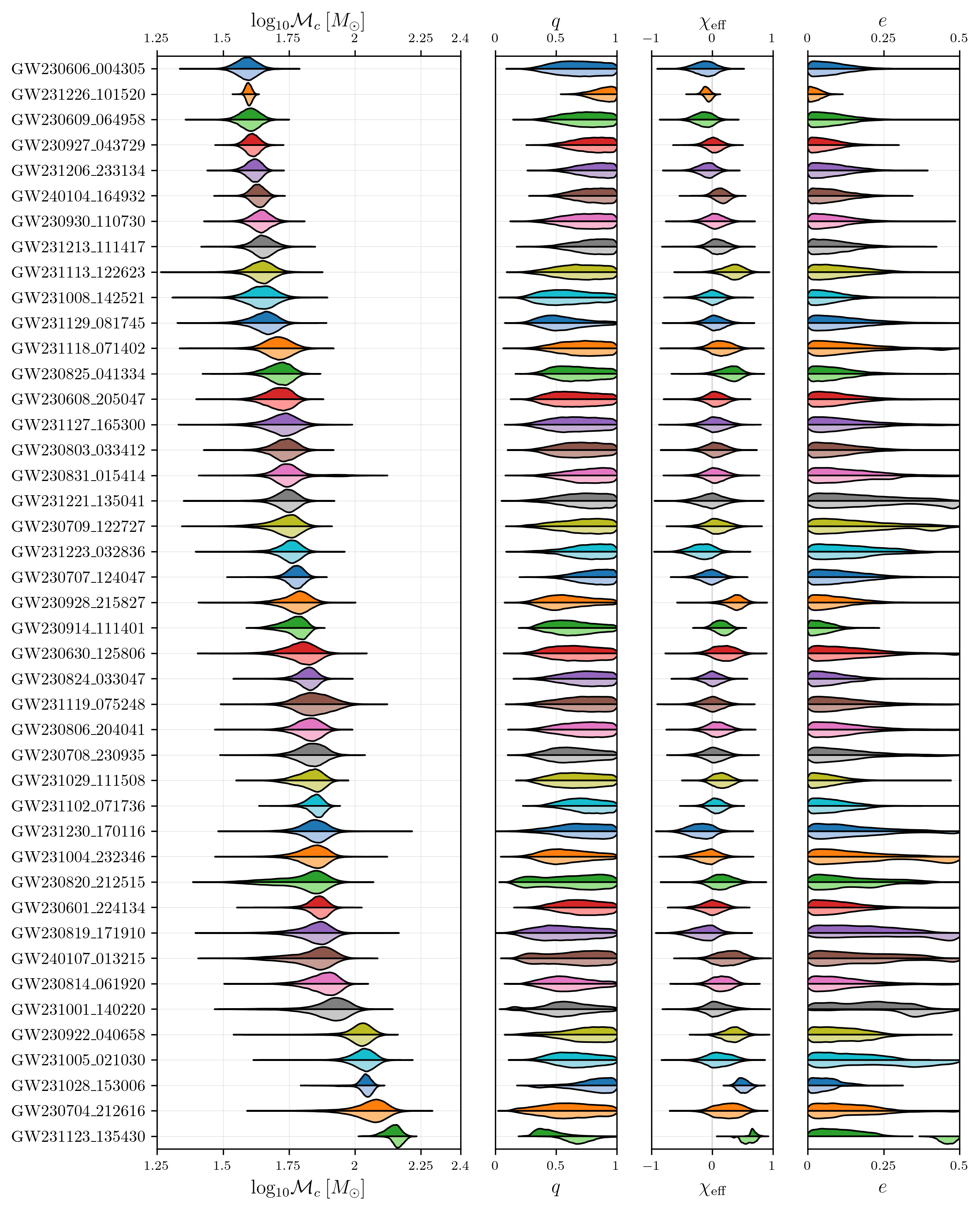} 
    \caption{The marginal probability distributions for the log-10 chirp mass $\mathcal{M}_c$ in the detector frame, mass ratio $q$, effective spin $\chi_\text{eff}$, and eccentricity $e$ of the events from O4a analyzed in this work. The upper half of each violin plot reflects the results obtained from the SEOBNRv5EHM analysis while the lower half shows the TEOBResumS-Dali result.}
    \label{fig:violin_D}
\end{figure*}

\subsection{Survey of Inferences with eccentricity}
\label{sec:survey}
Figures \ref{fig:violin_B} and \ref{fig:violin_D} show the results of our inferences, providing one-dimensional marginal
probability distributions for the (log of the) chirp mass, mass ratio,  effective spin,
and orbital eccentricity, for both waveform approximations SEOBNRv5EHM (top) and TEOBResumS-Dali (bottom). In this
figure, we plot the eccentricity as each code defines it as a parameter, not attempting to standardize between them. For
convenience in visualization, these results are sorted by increasing chirp mass. Most events are very consistent with
zero eccentricity (i.e., no peak away from $e=0$) and show good agreement between the two waveform
approximations. However, some events show modest to strong indications of orbital eccentricity, relative to a uniform
eccentricity prior.
Table \ref{tab:events} quantitatively summarizes the inferences presented in this figure. 

For almost all events and parameters, the posterior distributions derived using these two waveform approximations agree extremely well with each other. Qualitatively speaking, the most substantial differences between these two approximations arise when one infers a substantial eccentricity. The two waveform models agree somewhat more closely when restricting eccentricity to their expected regime of
validity ($e<0.5$).
For many events, one visually striking difference between the two models is the tendency for the eccentricity
distribution inferred with TEOBResumS-Dali to assign a small but substantial probability of significant eccentricity,
near or above $e\simeq 0.5$, where the model is expected to be less reliable.

\subsection{Quantifying Systematics between waveforms}

Jensen-Shannon (JS) divergences are calculated to quantify differences between the marginal posterior distributions inferred with each waveform. Given two posterior density distributions $p(x)$ and $q(x)$, the JS divergence is defined as
\begin{equation}
    \text{JSD}(p|q) = \frac{1}{2}D_\text{KL}\left(p |\frac{p+q}{2} \right) + \frac{1}{2}D_\text{KL}\left(q | \frac{p+q}{2} \right), 
\end{equation}
where $D_\text{KL}$ refers to the Kullback-Leibler (KL) divergence that is evaluated by applying kernel density estimation (KDE) towards the posterior samples. JS divergence values closer to 0 indicate a strong consistency between two posterior distributions compared to larger values near $\ln 2$ which reflect disagreement.

\begin{figure*}
    \centering
    \includegraphics[width=0.9\textwidth]{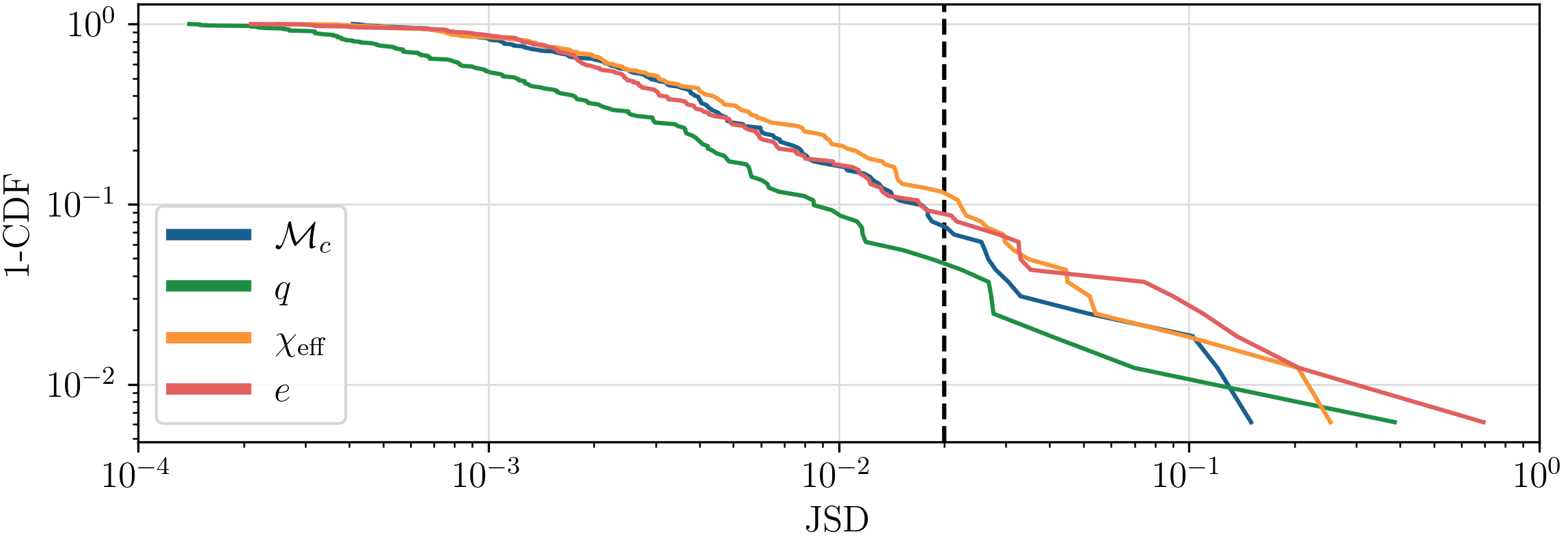}
    \includegraphics[width=0.9\textwidth]{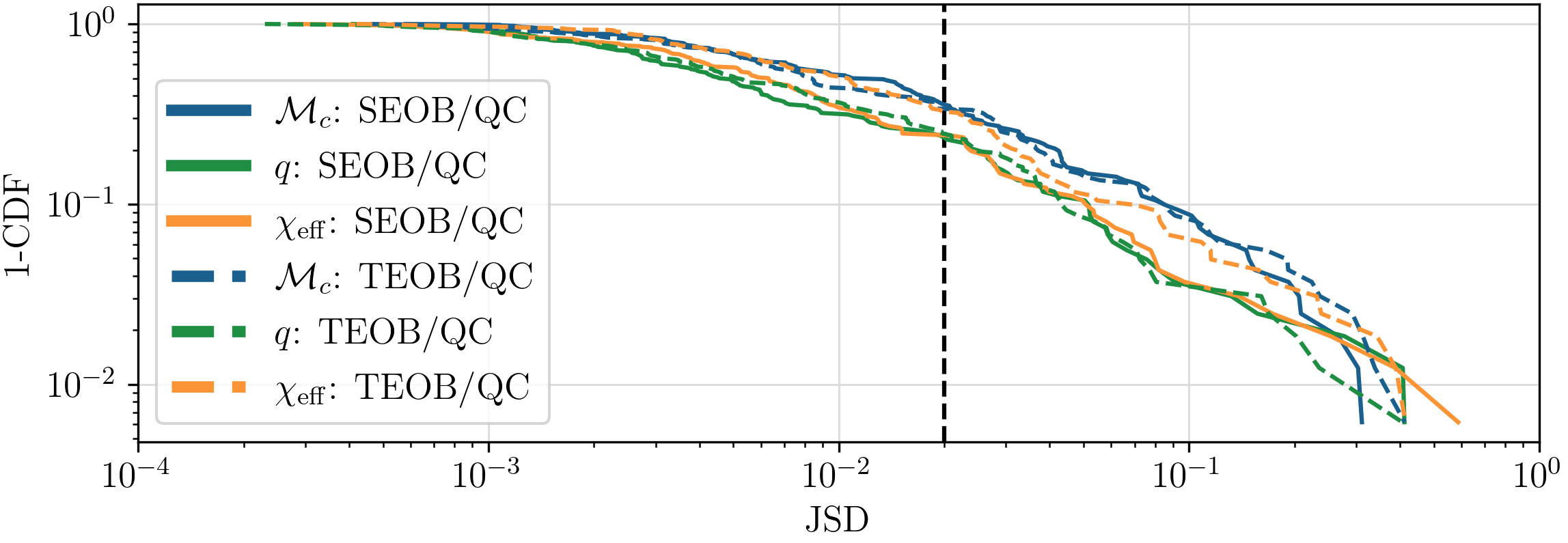}
    \caption{Fraction of events with JS divergences above a threshold, versus the JS divergence (JSD) using the entire set of event analyses shown for the chirp mass $\mathcal{M}_c$, mass ratio $q$, effective spin $\chi_\text{eff}$, and eccentricity parameters (top). Only samples with eccentricity values less than 0.5 are represented. The JS divergences compare SEOBNRv5EHM and TEOBResumS-Dali (top), and each of these waveforms with the public GWOSC data that uses quasi-circular (QC) waveforms (bottom), making no attempt to correct for any differences in distance prior between runs. The vertical dashed (black) line in the histogram indicates a JS divergence of 0.02.}
    \label{fig:hist}
\end{figure*}

For each event, we compare the JS divergences between SEOBNRv5EHM and TEOBResumS-Dali, as well as between each of these
waveforms and the latest versions of the GWOSC results which utilize quasi-circular (QC) waveform models \cite{ligo_scientific_collaboration_and_virgo_2022_6513631, ligo_scientific_collaboration_and_virgo_2023_8177023, ligo_scientific_collaboration_and_virgo_2025_17014085}.
Figure \ref{fig:hist} shows the fraction of events with a JS divergence for a specific one-dimensional posterior
distribution is above a specified threshold, showing this cumulative distribution for  different
pairs of waveforms on different panels (SEOB/TEOB on the top) and line styles (QC versus SEOB or TEOB on the bottom).
The vertical dashed line corresponds to a
JS divergence of 0.02, above which differences between one-dimensional marginal distributions are easily identified by
eye.
These cumulative distributions show that overall the event analyses are broadly consistent between each waveform, with
the exception of a few large JS divergences. The JS divergences between SEOBNRv5EHM and TEOBResumS-Dali appear to be
more consistent than either of these waveforms compared to the QC waveforms for these parameters -- keeping in mind that the QC analyses were performed with slightly different analysis settings (e.g., PSD) and code.
For the eccentric consistency check shown in the top panel, the curve corresponding to eccentricity exhibits the most frequent large JS divergences, relative to the other parameters shown here.

\begin{figure*}
    \centering
    \includegraphics[width=\textwidth]{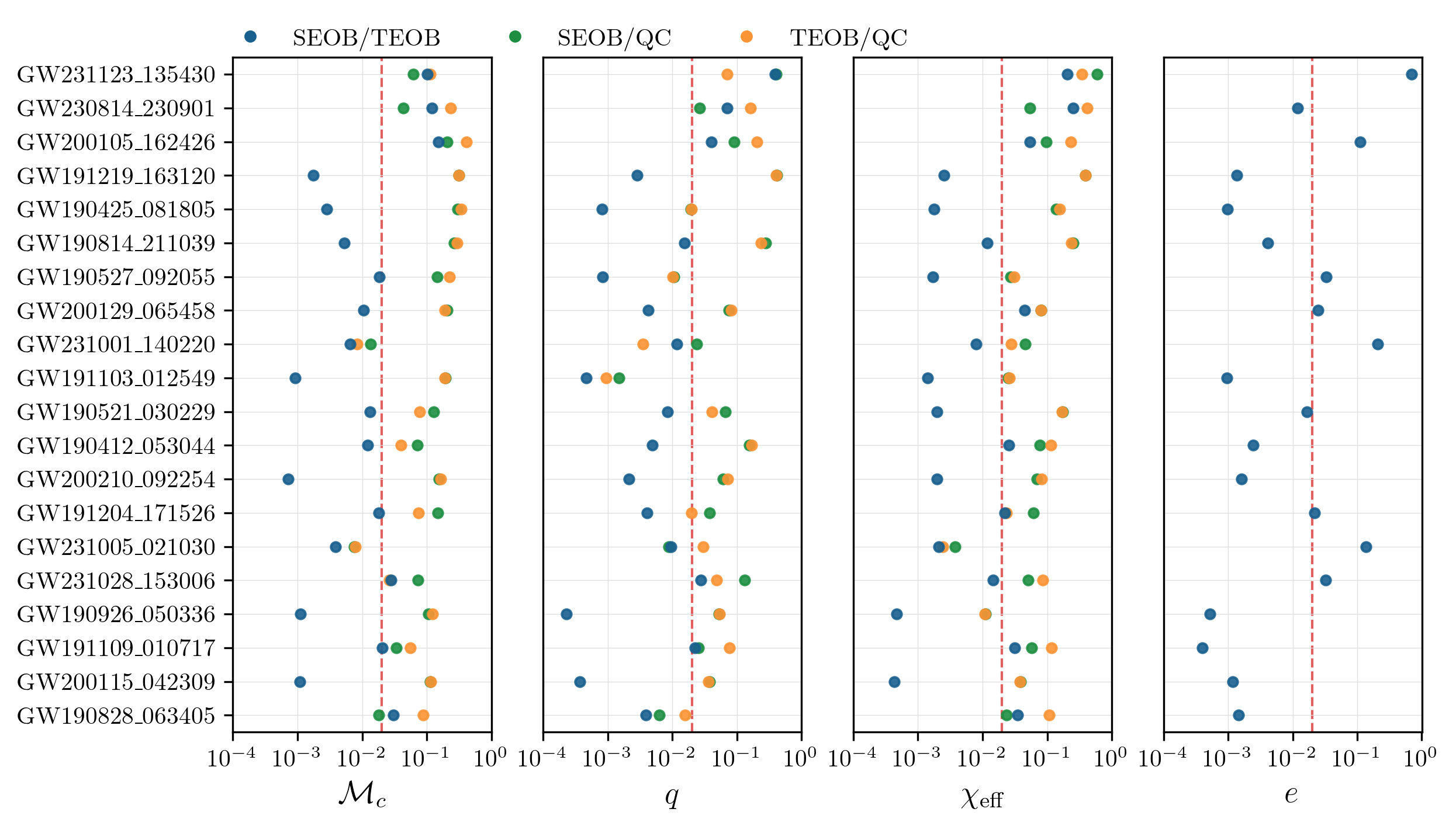}
    \caption{The JS divergences between the one-dimensional marginalized posteriors of the chirp mass $\mathcal{M}_c$, mass ratio $q$, effective spin $\chi_\text{eff}$, and eccentricity $e$ parameters for specific events with the largest JS divergences. Only samples with eccentricity values less than 0.5 are represented. The dashed vertical (red) line denotes a JS divergence of 0.02. The comparisons are between SEOBNRv5EHM and TEOBResumS-Dali (blue), and each of these waveforms with the public GWOSC data that uses quasi-circular (QC) waveforms (green, orange), again making no attempt to enforce consistency of distance prior between GWOSC and our analysis.}
    \label{fig:jsd}
\end{figure*}

Figure \ref{fig:jsd} presents a per-event, per-parameter JS divergence visualization that highlights substantial
differences between each waveform model comparison.  Focusing on the SEOB/TEOB comparisons in blue, only a handful of
events exhibit substantial systematics: notably, a few events whose conclusions about eccentricity differ between
waveforms --  the well-explored GW231123; GW200129, discussed later; GW200105, a NSBH binary
also with indications of eccentricity -- as well as a few other events that exhibit more modest differences in other
parameters (e.g., GW230814, GW231001, GW231123).  We will describe these differences in more detail for each set of
events later in this work in Sections \ref{sec:sub:ecc} and  \ref{sec:sub:systematics}.

As expected from the violin plots, these JS divergences quantitatively demonstrate that the SEOBNRv5EHM and TEOBResumS-Dali waveform models draw largely the same conclusions for every event and every parameter. However, for a select sample of high mass events, the eccentricity distribution differs in the TEOBResumS-Dali analysis compared to the SEOBNRv5EHM analysis. The TEOBResumS-Dali results contained a high eccentricity feature in the range $e\in[0.5, 0.8]$, which is likely pertaining to a waveform systematic for a few high mass systems in O3 that had an eccentricity prior limit extending to 0.9. Similar behavior is observed in O4 events where there is a peak in the eccentricity distribution approaching the upper prior limit ranging from $e\in[0.3, 0.5]$. In Ref. \cite{2024ApJ...972...65I}, regardless of reference frequency, a waveform degeneracy is observed at high eccentricities for high total detector-frame mass systems which is consistent with our interpretation of this waveform systematic. To summarize, our parameter inferences are quite consistent with one another and with previously published work, while inconsistencies arise in regions of the parameter space where more significant waveform systematics are  expected. 

%


\subsection{Events with substantial eccentricity}
\label{sec:sub:ecc}
\input{special_event_table.tex}

We compute a Bayes factor between the quasi-circular and eccentric hypotheses using a Savage-Dickey density ratio for every event and waveform model to quantify the support for eccentricity.
Assuming a uniform prior on eccentricity, $e\in[0, 0.5]$, the prior density at $e=0$ is $p(e=0)=2$.
The Bayes factor is calculated as:
\begin{equation*}
    \log_{10} \mathcal{B} = - \log_{10} \left(\frac{p(e=0|d)}{p(e=0)}\right) = -\log_{10} \left( 0.5 p(e=0|d)\right),
\end{equation*}
where $p(e=0|d)$ is the posterior density evaluated at the quasi-circular hypothesis given the data $d$.
In these calculations, we restrict the prior range on eccentricity for GWTC-3 events to be below $0.5$, to mitigate the high eccentricity
waveform systematics seen from the TEOBResumS-Dali analyses with large eccentricity.
This posterior density is estimated using a boundary-corrected KDE constructed by reflection at $e=0$. A positive log-10 Bayes factor value indicates evidence for the eccentric hypothesis. We employ a threshold $\log_{10} \mathcal{B} > 0.1$ to suggest some support for eccentricity.

\begin{figure*}
    \centering
    \includegraphics[width=\textwidth]{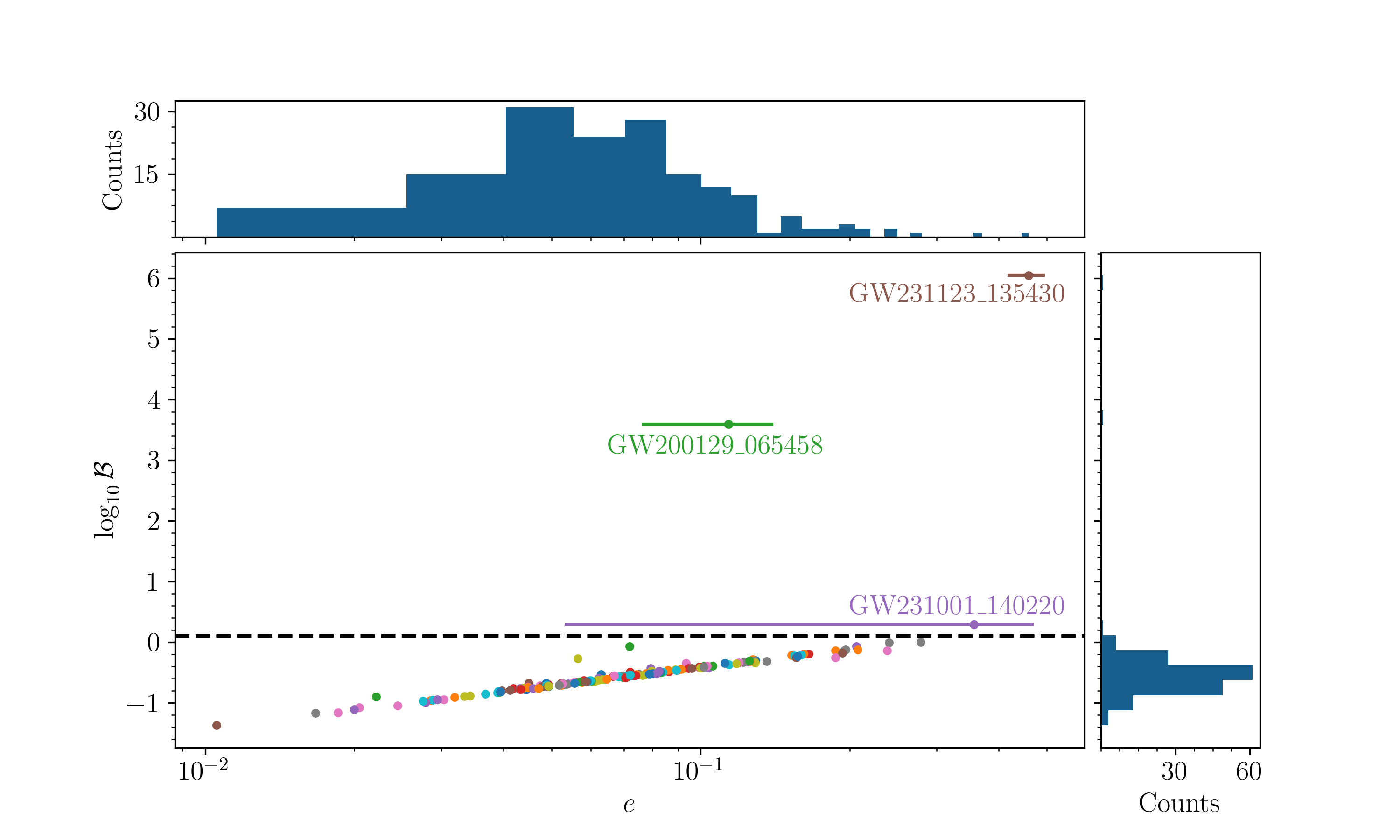} 
    \caption{The (maximum) log-10 Bayes factors compared against the median eccentricity $e$ values for every event analysis represented in Table \ref{tab:events}; we note that our table excludes GW200105.  For each event, we show the largest of the two Bayes factors for eccentricity. Each point is color coded by event. For events with a positive $\log_{10} \mathcal{B}$, the $90\%$ credible intervals for eccentricity are plotted and the corresponding event name is labeled. The dashed (black) line indicates our threshold used to denote minimum support for eccentricity ($\log_{10} \mathcal{B} > 0.1$).}
    \label{fig:lnb}
\end{figure*}

Because the Savage-Dickey ratio relies on posterior support at $e=0$, this approach will break down for event candidates with extremely substantial support for nonzero eccentricity. In that scenario, we use an alternative method, relying on RIFT's ability to directly integrate marginal likelihoods, to quantify extremely strong support for nonzero eccentricity. This Bayes factor is expressed as the ratio of the evidences for two hypotheses, eccentric ($\mathcal{H_\text{2}}$) and quasi-circular ($\mathcal{H_\text{1}}$):
\begin{equation*}
    \log_{10} \mathcal{B} = \log_{10} \frac{Z(d|\mathcal{H_\text{2}})}{Z(d|\mathcal{H_\text{1}})},
\end{equation*}
where the evidence refers to $Z(d|\mathcal{H})={\int d\bm{\lambda} {\cal L}(\bm{\lambda} ) p(\bm{\lambda}|\mathcal{H})}$. We use this method to compute the log-10 Bayes factor for GW200129\_065458 and GW231123\_135430 which have at least one result displaying strong support for eccentricity.
Figure \ref{fig:lnb} shows the highest log-10 Bayes factors obtained from either SEOBNRv5EHM or TEOBResumS-Dali analysis and the median eccentricity value associated with that event analysis. The vast majority of events yielded negative log-10 Bayes factors, with the exception of: GW200129\_065458, GW231001\_140220, and GW231123\_135430.

Table \ref{tab:spec_events} shows events identified via visual inspection to have eccentric features, as estimated by at least one of our parameter inferences. The log-10 Bayes factors calculated from the one-dimensional marginal eccentricity distributions of each analysis are also provided. 

\begin{figure*}
    \centering
    \includegraphics[width=0.8\textwidth]{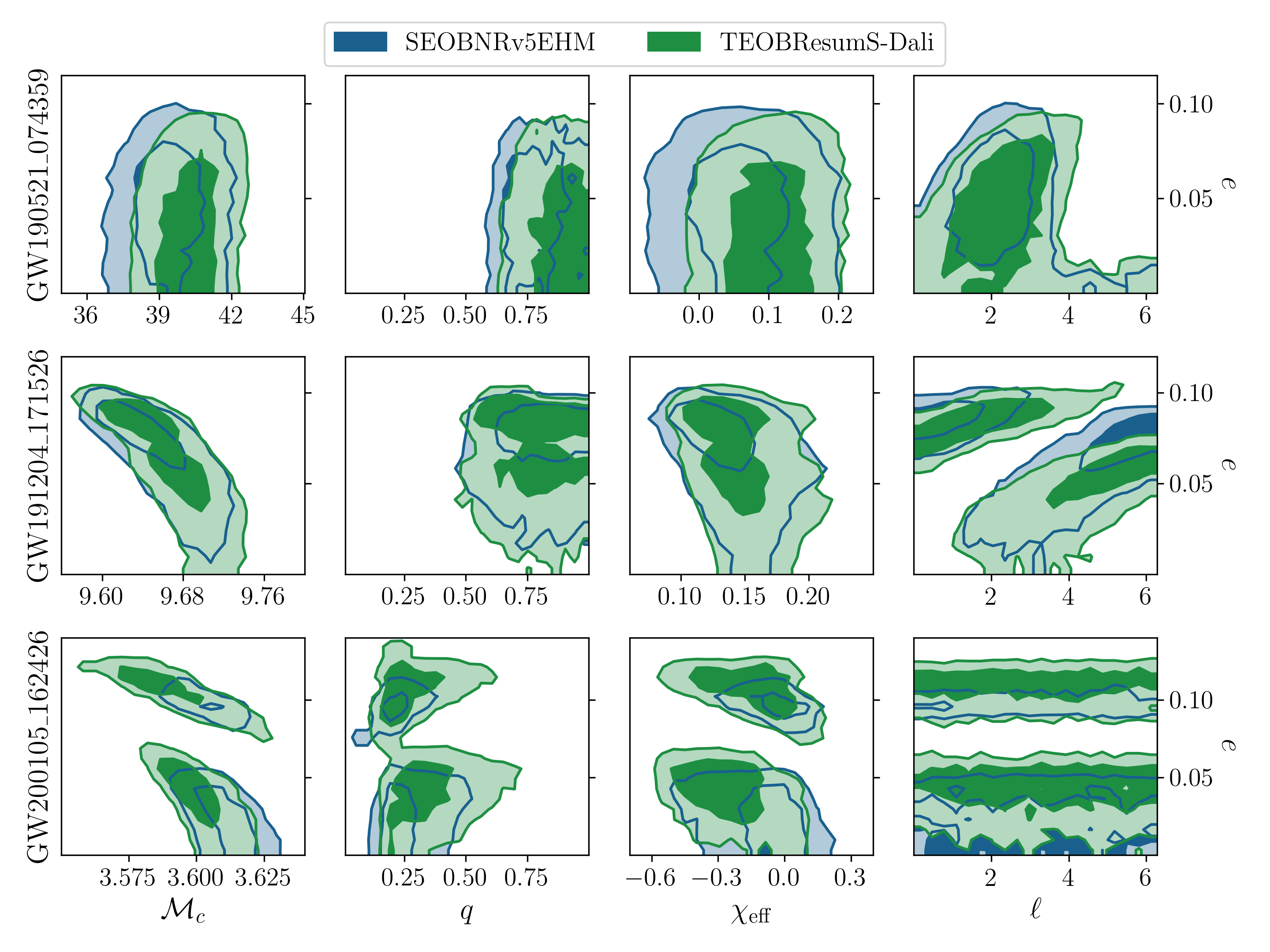}
    \caption{Two-dimensional marginal posterior distributions for eccentricity $e$ and select parameters: chirp mass $\mathcal{M}_c$, mass ratio $q$, effective spin $\chi_\text{eff}$, and mean anomaly $\ell$. Each row corresponds to an event consistent with eccentricity in our results and obtained negative log-10 Bayes factors for each analysis. The results from the SEOBNRv5EHM (blue) and TEOBResumS-Dali (green) analyses are shown. The contours represent the 50$\%$ and 90$\%$ credible intervals.}
    \label{fig:contours}
\end{figure*}

GW200105$\_$162426 has previously been identified as a candidate which may have modest indications of eccentricity. GW190521$\_$074359 and GW191204\_171526 have not been previously identified as eccentric candidates. For GW191204\_171526, we find nonzero eccentric features at $e\sim0.07$ with both waveform models. We also find a secondary nonzero peak in the eccentricity distribution at $e\sim0.3$ in the TEOBResumS-Dali analysis for GW190527\_092055, which we discuss in Appendix \ref{app:A}. As seen in Figure \ref{fig:violin_B} and the log-10 Bayes factors provided in Table \ref{tab:spec_events}, however, our analyses do not provide compelling evidence for eccentricity for any of these events. Figure \ref{fig:contours} illustrates the two-dimensional marginal posterior distributions for these four events in greater detail. Even though these marginal distributions hint at interesting structures involving eccentricity and its correlations with other parameters, the evidence remains marginal.
GW200105$\_$162426 has more extensively been analyzed and shown to have consistent evidence of eccentricity. For example, analyses using the waveform pyEFPE, which incorporates both eccentricity and precession, and IMRPhenomTEHM, an aligned-spin eccentric IMR waveform model, have both obtained similar results \cite{2025ApJ...995...47P, 2025arXiv250315393M}. 
Our conclusions contrast with previously presented results obtained with other methods, analysis settings, and waveform models. For example, although we do not identify GW190521$\_$030229 as an eccentric candidate, previous investigations have previously found evidence of eccentricity for GW190521$\_$030229 when using the SEOBNRE waveform with \textsc{Bilby} while more recently lacked support when utilizing SEOBNRv4EHM with \texttt{DINGO} \cite{2022ApJ...940..171R, 2025PhRvD.112j4045G}. 

\begin{figure*}
    \centering
    \begin{minipage}[b]{0.49\textwidth}
        \centering
        \includegraphics[width=\textwidth]{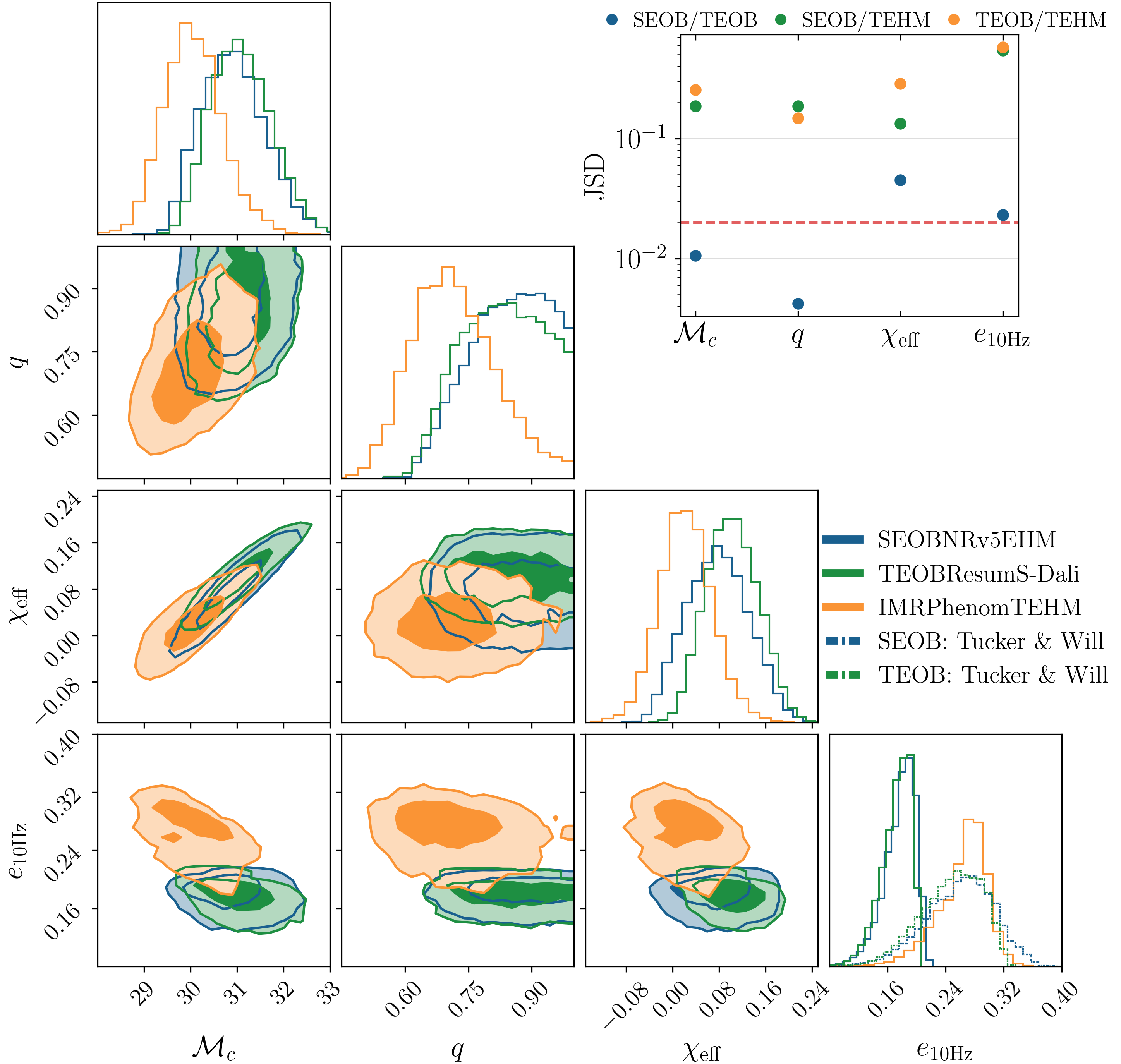}
        \label{fig:200129_js}
    \end{minipage}
    \hfill
    \begin{minipage}[b]{0.5\textwidth}
        \centering
        \includegraphics[width=\textwidth]{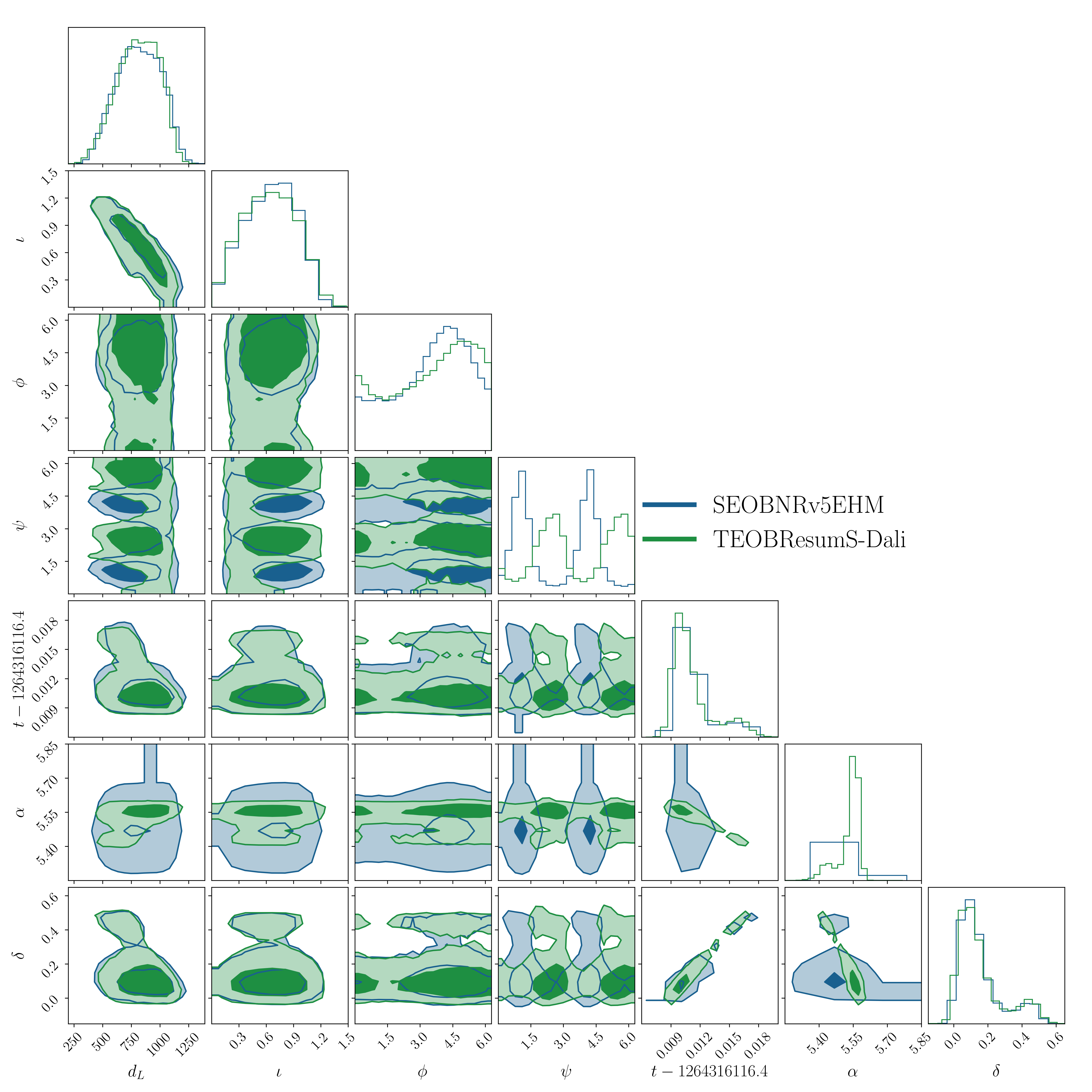}
        \label{fig:200129_ext}
    \end{minipage}
    \caption{\textit{Left}: One- and two-dimensional posterior distributions for a subset of intrinsic parameters (chirp mass $\mathcal{M}_c$, mass ratio $q$, effective spin $\chi_\text{eff}$, eccentricity $e$, and mean anomaly $\ell$) for GW200129$\_$065458 using SEOBNRv5EHM (blue), TEOBResumS-Dali (green), and an independent IMRPhenomTEHM (orange) result. The IMRPhenomTEHM result is a publicly available result obtained using a uniform prior on eccentricity, at a reference frequency of 10 Hz, and utilizes data from the \texttt{gw$\_$subtract} method \cite{2025ApJ...995...47P, planas_llompart_2025_15576673}. Given that our results are inferred at a reference frequency of 20 Hz, our eccentricity values are evolved according to Eq.~\ref{eq:ecc}. We also apply the PN-corrected analytic map for eccentricity given in Eq.~3.3 of Tucker \& Will \cite{PhysRevD.104.104023}, and the resulting posteriors (dash-dotted) are shown in the one-dimensional eccentricity panel. The upper right panel shows the JS divergences (JSD) for comparisons between each of these waveforms for each of these parameters: SEOBNRv5EHM and TEOBResumS-Dali (blue), and each with IMRPhenomTEHM (green, orange). The dashed (red) line corresponds to a JSD value of 0.02.
    \textit{Right}: One- and two-dimensional posterior distributions for extrinsic parameters (luminosity distance $d_{L}$, inclination $\iota$, phase $\phi$, polarization $\psi$, geocentric time $t$, right ascension $\alpha$, and declination $\delta$) for this event using our results. The contours correspond to the $50\%$ and $90\%$ confidence intervals for both panels in this figure.}
    \label{fig:200129_full}
\end{figure*}

We identify distinct evidence of nonzero eccentricity recovered across both waveform models for GW200129$\_$065458 and with positive log-10 Bayes factors obtained. For SEOBNRv5EHM, we find at 90\% confidence $e=0.12^{+0.03}_{-0.04}$ and $\log_{10} \mathcal{B}=3.12$. When analyzing these events with TEOBResumS-Dali, we measure the eccentricity of GW200129$\_$065458 to be within $e=0.11^{+0.03}_{-0.04}$ at 90\% confidence with $\log_{10} \mathcal{B}=3.59$. For these measurements at 20 Hz, nine orbits are completed in the signal. Figure \ref{fig:200129_full} shows a dual panel consisting of a comparison of our GW200129$\_$065458 analyses to an independent analysis using IMRPhenomTEHM with a uniform prior on eccentricity on the left, and the corresponding inferred extrinsic parameters from our analyses on the right \cite{2025ApJ...995...47P}. The data used for the IMRPhenomTEHM analysis utilizes the \texttt{gw$\_$subtract} method and infers eccentricity at a reference frequency of 10 Hz \cite{2025ApJ...995...47P, planas_llompart_2025_15576673}. Given that our analyses infer eccentricity at a reference frequency of 20 Hz, for the purpose of comparison, we evaluate our eccentricity values at a standard value (10 Hz) that matches the reference frequency in the IMRPhenomTEHM analysis. We adopt a formula from Ref. \cite{2025arXiv250722862M} for the eccentricity evolution at the lowest post-Newtonian (PN) order, 
\begin{equation}
\begin{aligned}
    e_{10\text{Hz}}
    &= e\left(\frac{f_{\text{ref}}}{10\text{Hz}}\right)^{19/18} \\
    &\times \left[1+\frac{19}{9}e\left(1-\left(\frac{f_{\text{ref}}}{10\text{Hz}}\right)^{19/18}\right)\right],
\end{aligned}
\label{eq:ecc}
\end{equation}
derived from Ref. \cite{1964PhRv..136.1224P} which is valid for small eccentricities in the inspiral regime.
Additionally, we apply the PN-corrected eccentricity map for non-spinning inspiralling binaries given in Eq.~3.3 of Tucker \& Will \cite{PhysRevD.104.104023}.
Our results reveal a significant nonzero feature of small eccentricity. From the left panel, as evident by the posterior distributions obtained by applying Eq.~\ref{eq:ecc}, the SEOBNRv5EHM and TEOBResumS-Dali results are much more consistent than when compared to the independent IMRPhenomTEHM analysis. 
In contrast, when evolved using the PN-corrected analytic map from Tucker \& Will \cite{PhysRevD.104.104023}, the SEOBNRv5EHM and TEOBResumS-Dali eccentricity posteriors more closely resemble the IMRPhenomTEHM result in the one-dimensional eccentricity posterior panel.
The JS divergences indicate that the TEOBResumS-Dali result appeared to disagree with the IMRPhenomTEHM results the most, roughly across the majority of the intrinsic parameters shown. For the extrinsic parameters, the SEOBNRv5EHM and TEOBResumS-Dali results are generally in agreement, in the right panel, except for the polarization parameter. Therefore, we consider these results as evidence consistent with eccentricity. 

Thus far, GW200129$\_$065458 has had previous claims of eccentricity at a reference frequency of 10 Hz. Using deglitched data from the \texttt{gw\_subtract} method, Gupte et al. \cite{2025PhRvD.112j4045G} use SEOBNRv4EHM and reported a mean $e=0.27^{+0.10}_{-0.12}$ while another subsequent analysis in Planas et al. \cite{2025PhRvD.112l3004P} finds a median $e=0.24^{+0.06}_{-0.07}$ with IMRPhenomTEHM. Both measurements are consistent with each other but are slightly higher than the result we obtain. We further discuss the impact of analysis settings on GW200129$\_$065458 in Section \ref{sec:gw200129}.

\begin{figure}
    \centering
    \includegraphics[width=0.5\textwidth]{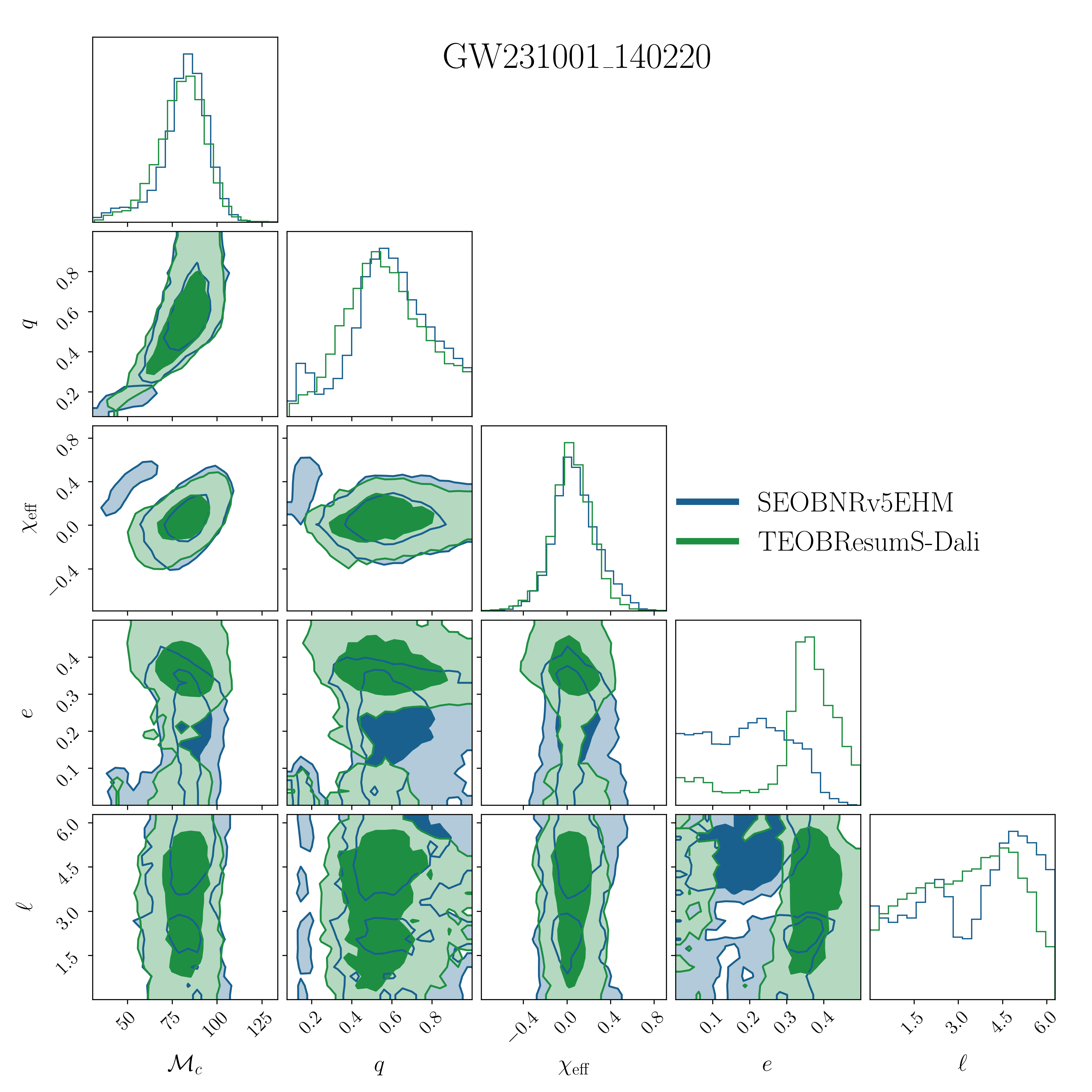}
    \caption{One- and two-dimensional posterior distributions for a subset of intrinsic parameters (chirp mass $\mathcal{M}_c$, mass ratio $q$, effective spin $\chi_\text{eff}$, eccentricity $e$, and mean anomaly $\ell$) for GW231001$\_$140220 using SEOBNRv5EHM (blue) and TEOBResumS-Dali (green).}
    \label{fig:231001}
\end{figure}

GW231001$\_$140220 also contains a prominent feature of moderate eccentricity solely in the TEOBResumS-Dali result. This eccentric peak is characterized by $e=0.36^{+0.11}_{-0.30}$ for GW231001$\_$140220 at 90\% confidence with $\log_{10} \mathcal{B}=0.30$. Figure \ref{fig:231001} shows some railing approaching the maximum prior limit ($e = 0.5$) in the eccentricity distribution for the TEOBResumS-Dali analysis. The high eccentricity feature may be attributed in part to waveform systematics in the TEOBResumS-Dali approximation, mentioned in the previous section, for higher mass systems ($\mathcal{M}_c=81.27^{+18.50}_{-25.52}$) with large eccentricities. Conversely, SEOBNRv5EHM result shows a broader eccentricity distribution with no strong evidence of eccentricity, having a negative Bayes factor of $\log_{10} \mathcal{B}=-0.12$. GW231001$\_$140220 has yet to be identified as a candidate for eccentricity.
Xu et al. \cite{2025arXiv251219513X} did not find that this event satisfied their criterion for eccentric support with IMRPhenomTEHM. More recently, Gupte et al. \cite{Gupte_2026} have used SEOBNRv5EHM and found positive Bayes factors for their aligned spin analysis of GW231001$\_$140220, but argued their positive Bayes factor was not significant enough when considered in isolation.

\begin{figure}
    \centering
    \includegraphics[width=0.5\textwidth]{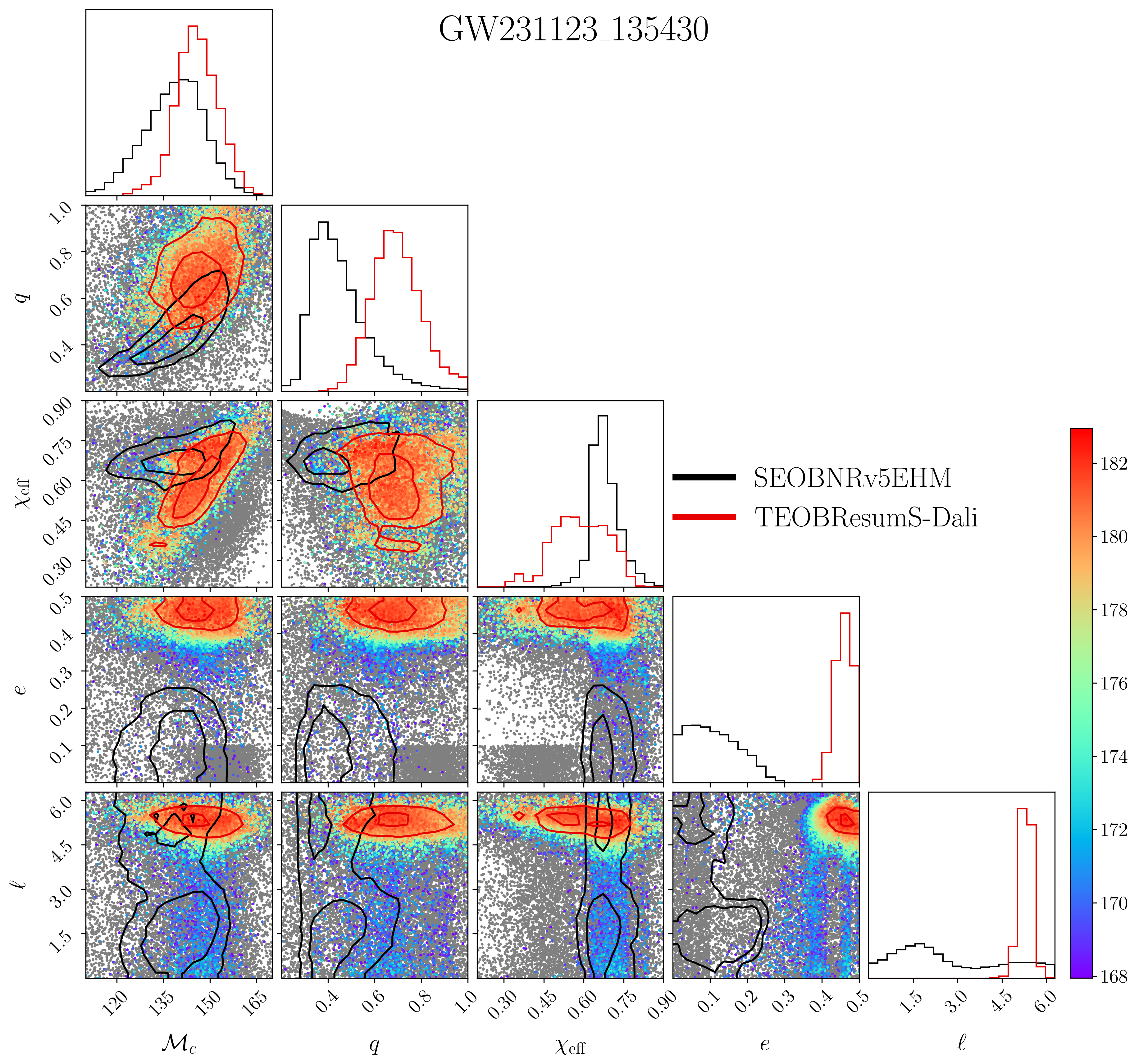}
    \caption{One- and two-dimensional posterior distributions for the chirp mass $\mathcal{M}_c$, mass ratio $q$, effective spin $\chi_\text{eff}$, eccentricity $e$, and mean anomaly $\ell$ parameters for GW231123$\_$135430 with SEOBNRv5EHM (black) and TEOBResumS-Dali (red). The $\ln$ marginal likelihoods from the TEOBResumS-Dali result are overplotted. The contours denote the 50\% and 90\% confidence intervals.}
    \label{fig:231123_compare}
\end{figure}

There is an additional event which yielded a positive log-10 Bayes factor, GW231123\_135430. For GW231123\_135430, there are significant observed differences in the intrinsic parameters and most significantly in the eccentricity distributions between the two waveform models. Therefore, the log-10 Bayes factor is calculated differently for this event. Figure \ref{fig:231123_compare} compares the results from both waveform models and shows the $\ln$ marginal likelihoods of the TEOBResumS-Dali result. The SEOBNRv5EHM analysis resulted in a marginal log-10 Bayes factor suggesting no indication for eccentricity with $\log_{10} \mathcal{B}=0.04$. In comparison, the TEOBResumS-Dali analysis has a narrow eccentricity distribution that rails against the upper limit on eccentricity at $e=0.5$, resulting in $\log_{10} \mathcal{B}=6.05$ and a median $e=0.46^{+0.03}_{-0.04}$ at $90\%$ confidence. Due to the marginal SEOBNRv5EHM log-10 Bayes factor that is less than 0.1, we do not consider this event to have evidence of eccentricity. Waveform systematics for this event are further discussed in the next section. Thus far, Xu et al. \cite{2025arXiv251219513X} has utilized IMRPhenomTEHM with a larger uniform eccentricity prior up to 0.65, measuring a higher median eccentricity at 10 Hz ($e=0.57^{+0.03}_{-0.05}$ at 90\% confidence intervals) and finding their strongest aligned-spin eccentric support with their Bayes factors for this event. Meanwhile, Gupte et al. \cite{Gupte_2026} found a positive Bayes factor for eccentricity in this event, but did not consider it significant enough to claim evidence for eccentricity.

Unfortunately, several of these events with the largest potential indication of eccentricity may have the most substantial systematics between conclusions derived using these two waveform models. In Appendix \ref{app:A}, we provide additional figures for events with eccentric features, such as GW200129\_065458 and GW231001\_140220, and discuss a few additional events we do not consider as eccentric.

\subsection{Events with substantial systematics}
\label{sec:sub:systematics}

We identify a few events to have significant systematics between the SEOBNRv5EHM and TEOBResumS-Dali waveform models in our analyses:  GW230814\_230901, GW231001\_140220, and GW231123\_135430. This selection is based on the JS divergences computed between the posterior distributions for intrinsic parameters inferred from each waveform, with at least one JS divergence above 0.2, and supported by visual inspection. Figures \ref{fig:violin_D} and \ref{fig:jsd} capture these distinct differences for these select events particularly in the chirp mass, mass ratio, effective spin, and eccentricity. 

\begin{figure}
    \centering
    \includegraphics[width=0.5\textwidth]{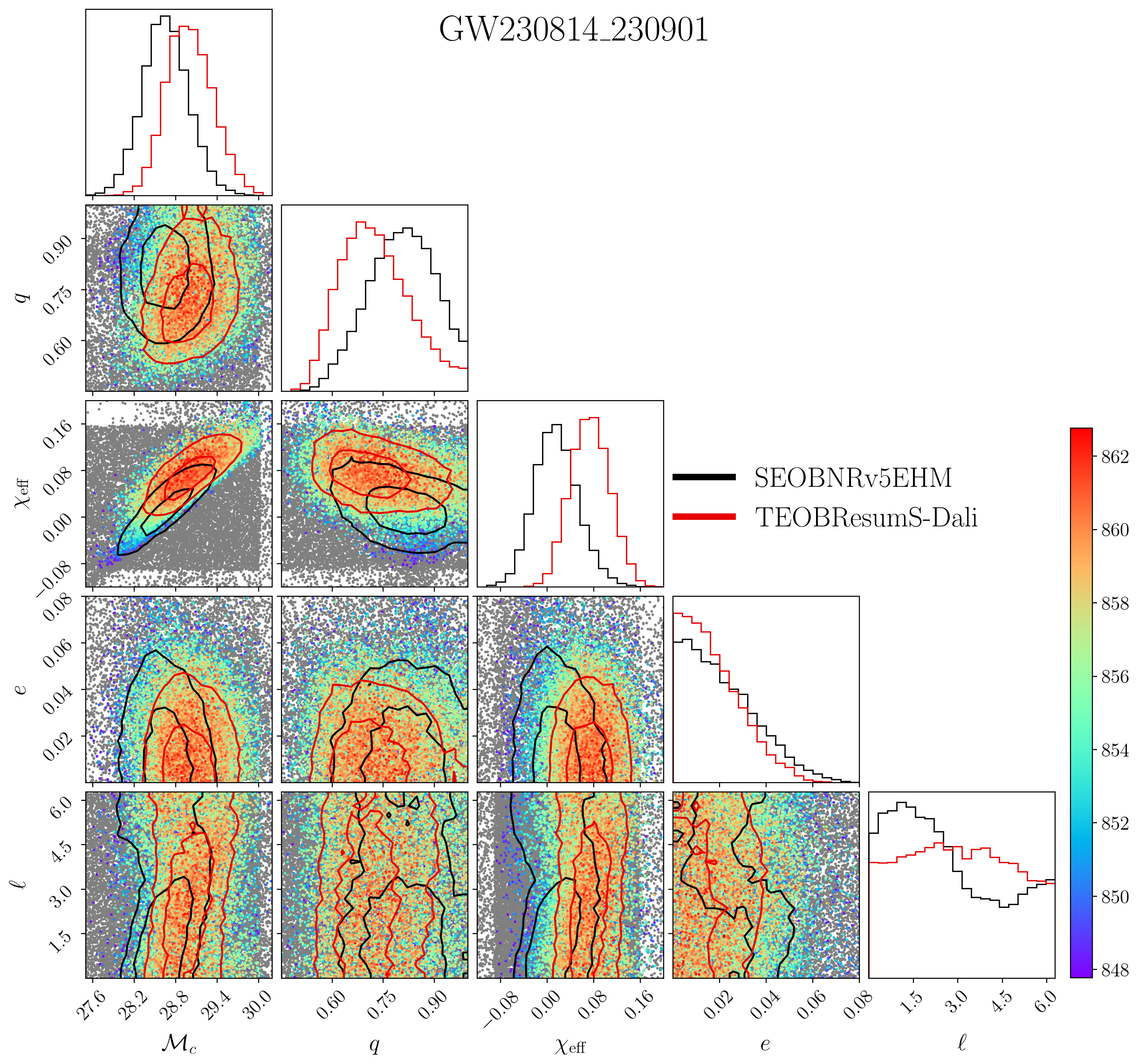}
    \caption{One- and two-dimensional posterior distributions for the chirp mass $\mathcal{M}_c$, mass ratio $q$, effective spin $\chi_\text{eff}$, eccentricity $e$, and mean anomaly $\ell$ parameters for GW230814$\_$230901 using SEOBNRv5EHM (black) and TEOBResumS-Dali (red). The $\ln$ marginal likelihoods from the TEOBResumS-Dali are overplotted. The contours denote the 50\% and 90\% confidence intervals.}
    \label{fig:230814}
\end{figure}

GW230814\_230901 stands out for the moderate differences in the posterior samples for the effective spin, mass ratio, and chirp mass parameters. Figure \ref{fig:230814} provides a comparison between the two analyses across each parameter with the likelihood values from the TEOBResumS-Dali result overplotted. The effective spin parameter shows the largest difference with a JSD of 0.253, while the chirp mass and mass ratio reflect smaller discrepancies with JSD values of 0.121 and 0.07, respectively. 

For GW231001\_140220, the analyses largely agree with each other except in the eccentricity. Figure \ref{fig:231001} shows how for the eccentricity, the TEOBResumS-Dali result has a distinct nonzero peak that slightly rails against the maximum prior while the SEOBNRv5EHM samples appear much more broadened. The differences in the posterior samples for the eccentricity are quantified with a moderate JSD of 0.206. SEOBNRv5EHM also shows a subtle secondary peak in the mass ratio, but it does not produce a significant JS divergence.

GW231123\_135430 appears to display the most substantial waveform systematics in our entire sample of events. Figures \ref{fig:violin_D} and \ref{fig:231123_compare} clearly demonstrate the contrast between the two analyses across each intrinsic parameter. Most significantly, from the eccentricity distributions we obtain a JSD of 0.693, as the SEOBNRv5EHM result rails against the lower bound while the TEOBResumS-Dali result concentrates towards the upper bound. The likelihoods obtained with TEOBResumS-Dali shown in  Figure \ref{fig:231123_compare} further highlights the substantial differences in interpretation for this event using different waveforms. We find moderate JSD values of 0.387 for the mass ratio and 0.205 for the effective spin. We also obtain a mild difference in the chirp mass with a JSD of 0.103.

When comparing each waveform to public, mixed GWTC-4 results (Figure \ref{fig:jsd}), we find consistent conclusions from the JS divergences that span from mild to substantial in the same pattern of intrinsic parameters. For GW230814\_230901, the TEOBResumS-Dali analysis showed a stronger disagreement with the GWOSC result across each parameter when compared to either waveform. We also find that in the effective spin for GW231001\_140220, the SEOBNRv5EHM result slightly differs from the GWOSC result, comparable to the difference with TEOBResumS-Dali. With GW231123\_135430, notably both of our analyses contrasted most with the GWOSC results in the effective spin parameter which is indicated by JSD values of 0.587 for SEOBNRv5EHM and 0.344 for TEOBResumS-Dali. As evident by Figure \ref{fig:jsd}, we also note that other events have systematic differences with generally moderate JS divergences when comparing SEOBNRv5EHM or TEOBResums-Dali to the GWOSC results. These events largely comprise of low mass systems including BNS and NSBH candidates (e.g. GW200105\_162426, GW190425\_081805, GW191219\_163120, and GW190814\_211039).

\subsection{GW200129 is sensitive to analysis settings}
\label{sec:gw200129}
\begin{figure*}
    \centering
    \begin{minipage}[b]{0.49\textwidth}
        \centering
        \includegraphics[width=\textwidth]{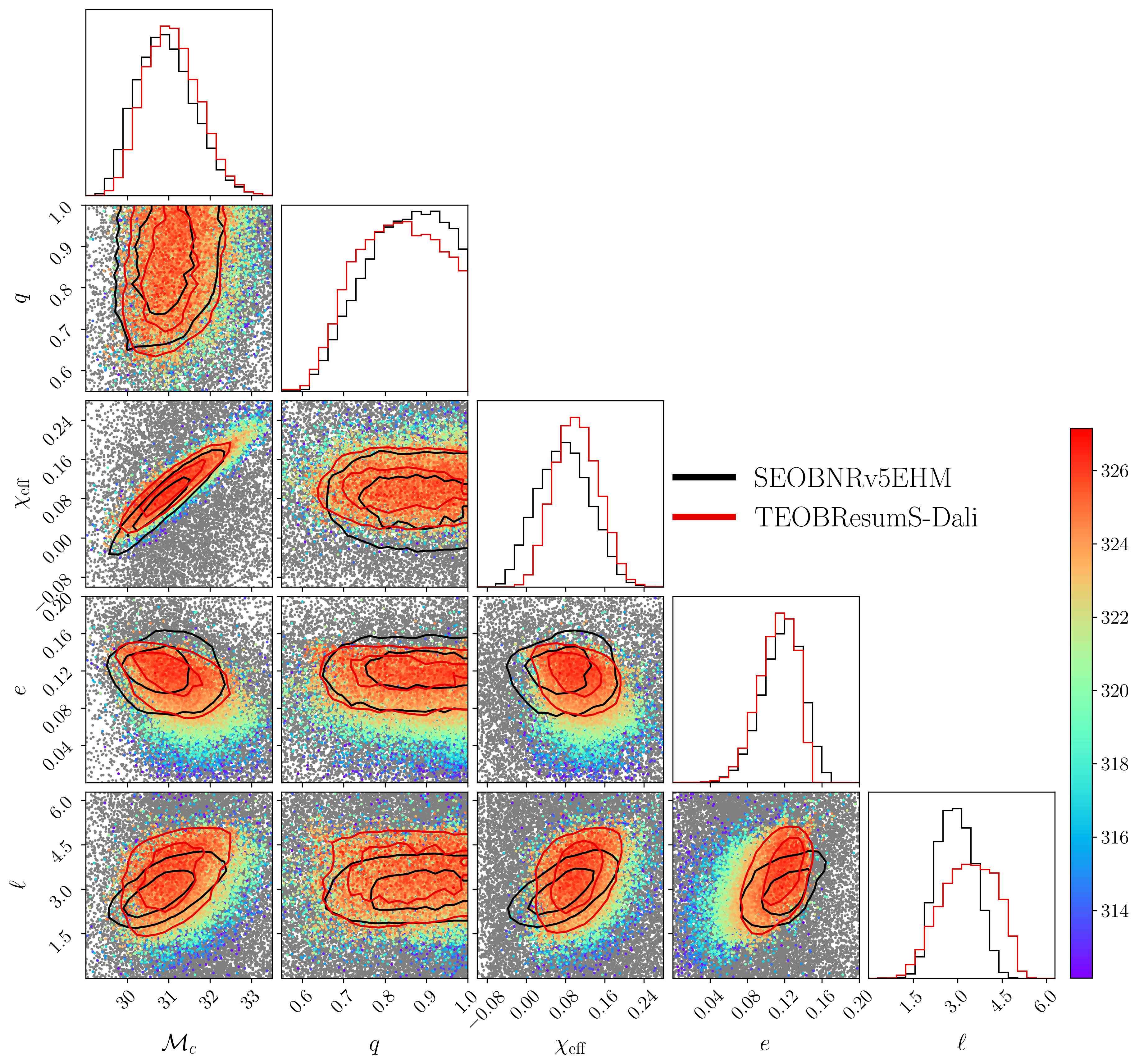}
    \end{minipage}
    \hfill
    \begin{minipage}[b]{0.49\textwidth}
        \centering
        \includegraphics[width=\textwidth]{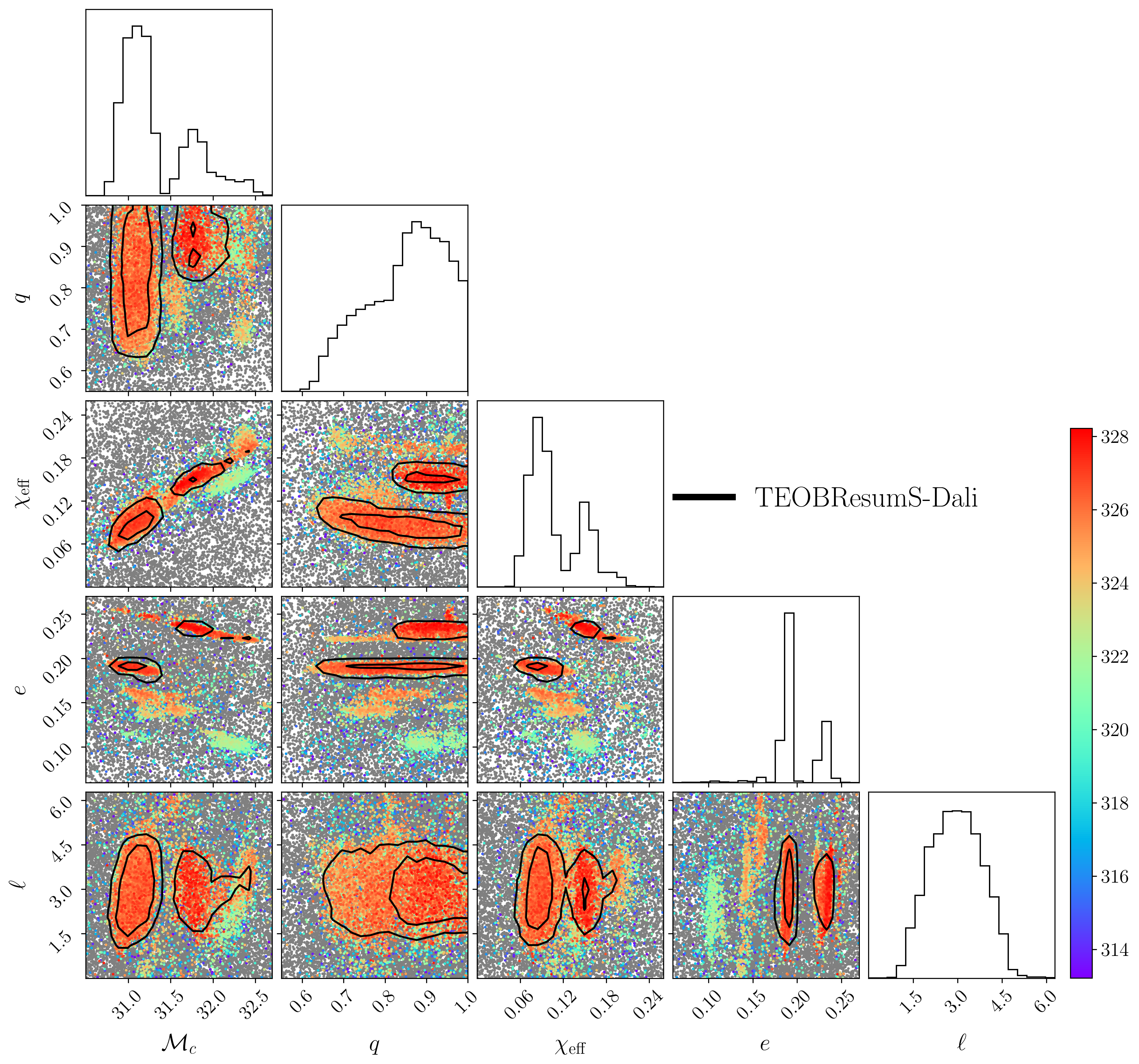}
    \end{minipage}
    \caption{One- and two-dimensional posterior distributions for a subset of intrinsic parameters (chirp mass $\mathcal{M}_c$, mass ratio $q$, effective spin $\chi_\text{eff}$, eccentricity $e$, and mean anomaly $\ell$) for GW200129$\_$065458 using SEOBNRv5EHM (black) and TEOBResumS-Dali (red). On the left is a comparison of our results at a reference frequency of 20 Hz and sampling rate of 4096 Hz. The right panel shows a comparison of runs at a reference frequency of 10 Hz, with an increased sampling rate of 16384 Hz. The colored points in both panels represent the values of the $\ln$ marginal likelihood from the TEOBResumS-Dali result. The contours correspond to the $50\%$ and $90\%$ confidence intervals for each joint distribution.}
    \label{fig:200129_panel}
\end{figure*}

Previous investigations of GW200129\_065458 have both identified potential indications of eccentricity and demonstrated that these conclusions can be sensitive to their analysis approach, emphasizing sensitivity to the deglitching strategy \cite{2025PhRvD.112j4045G, 2025PhRvD.112l3004P}. 
Notably, Payne et al. \cite{PhysRevD.106.104017} demonstrated the impact of glitch mitigation strategies on the inference of spin-precession in this event.
In this work, corroborating and building on previously published results by Wagner \cite{gwastro-mergers-WagnerThesisPhD}, we demonstrate our conclusions about GW200129\_065458 are sensitive to additional analysis choices, including the sampling frequency. Figure \ref{fig:200129_panel} has two panels, showing the likelihoods and posteriors obtained using the default settings ($f_\text{ref}$=20 Hz and sampling rate of 4096 Hz) as previously reported on the left; and on the right, a TEOBResumS-Dali result performed with a reference frequency at 10 Hz and sampling rate at 16384 Hz. The results clearly differ across parameters, mainly the TEOBResumS-Dali result in the panel on the right contains a bimodal shape in the chirp mass, effective spin, and eccentricity parameters at reference frequencies lower than our default analysis settings. The eccentricity posterior distributions are also more narrow compared to our default analysis settings result. 

\begin{figure}
    \centering
    \includegraphics[width=0.5\textwidth]{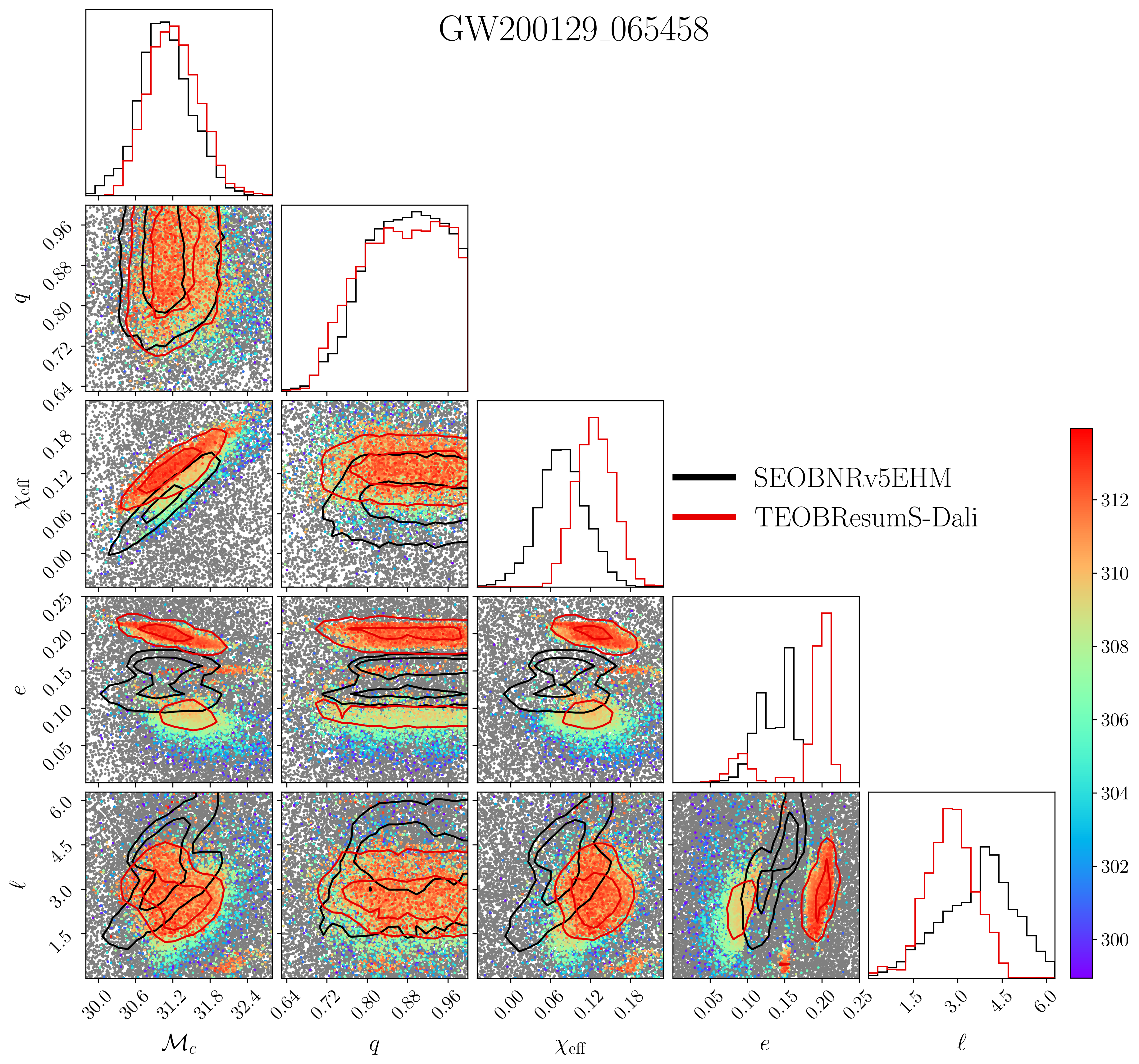}
    \caption{One- and two-dimensional posterior distributions for the chirp mass $\mathcal{M}_c$, mass ratio $q$, effective spin $\chi_\text{eff}$, eccentricity $e$, and mean anomaly $\ell$ parameters for GW200129$\_$065458 using a reference frequency of 15 Hz with SEOBNRv5EHM (black) and TEOBResumS-Dali (red). The $\ln$ marginal likelihoods from the TEOBResumS-Dali result are overplotted. The contours denote the 50\% and 90\% confidence intervals.}
    \label{fig:200129_15Hz}
\end{figure}

We also lower the reference frequency to 15 Hz, retain a sampling rate of 4096 Hz, and narrow the eccentricity prior to an upper bound of $e=0.3$. Figure \ref{fig:200129_15Hz} shows how we recover eccentricity posteriors that are more broad than the 16 \unit{kHz} sampling rate analysis. A bimodal feature is seen in the eccentricity posterior distributions for the TEOBResumS-Dali result while a more narrow bimodal shape is slightly formed in the SEOBNRv5EHM result. Apart from eccentricity, this analysis comparison is more consistent with our main analysis for this event. These comparisons further illustrate that the eccentricity and correlated parameters inferred can vary with the analysis settings. In particular, lower reference frequencies and higher sampling rates can introduce additional structure in the posteriors, such as bimodal features that are not present under our default settings. Therefore, we adopt the configuration used in our main analysis which yields more well-constrained posterior distributions across each parameter and waveform model. This highlights the importance of assessing certain analysis settings when interpreting potential signatures of eccentricity.

\section{Discussion}
\label{sec:discuss}

We identified several events as good candidates for eccentricity in O3 and O4a. Our candidate list omits events highlighted in other studies (e.g., Romero-Shaw et al. \cite{2022ApJ...940..171R}; Iglesias et al. \cite{2024ApJ...972...65I}; Gupte et al. \cite{2025PhRvD.112j4045G}; Planas et al. \cite{2025PhRvD.112l3004P}; Xu et al. \cite{2025arXiv251219513X}; Gupte et al. \cite{Gupte_2026}) and includes others not previously highlighted. 

For example, in Romero-Shaw et al. \cite{2022ApJ...940..171R}, four eccentric candidates are reported using the aligned-spin eccentric model SEOBNRE at a reference frequency of 10 Hz: GW190521\_030229, GW190620\_030421, GW191109\_010717, and GW200208\_222617.
We do not analyze GW200208\_22617 \cite{2025PhRvD.112f3052R,2025arXiv250722862M} solely because of the FAR threshold adopted in our study.
For those three remaining events, GW190620\_030421, GW190521\_030229, and GW191109\_010717, we find no evidence consistent with eccentricity.
Romero-Shaw et al. \cite{2022ApJ...940..171R} did not find evidence for eccentricity in GW190521\_074359.
In our analysis, we found signatures consistent with eccentricity for GW190521\_074359 at a reference frequency of 20 Hz, but the log-10 Bayes factors indicate low support. 

We identify at least tentative indications of eccentricity in one of the three candidates identified by Gupte et al. \cite{2025PhRvD.112j4045G} using SEOBNRv4EHM, omitting GW200208\_222617.
Specifically, GW200129\_065458 is also identified to be a candidate with eccentricity, as discussed in Section \ref{sec:sub:ecc}.
GW190701\_203306 is reported to contain signatures of eccentricity and their inferred eccentricity distribution rails against their upper prior, but we do not find evidence or similar behavior in our analysis.
Appendix \ref{app:A} further discusses our analysis and conclusions about this event.
This study did not find evidence of eccentricity for GW190521\_074359 at a reference frequency of 5.5 Hz, which we conclude at 20 Hz in our analysis.
Marginal support for eccentricity was seen for GW190620\_030421 and GW191109\_010717, lacking significant support with their Bayes factors.
However, we have not observed such signatures in our analysis.

Among the four events favoring the eccentric hypothesis in Planas et al. \cite{2025PhRvD.112l3004P} using IMRPhenomTEHM, we only find evidence for eccentricity in GW200129\_065458. This study claims evidence for eccentricity in GW200129\_065458 and GW200208\_222617, and finds potential eccentric features in GW190701\_203306 and GW190929\_012149. For GW200129\_065458, this study finds that even with various glitch mitigation techniques, the eccentric hypothesis is favored over the quasi-circular hypotheses. In particular, our comparison to their result using a uniform eccentric prior with \texttt{gw\_subtract} data was discussed in the previous section. The corresponding Bayes factors for this result show even stronger support for eccentricity than our analysis. This study does not support a strong claim of eccentricity for GW190701\_203306 because although a marginal eccentric feature is present, the Bayes factors show weak support for eccentricity and this high mass event is affected by a known glitch. For similar reasons, GW190929\_012149 is not identified as an eccentric candidate despite exhibiting slightly stronger support in their Bayes factors. We compare the uniform eccentricity run with \texttt{nlive=2000} for GW190701\_203306 in Appendix \ref{app:A}. We find no indications of eccentricity for this event and GW190929\_012149.

Investigations on O4a using IMRPhenomTEHM in Xu et al. \cite{2025arXiv251219513X} have identified several additional candidates: GW230706\_104333, GW230712\_090405, GW231114$\_$043211, GW231123\_135430, GW231221\_135041, GW231223\_032836, and GW231224\_024321. We find no such evidence of eccentricity for these events except GW231123\_135430, which similarly exhibits a distinct eccentric feature in this study. Although they obtain their highest Bayes factor supporting the aligned-spin eccentric hypothesis for this event, this support is significantly reduced when compared against a precessing, quasi-circular hypothesis. Conversely, Gupte et al. \cite{Gupte_2026} have claimed no strong support for eccentricity in O4a with SEOBNRv5EHM. Instead, this study focuses on a subset of 9 events with the highest Bayes factors which notably includes GW231001\_140220 as the highest among them at a reference frequency of 10 Hz. We obtain a negative Bayes factor with our SEOBNRv5EHM analysis at a reference frequency at 13.33 Hz, as previously reported, but find a positive Bayes factor for this event with TEOBResumS-Dali.

GW200129\_065458 had the strongest potential indications of eccentricity within our suite of analyses. While this event has been highlighted in previous work as a potentially eccentric event \cite{2025PhRvD.112j4045G}, these same studies and others have highlighted the numerous challenges associated with building confidence in results derived from this deglitched event. Indeed, several studies including those mentioned have found quantitatively different results when using different analysis and deglitching settings, though the same qualitative conclusions generally remain. Further studies are needed to evaluate the impact of glitches and effectiveness of various mitigation techniques on parameter estimation with eccentricity.

GW231123$\_$135430 had strong but ambiguous indications of eccentricity, with significant support only seen with our
(nonprecessing, eccentric) TEOBResumS-Dali analysis. 
Previously, Jan et al. \cite{2025arXiv251220060J} also analyzed this event with this model family, including both nonprecessing and precessing forms of the model. 
Our analyses are consistent with their nonprecessing eccentric results; however, their analysis strongly favors precession over eccentricity. 
In Appendix B of Zeeshan et al. \cite{2026arXiv260211030Z}, which uses the SEOBNRv5EHM results presented in this analysis, several events display waveform systematics when compared to quasi-circular mixed and precessing waveforms (SEOBNRv4PHM/SEOBNRv5PHM) from GWTC-2.1 and GWTC-4 results, including GW231123$\_$135430. 
Similarly to a previous discussion where we compare our results to the quasi-circular mixed results (Section \ref{sec:sub:systematics}), the precessing model comparison also finds significant disagreements in GW231123$\_$135430, particularly in the effective spin parameter. 
Additionally, this precessing comparison notes modest shifts in the effective spin distributions for GW190708\_232457, GW190720\_000836, and the asymmetric system GW190412\_053044 \cite{LIGO-O3-GW190412}.

We also identified several events as exhibiting substantial model systematics, either between the two eccentric waveforms used in this study or relative to previously-published quasi-circular results. Several events were flagged as possessing potential systematics between the different quasi-circular models in the GWTC-4 analysis including: GW230624$\_$113103, GW231028$\_$153006, GW231118$\_$005626, GW231118$\_$090602, and GW231123$\_$135430 \cite{LIGO-O4a-cbc-catalog_results}. Xu et al. \cite{2025arXiv251219513X} identified potential waveform systematics in three of these events (GW231028$\_$153006, GW231118$\_$090602, and GW231123$\_$135430) as well as additional O4a events with their quasi-circular analyses: GW230628$\_$231200, GW230814$\_$230901, GW230927$\_$153832, and GW231226$\_$101520. Similar to this study, we also see the most systematics with the highest-mass system GW231123$\_$135430 and some discrepancies with GW231028$\_$153006. Figure \ref{fig:jsd} confirms how for most events, the differences between each of our analyses with quasi-circular models from GWTC-4 appear more dominant over the two eccentric analyses.

\section{Conclusions}
\label{sec:conclude}

In this work, we perform parameter inferences using two models for GW radiation from a merging compact binary which incorporates orbital eccentricity, applying our method systematically to 162 high-significance event candidates from O3 and O4a.  We do not present new results for the thoroughly-analyzed event GW200105.  For binary black hole events, excepting only GW200129\_065458, we find that no candidate exhibits significant evidence for eccentricity. We identify three events to have at least one positive Bayes factor that favors the eccentric hypothesis: GW200129\_065458, GW231001\_140220, and GW231123\_135430. Two additional events, GW190521\_074359 and GW191204\_171526, showed potential eccentric features but lacked support with negative Bayes factors. From the three candidates, GW200129\_065458 provides the strongest support for eccentricity. Although we find indications of features in the eccentricity posterior distributions for these events, we cannot conclusively say they arise from eccentricity. 

We further examine events with the most substantial waveform systematics, namely GW230814\_230901, GW231001\_140220, and GW231123\_135430. The latter two events also exhibit eccentric features. Among the entire sample of events analyzed, GW231123\_135430 displayed the largest discrepancies across multiple parameters. The SEOBNRv5EHM and TEOBResumS-Dali results are generally consistent with each other, except for a few large outliers in the eccentricity. Most events continue to remain consistent with zero eccentricity, with the largest differences arising from inferring high eccentricities. In particular, TEOBResumS-Dali tended to show non-negligible support for large eccentricity in higher-mass events.

As discussed in Section \ref{sec:discuss}, for several events our inferences exhibit somewhat or substantially less support for eccentricity than other previously reported investigations. In some cases, these differences persist despite adopting the same waveform model. While head-to-head investigations involving multiple codes and waveforms are outside the scope of this systematic investigation using a single consistent framework, we anticipate that some of these differences may reflect different analysis settings choices. As our investigations demonstrate by concrete example, at times eccentric parameter inferences can be sensitive to analysis settings. We anticipate other settings such as data conditioning could also contribute to discrepancies between different approaches, particularly for high eccentricity or high mass sources where these differences could have an outsized impact (A. Jan, private communication). 

By analyzing the vast majority of events in the current catalog of gravitational-wave events for eccentricity in a consistent framework, we establish a comprehensive catalog of eccentric parameter estimation using the RIFT pipeline. This can inform population-level studies and is subsequently used in Zeeshan et al. \cite{2026arXiv260211030Z}.

\section*{Acknowledgements}
We thank the anonymous referee for their helpful comments on this manuscript. This material is based upon work supported by the NSF's LIGO Laboratory, a major facility fully funded by the National
Science Foundation. The authors acknowledge the computational resources provided by the LIGO Laboratory's CIT cluster,
which is supported by National Science Foundation Grants PHY-0757058 and PHY-0823459. NM acknowledges support from NSF Grant No. PHY-2309172. ROS acknowledges support from NSF
Grant No. AST-1909534, AST-2206321, PHY-2012057, PHY-2309172 and the Simons Foundation.

\appendix
\section{Supplemental results for eccentric features in select events}
\label{app:A}

\begin{figure*}
    \centering
    \begin{minipage}[b]{0.49\textwidth}
        \centering
        \includegraphics[width=\textwidth]{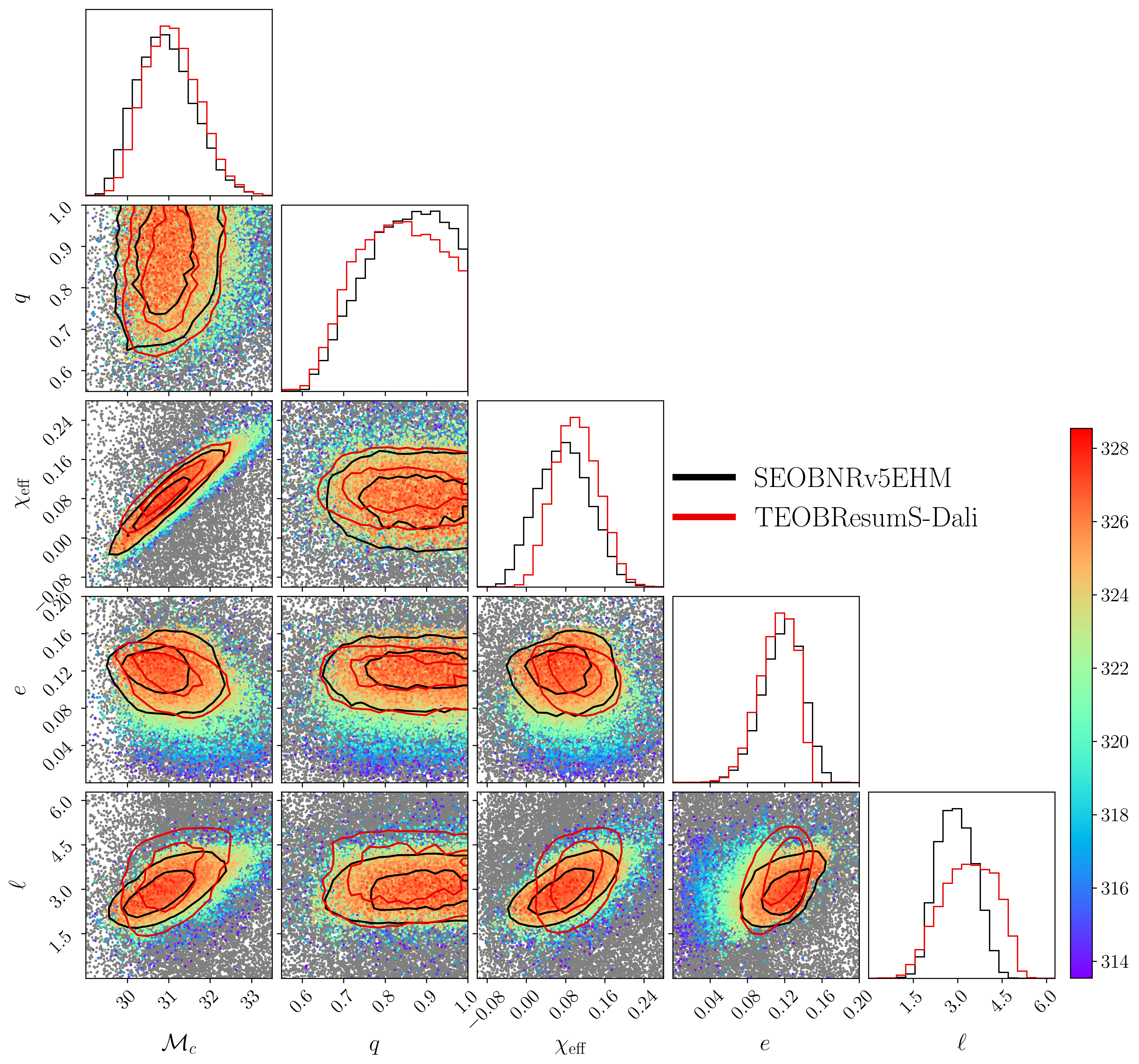}
        \label{fig:200129_v5EHM}
    \end{minipage}
    \hfill
    \begin{minipage}[b]{0.49\textwidth}
        \centering
        \includegraphics[width=\textwidth]{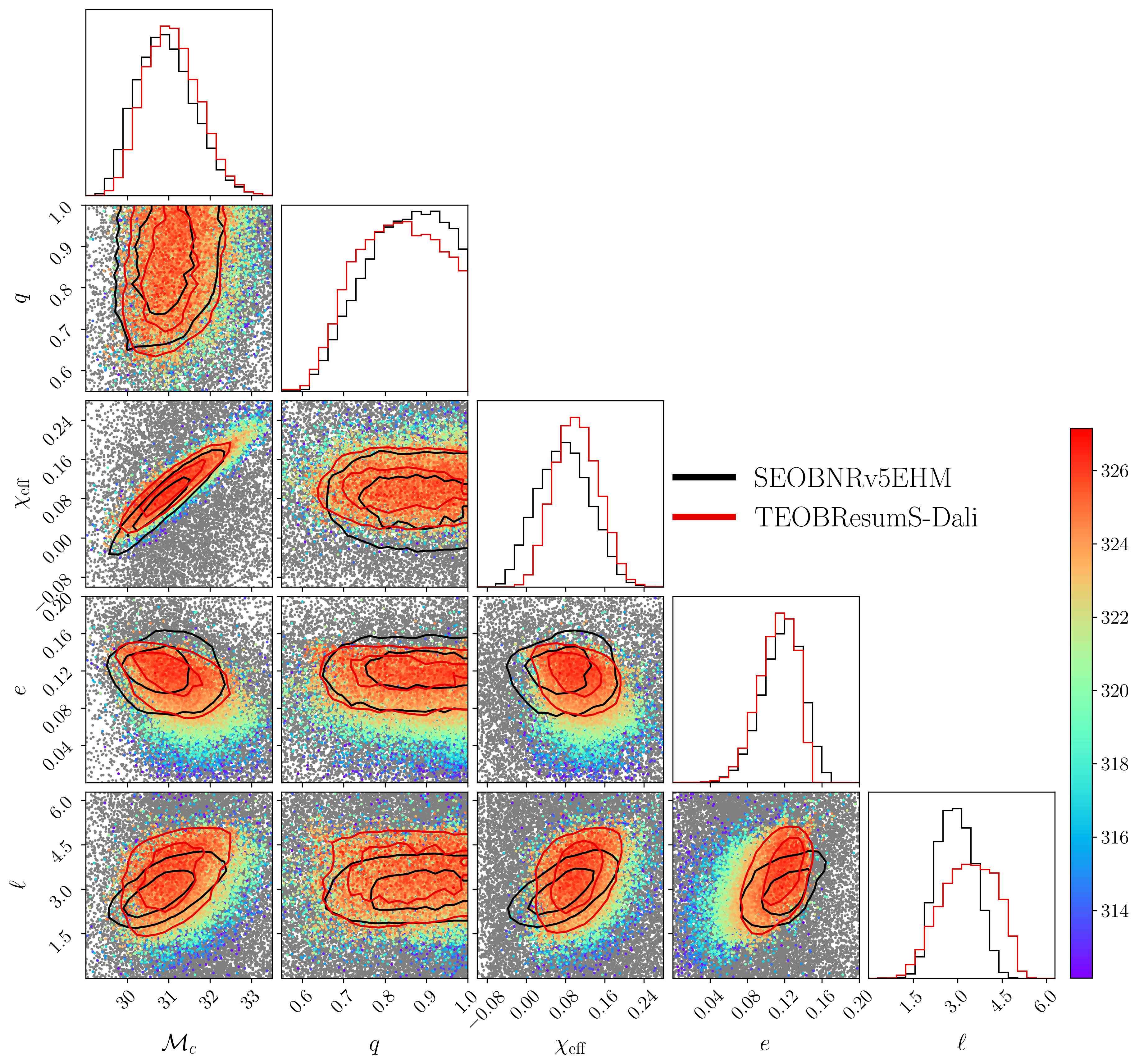}
        \label{fig:200129_Dali}
    \end{minipage}
    \caption{One- and two-dimensional posterior distributions for a subset of intrinsic parameters (chirp mass $\mathcal{M}_c$, mass ratio $q$, effective spin $\chi_\text{eff}$, eccentricity $e$, and mean anomaly $\ell$) for GW200129$\_$065458 using SEOBNRv5EHM (black) and TEOBResumS-Dali (red). The colored points represent the values of the $\ln$ marginal likelihoods for both SEOBNRv5EHM (left) and TEOBResumS-Dali (right). The contours correspond to the $50\%$ and $90\%$ confidence intervals for each joint distribution.}
    \label{fig:200129}
\end{figure*}

\begin{figure*}
    \centering
    \begin{minipage}[b]{0.49\textwidth}
        \centering
        \includegraphics[width=\textwidth]{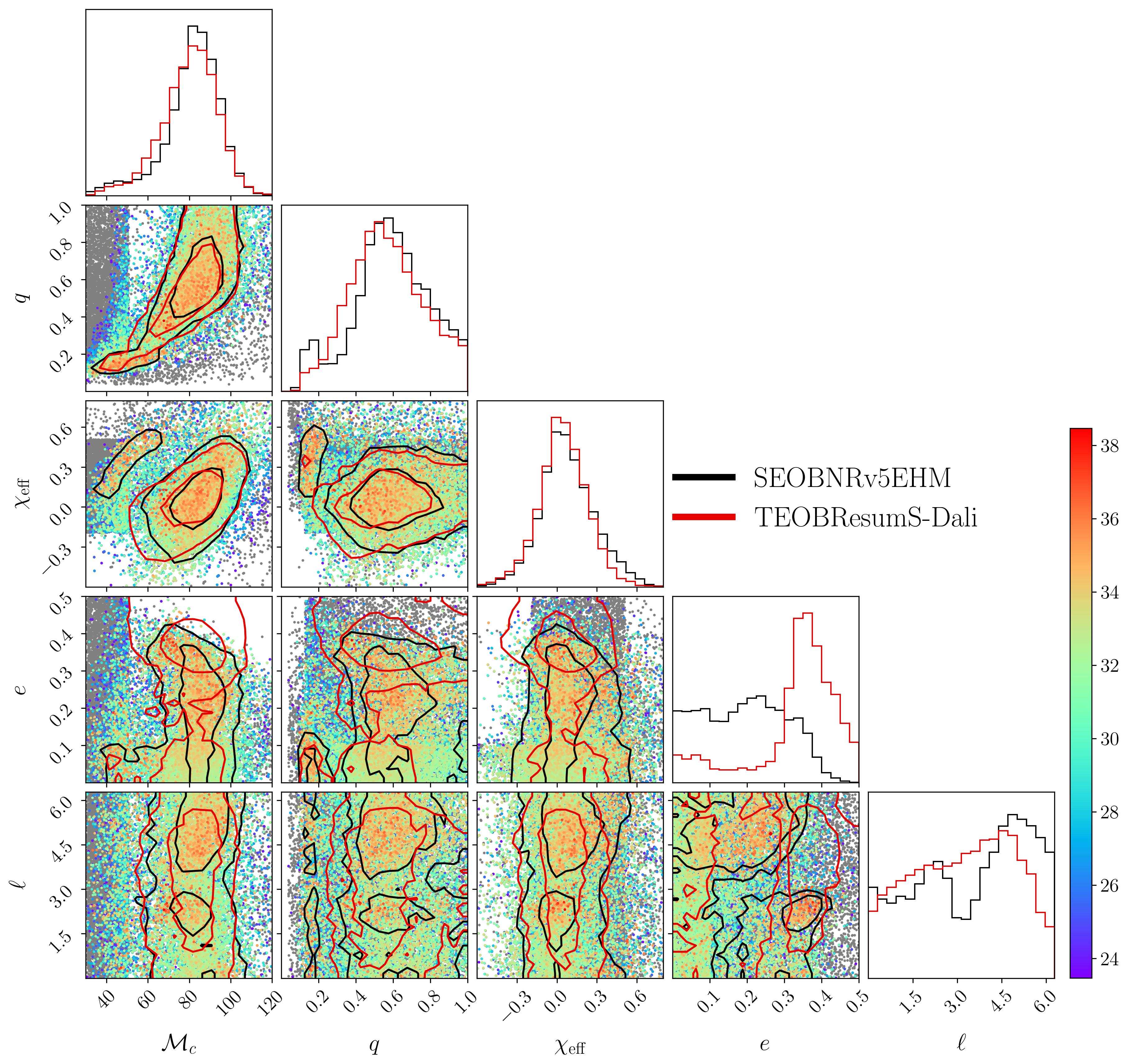} 
    \label{fig:231001_v5EHM}
    \end{minipage}
    \hfill
    \begin{minipage}[b]{0.49\textwidth}
        \centering
        \includegraphics[width=\textwidth]{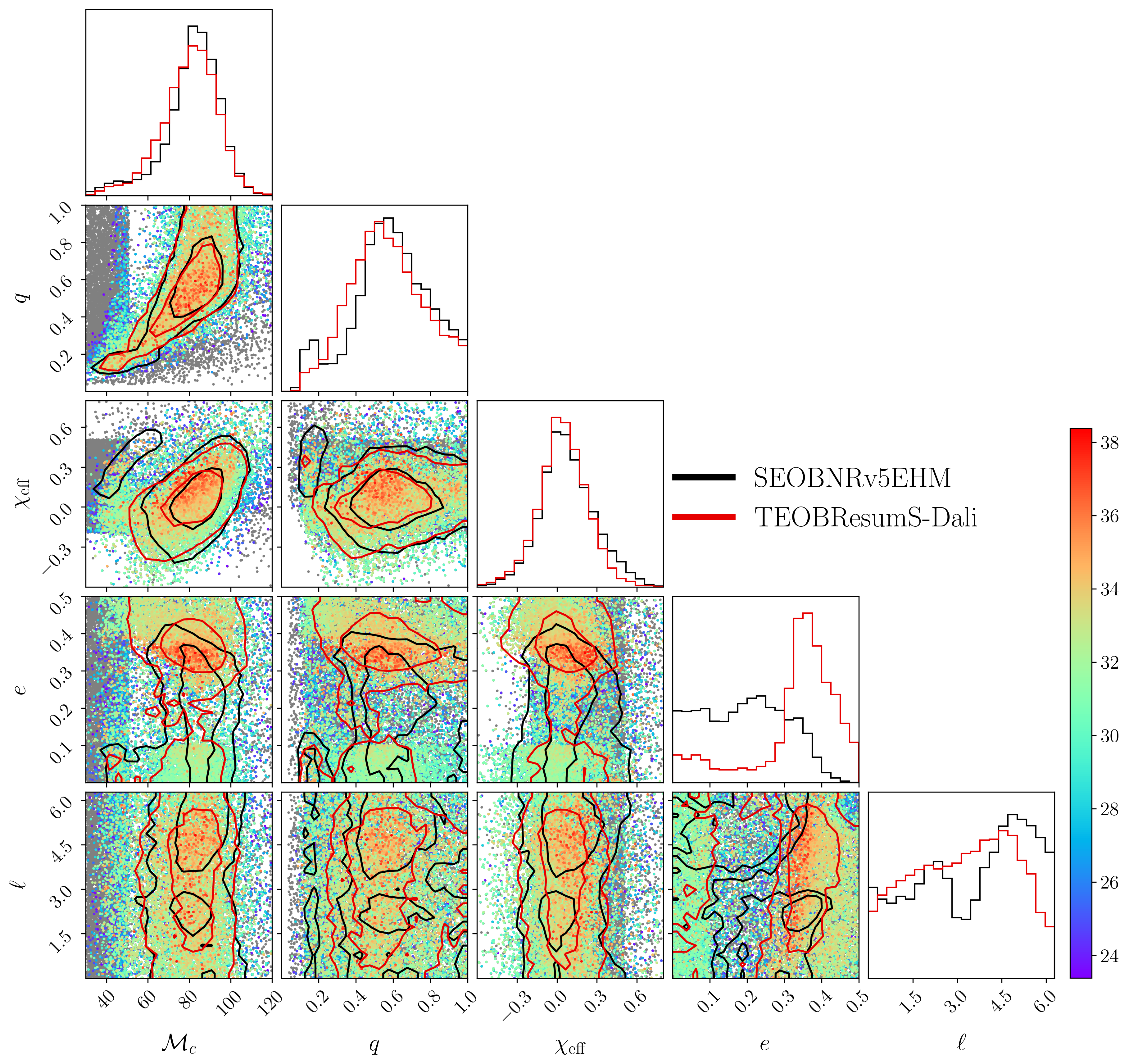}
        \label{fig:231001_Dali}
    \end{minipage}
    \caption{One- and two-dimensional posterior distributions for a subset of intrinsic parameters (chirp mass $\mathcal{M}_c$, mass ratio $q$, effective spin $\chi_\text{eff}$, eccentricity $e$, and mean anomaly $\ell$) for GW231001$\_$140220 using SEOBNRv5EHM (black) and TEOBResumS-Dali (red). The colored points represent the values of the $\ln$ marginal likelihoods for both SEOBNRv5EHM (left) and TEOBResumS-Dali (right). The contours correspond to the $50\%$ and $90\%$ confidence intervals for each joint distribution.}
    \label{fig:231001_rift}
\end{figure*}

Figures \ref{fig:200129} and \ref{fig:231001_rift} reflect inferences obtained from each waveform model for GW200129\_065458 and GW231001\_140220, respectively. The $\ln$ marginal likelihood values from each analysis are represented in each panel. For GW200129\_065458, the SEOBNRv5EHM and TEOBResumS-Dali results at 20 Hz are quite consistent with well-defined likelihood peaks. Alternatively, the waveform systematics in GW231001\_140220 are notable as both waveforms produced differing eccentricity distributions. 

\begin{figure}
    \centering
    \includegraphics[width=0.5\textwidth]{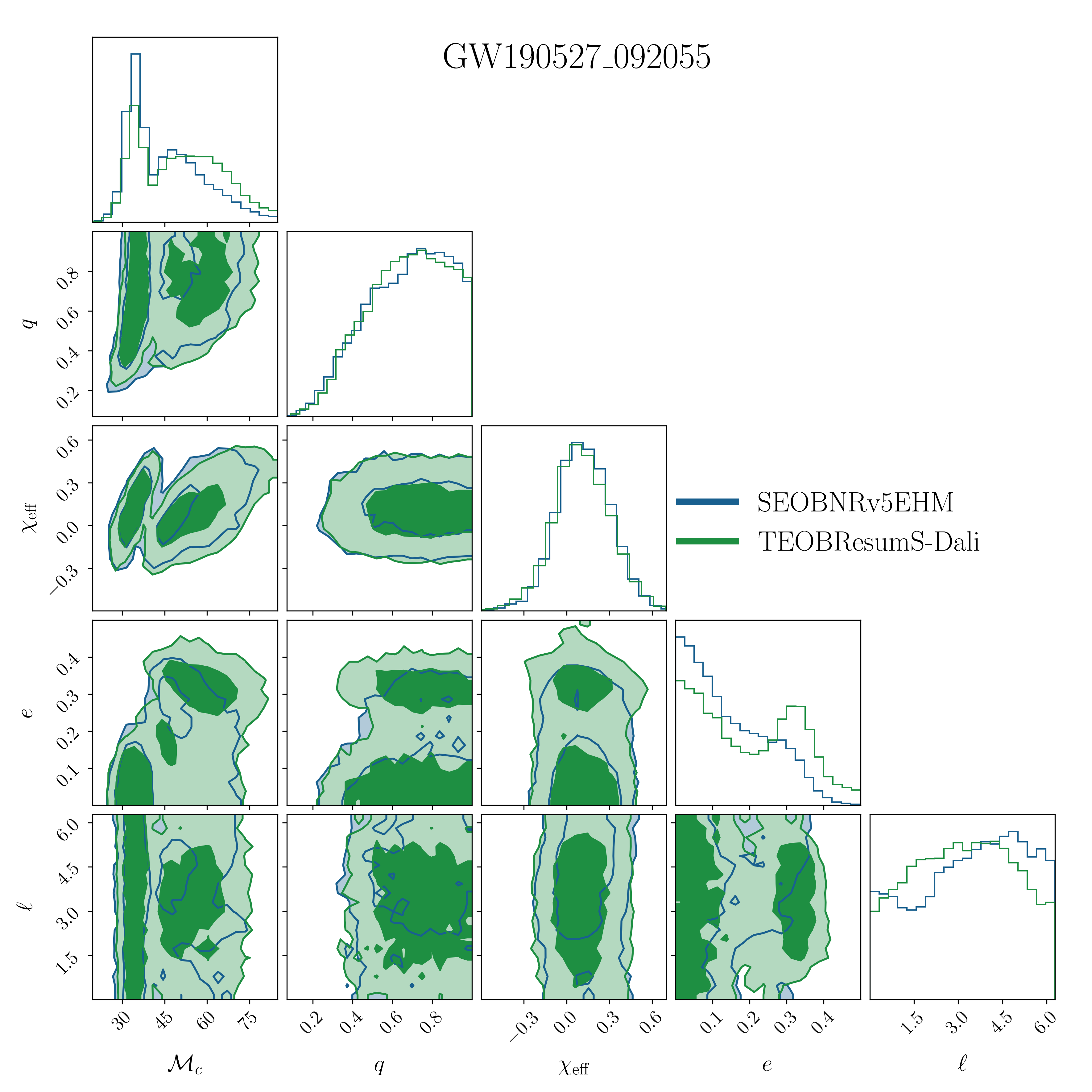}
    \caption{One- and two-dimensional posterior distributions for a subset of intrinsic parameters (chirp mass $\mathcal{M}_c$, mass ratio $q$, effective spin $\chi_\text{eff}$, eccentricity $e$, and mean anomaly $\ell$) for GW190527\_092055 using SEOBNRv5EHM (blue) and TEOBResumS-Dali (green). The contours correspond to the $50\%$ and $90\%$ confidence intervals for each joint distribution.}
    \label{fig:190527}
\end{figure}

We build on a previous RIFT analysis of GW190527\_092055 and find a small, secondary peak in the eccentricity distribution of the TEOBResumS-Dali analysis \cite{gwastro-RIFT_FinerNet}. Figure \ref{fig:190527} shows a comparison of our results for GW190527\_092055. Similar to Ref. \cite{gwastro-RIFT_FinerNet}, we observe a strong bimodal shape in the detector-frame chirp mass. In only the TEOBResums-Dali result, we identify an eccentric feature at $e\sim0.3$. Given that the eccentricity distributions have concentrated support around the lower bound ($e=0$) and the Bayes factors do not favor the eccentric hypothesis, we interpret this feature as arising from waveform systematics rather than evidence consistent with eccentricity.

\begin{figure}
    \centering
    \includegraphics[width=0.5\textwidth]{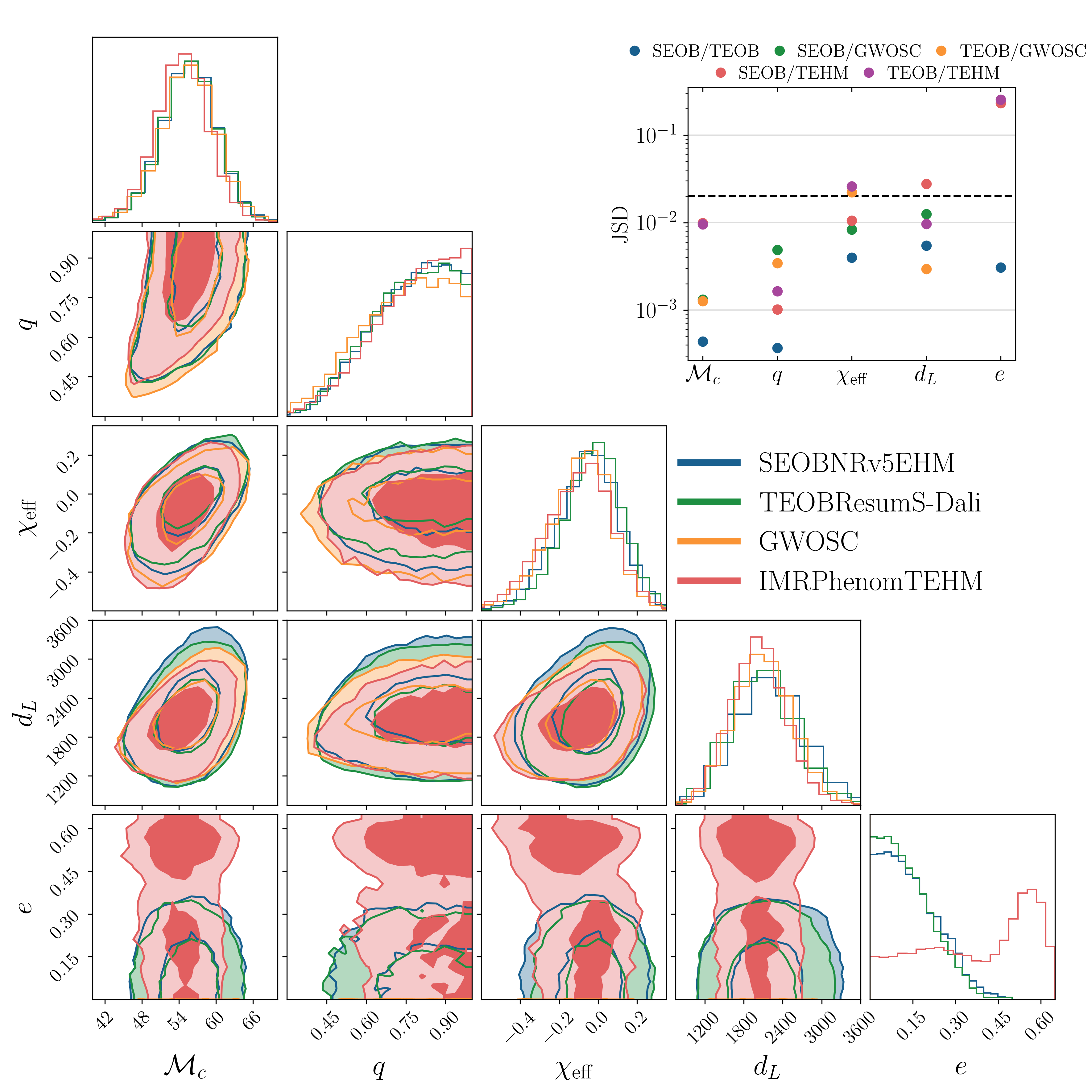}
    \caption{One- and two-dimensional posterior distributions for a few parameters (chirp mass $\mathcal{M}_c$, mass ratio $q$, effective spin $\chi_\text{eff}$, and luminosity distance $d_{L}$) for GW190701\_203306. Our SEOBNRv5EHM (blue) and TEOBResumS-Dali (green) analyses are compared to the public GWOSC mixed (orange) and IMRPhenomTEHM (red) results \cite{2025PhRvD.112l3004P, planas_llompart_2025_15259941, ligo_scientific_collaboration_and_virgo_2022_6513631}. The upper right panel shows the JS divergences (JSD) for these parameters between each result. The dashed (black) line represents a JS divergence of 0.02.}
    \label{fig:190701}
\end{figure}

We also find no evidence of eccentricity in GW190701\_203306. Gupte et al. \cite{2025PhRvD.112j4045G} and Planas et al. \cite{2025PhRvD.112l3004P} have reported finding potential signatures of eccentricity for this event, where the former favors and the latter slightly favors the eccentric hypothesis. Figure \ref{fig:190701} shows a comparison of our SEOBNRv5EHM and TEOBResumS-Dali results to the public GWTC-2.1 and independent IMRPhenomTEHM analyses for GW190701\_203306 \cite{2025PhRvD.112l3004P, planas_llompart_2025_15576673, ligo_scientific_collaboration_and_virgo_2022_6513631}. To note, the IMRPhenomTEHM result being compared is the result with a uniform prior on eccentricity and the highest number of live points (\texttt{nlive=2000}) used. Across these select few parameters, our results are largely consistent with one another and with the public GWOSC mixed, quasi-circular waveform result. The JS divergences in the upper panel show how the results are also agreeable compared to IMRPhenomTEHM across parameters except eccentricity. In the effective spin, TEOBResums-Dali appeared to slightly differ with both the GWOSC and IMRPhenomTEHM results. There is also a slight JS divergence greater than 0.02 between SEOBNRv5EHM and this waveform for the luminosity distance. The largest JS divergence appears to be in the eccentricity where our results are consistent with one another but differ compared to the independent IMRPhenomTEHM result, which has a dominant peak that rails against their upper bound ($e=0.65$). We also note that this event is known to contain a glitch and our analyses utilize the glitch mitigation performed on this event. Gupte et al. \cite{Gupte_2026} have explored glitch marginalization techniques towards GW190701\_203306 and found it decreases the Bayes factor support of the eccentric hypothesis. Hence, further work on the impact of glitch mitigation is needed to avoid biasing eccentric parameter estimation.

A standardized definition of eccentricity and mean anomaly can be measured with \texttt{gw\_eccentricity} to directly compare the eccentricity values between the differing waveform models utilized in our analysis \cite{Shaikh:2023ypz, Shaikh:2025tae}. We do not report post-processed eccentricities in our results due to the inability to obtain a sufficiently pure sample of post-processed posteriors, defined as at least 99\% of the posterior samples, with each waveform across the entire dataset of analyzed events. Reporting only a subset of post-processed posterior samples would introduce bias into each posterior distribution.

\bibliographystyle{apsrev4-2}
\bibliography{references,%
ros,%
gw-astronomy-mergers,%
gw-astronomy-mergers-nr,%
gw-astronomy-mergers-approximations,%
LIGO-publications}

\end{document}

%% file: symbols.tex
\def\RIT{Center for Computational Relativity and Gravitation, Rochester Institute of Technology, Rochester, New York 14623, USA}

%% file: event_table_reruns_log.tex
    \setlength\tabcolsep{3pt}
    \begin{longtable*}[t]{l c c c c c c c c c c}
    \caption{A table showing the total events analyzed. The columns contain the minimum FAR \cite{LIGO-O4a-cbc-catalog_results, LIGO-O3-O3b-catalog, GWTC3_pop}, key source parameters, starting frequency ($f_\mathrm{start}$), sampling rate ($s$), and log-10 of the Bayes factor ($\log_{10} \mathcal{B}$) associated with each event analysis.  The values represented in each row correspond to the analysis (SEOBNRv5EHM or TEOBResumS-Dali) that resulted in the highest log-10 Bayes factor. Only samples with eccentricity values less than 0.5 are represented. We report detector-frame (redshifted) mass parameters throughout our analysis. The inferred properties provide the median values and 90\% confidence intervals.}\label{tab:events}
    \\ 
    \hline \hline
    Name & FAR$_{\mathrm{min}}$\ & $\mathcal{M}_c$ & $m_1$ & $m_2$ & $\chi_{\mathrm{eff}}$ & $e$ & $f_{\mathrm{start}}$\ & $s$\ & $\log_{10} \mathcal{B}$ & Approx.\\
    & [yr$^{-1}$] & [$M_{\odot}$] & [$M_{\odot}$] & [$M_{\odot}$] & & & [Hz] & [Hz] & & \\
    \hline
    \endfirsthead
    \caption{\it{(Continued)}}
    \\
    \hline \hline
    Name & FAR$_{\mathrm{min}}$\ & $\mathcal{M}_c$ & $m_1$ & $m_2$ & $\chi_{\mathrm{eff}}$ & $e$ & $f_{\mathrm{start}}$\ & $s$\ & $\log_{10} \mathcal{B}$ & Approx.\\
    & [yr$^{-1}$] & [$M_{\odot}$] & [$M_{\odot}$] & [$M_{\odot}$] & & & [Hz] & [Hz] & & \\
    \hline
    \endhead
    \hline
    \endfoot
    \hline
    \endlastfoot
        
 $\mathrm{GW190408\_181802}$ & $<1\times10^{-5}$ & $23.67_{-1.24}^{+1.22}$ & $31.62_{-4.22}^{+6.74}$ & $23.57_{-4.43}^{+3.63}$ & $-0.04_{-0.17}^{+0.14}$ & $0.04_{-0.04}^{+0.07}$ & 20 & 4096 & -0.78 & SEOB\\
        
 $\mathrm{GW190412\_053044}$ & $<1\times10^{-5}$ & $15.14_{-0.20}^{+0.23}$ & $34.87_{-3.93}^{+4.38}$ & $9.40_{-0.84}^{+0.99}$ & $0.24_{-0.08}^{+0.08}$ & $0.03_{-0.03}^{+0.05}$ & 20 & 4096 & -0.96 & SEOB\\
        
 $\mathrm{GW190413\_052954}$ & $8.17\times10^{-1}$ & $40.19_{-6.02}^{+6.86}$ & $53.99_{-10.00}^{+14.90}$ & $40.13_{-10.78}^{+9.85}$ & $0.03_{-0.30}^{+0.28}$ & $0.10_{-0.09}^{+0.17}$ & 10 & 4096 & -0.39 & SEOB\\
        
 $\mathrm{GW190413\_134308}$ & $1.81\times10^{-1}$ & $58.95_{-10.70}^{+9.60}$ & $80.14_{-14.22}^{+18.16}$ & $58.62_{-20.43}^{+15.01}$ & $0.02_{-0.29}^{+0.29}$ & $0.17_{-0.15}^{+0.23}$ & 10 & 4096 & -0.19 & SEOB\\
        
 $\mathrm{GW190421\_213856}$ & $2.83\times10^{-3}$ & $46.15_{-6.17}^{+6.88}$ & $61.50_{-9.39}^{+13.19}$ & $46.72_{-13.13}^{+10.18}$ & $-0.08_{-0.26}^{+0.22}$ & $0.06_{-0.05}^{+0.11}$ & 20 & 4096 & -0.65 & TEOB\\
        
 $\mathrm{GW190425\_081805}$ & $3.38\times10^{-2}$ & $1.49_{-0.00}^{+0.00}$ & $2.04_{-0.31}^{+0.68}$ & $1.43_{-0.32}^{+0.24}$ & $0.02_{-0.03}^{+0.11}$ & $0.01_{-0.01}^{+0.01}$ & 20 & 8192 & -1.37 & SEOB\\
        
 $\mathrm{GW190426\_152155}$ & $9.12\times10^{-1}$ & $2.59_{-0.02}^{+0.01}$ & $6.23_{-2.76}^{+3.38}$ & $1.56_{-0.43}^{+1.01}$ & $-0.06_{-0.34}^{+0.30}$ & $0.03_{-0.03}^{+0.05}$ & 20 & 8192 & -0.95 & SEOB\\
        
 $\mathrm{GW190426\_190642}$ & 4.1 & $137.43_{-20.28}^{+23.24}$ & $183.22_{-32.20}^{+52.48}$ & $138.63_{-39.54}^{+31.64}$ & $0.27_{-0.35}^{+0.36}$ & $0.24_{-0.22}^{+0.23}$ & 3 & 4096 & -0.01 & TEOB\\
        
 $\mathrm{GW190503\_185404}$ & $<1\times10^{-5}$ & $37.65_{-5.54}^{+5.35}$ & $53.06_{-9.61}^{+11.83}$ & $35.95_{-10.49}^{+9.72}$ & $-0.01_{-0.25}^{+0.21}$ & $0.04_{-0.04}^{+0.08}$ & 20 & 4096 & -0.75 & TEOB\\
        
 $\mathrm{GW190512\_180714}$ & $<1\times10^{-5}$ & $18.65_{-0.66}^{+0.70}$ & $28.75_{-6.41}^{+8.16}$ & $16.28_{-3.44}^{+4.54}$ & $0.03_{-0.14}^{+0.14}$ & $0.04_{-0.04}^{+0.07}$ & 20 & 4096 & -0.83 & SEOB\\
        
 $\mathrm{GW190513\_205428}$ & $<1\times10^{-5}$ & $30.06_{-2.33}^{+4.10}$ & $46.97_{-10.89}^{+13.46}$ & $26.47_{-6.95}^{+8.50}$ & $0.16_{-0.18}^{+0.22}$ & $0.06_{-0.06}^{+0.06}$ & 20 & 4096 & -0.53 & TEOB\\
        
 $\mathrm{GW190514\_065416}$ & 2.8 & $48.66_{-9.47}^{+9.74}$ & $66.24_{-12.23}^{+23.03}$ & $48.29_{-17.44}^{+12.78}$ & $-0.05_{-0.35}^{+0.31}$ & $0.19_{-0.17}^{+0.25}$ & 10 & 4096 & -0.13 & SEOB\\
        
 $\mathrm{GW190517\_055101}$ & $3.47\times10^{-4}$ & $35.90_{-2.91}^{+3.36}$ & $48.59_{-7.30}^{+11.87}$ & $35.43_{-7.51}^{+6.89}$ & $0.57_{-0.15}^{+0.17}$ & $0.07_{-0.06}^{+0.12}$ & 20 & 4096 & -0.58 & SEOB\\
        
 $\mathrm{GW190519\_153544}$ & $<1\times10^{-5}$ & $65.66_{-10.24}^{+7.70}$ & $94.11_{-13.98}^{+15.97}$ & $61.37_{-18.86}^{+16.39}$ & $0.38_{-0.21}^{+0.17}$ & $0.09_{-0.08}^{+0.14}$ & 10 & 4096 & -0.48 & SEOB\\
        
 $\mathrm{GW190521\_030229}$ & $<1\times10^{-5}$ & $110.93_{-15.60}^{+12.50}$ & $151.75_{-20.38}^{+22.64}$ & $109.38_{-33.20}^{+24.57}$ & $0.06_{-0.23}^{+0.24}$ & $0.21_{-0.18}^{+0.24}$ & 5 & 4096 & -0.07 & SEOB\\
        
 $\mathrm{GW190521\_074359}$ & $1.0\times10^{-2}$ & $39.42_{-2.32}^{+2.33}$ & $51.44_{-5.40}^{+6.38}$ & $40.23_{-6.32}^{+5.60}$ & $0.06_{-0.12}^{+0.12}$ & $0.04_{-0.04}^{+0.04}$ & 20 & 4096 & -0.67 & SEOB\\
        
 $\mathrm{GW190527\_092055}$ & $2.28\times10^{-1}$ & $50.13_{-19.37}^{+24.91}$ & $71.38_{-27.84}^{+38.10}$ & $46.80_{-23.87}^{+30.17}$ & $0.10_{-0.32}^{+0.34}$ & $0.19_{-0.17}^{+0.22}$ & 20 & 4096 & -0.25 & TEOB\\
        
 $\mathrm{GW190602\_175927}$ & $<1\times10^{-5}$ & $74.24_{-14.10}^{+12.11}$ & $104.24_{-18.19}^{+21.60}$ & $72.41_{-27.96}^{+20.06}$ & $0.17_{-0.25}^{+0.25}$ & $0.10_{-0.09}^{+0.18}$ & 10 & 4096 & -0.40 & SEOB\\
        
 $\mathrm{GW190620\_030421}$ & $1.12\times10^{-2}$ & $58.03_{-9.20}^{+9.23}$ & $84.82_{-16.84}^{+20.79}$ & $53.90_{-16.96}^{+16.79}$ & $0.40_{-0.24}^{+0.19}$ & $0.09_{-0.08}^{+0.16}$ & 10 & 4096 & -0.44 & SEOB\\
        
 $\mathrm{GW190630\_185205}$ & $<1\times10^{-5}$ & $29.01_{-1.64}^{+1.81}$ & $41.31_{-7.10}^{+9.85}$ & $27.38_{-5.97}^{+5.88}$ & $0.08_{-0.15}^{+0.15}$ & $0.03_{-0.03}^{+0.05}$ & 20 & 4096 & -0.95 & SEOB\\
        
 $\mathrm{GW190701\_203306}$ & $5.71\times10^{-3}$ & $55.73_{-7.68}^{+7.05}$ & $73.46_{-10.90}^{+14.59}$ & $56.99_{-16.34}^{+11.00}$ & $-0.05_{-0.29}^{+0.24}$ & $0.13_{-0.12}^{+0.19}$ & 10 & 4096 & -0.30 & SEOB\\
        
 $\mathrm{GW190706\_222641}$ & $<1\times10^{-5}$ & $77.50_{-14.72}^{+11.58}$ & $110.45_{-17.03}^{+21.18}$ & $72.83_{-25.42}^{+21.41}$ & $0.30_{-0.30}^{+0.24}$ & $0.13_{-0.11}^{+0.20}$ & 10 & 4096 & -0.29 & SEOB\\
        
 $\mathrm{GW190707\_093326}$ & $<1\times10^{-5}$ & $9.87_{-0.10}^{+0.11}$ & $13.05_{-1.55}^{+3.40}$ & $9.90_{-1.86}^{+1.29}$ & $-0.06_{-0.07}^{+0.11}$ & $0.04_{-0.04}^{+0.07}$ & 20 & 4096 & -0.82 & SEOB\\
        
 $\mathrm{GW190708\_232457}$ & $3.09\times10^{-4}$ & $15.42_{-0.26}^{+0.25}$ & $20.32_{-2.36}^{+6.07}$ & $15.49_{-3.28}^{+2.00}$ & $0.01_{-0.08}^{+0.11}$ & $0.05_{-0.04}^{+0.06}$ & 20 & 4096 & -0.69 & SEOB\\
        
 $\mathrm{GW190719\_215514}$ & $6.31\times10^{-1}$ & $38.06_{-6.00}^{+37.16}$ & $58.96_{-15.98}^{+63.79}$ & $35.14_{-13.56}^{+33.04}$ & $0.30_{-0.32}^{+0.32}$ & $0.12_{-0.11}^{+0.22}$ & 10 & 4096 & -0.33 & SEOB\\
        
 $\mathrm{GW190720\_000836}$ & $<1\times10^{-5}$ & $10.32_{-0.13}^{+0.11}$ & $15.17_{-3.05}^{+7.36}$ & $9.37_{-2.65}^{+2.19}$ & $0.15_{-0.09}^{+0.16}$ & $0.04_{-0.04}^{+0.06}$ & 20 & 4096 & -0.80 & SEOB\\
        
 $\mathrm{GW190725\_174728}$ & $4.58\times10^{-1}$ & $8.94_{-0.16}^{+0.15}$ & $14.52_{-3.87}^{+9.77}$ & $7.41_{-2.47}^{+2.48}$ & $-0.03_{-0.15}^{+0.34}$ & $0.05_{-0.04}^{+0.07}$ & 20 & 4096 & -0.71 & SEOB\\
        
 $\mathrm{GW190727\_060333}$ & $<1\times10^{-5}$ & $44.20_{-5.01}^{+5.34}$ & $57.49_{-7.76}^{+11.14}$ & $45.69_{-11.16}^{+8.18}$ & $0.09_{-0.24}^{+0.24}$ & $0.13_{-0.12}^{+0.20}$ & 10 & 4096 & -0.31 & SEOB\\
        
 $\mathrm{GW190728\_064510}$ & $<1\times10^{-5}$ & $10.11_{-0.11}^{+0.09}$ & $14.14_{-2.33}^{+10.37}$ & $9.59_{-3.51}^{+1.79}$ & $0.12_{-0.07}^{+0.24}$ & $0.03_{-0.03}^{+0.06}$ & 20 & 4096 & -0.89 & SEOB\\
        
 $\mathrm{GW190731\_140936}$ & $3.35\times10^{-1}$ & $47.95_{-8.93}^{+7.60}$ & $64.62_{-10.84}^{+14.24}$ & $48.06_{-16.72}^{+11.55}$ & $0.09_{-0.27}^{+0.26}$ & $0.11_{-0.10}^{+0.19}$ & 10 & 4096 & -0.37 & SEOB\\
        
 $\mathrm{GW190803\_022701}$ & $7.32\times10^{-2}$ & $43.33_{-6.13}^{+6.76}$ & $58.01_{-9.69}^{+13.87}$ & $43.81_{-12.55}^{+9.71}$ & $0.02_{-0.27}^{+0.26}$ & $0.07_{-0.06}^{+0.14}$ & 20 & 4096 & -0.57 & SEOB\\
        
 $\mathrm{GW190805\_211137}$ & $6.28\times10^{-1}$ & $62.53_{-10.31}^{+9.57}$ & $85.69_{-15.31}^{+21.71}$ & $61.50_{-19.52}^{+15.02}$ & $0.40_{-0.40}^{+0.27}$ & $0.16_{-0.15}^{+0.24}$ & 10 & 4096 & -0.19 & SEOB\\
        
 $\mathrm{GW190814\_211039}$ & $<1\times10^{-5}$ & $6.39_{-0.04}^{+0.03}$ & $22.48_{-3.23}^{+3.04}$ & $2.88_{-0.23}^{+0.32}$ & $-0.09_{-0.17}^{+0.13}$ & $0.02_{-0.02}^{+0.03}$ & 20 & 8192 & -0.90 & TEOB\\
        
 $\mathrm{GW190828\_063405}$ & $<1\times10^{-5}$ & $34.45_{-2.54}^{+2.65}$ & $43.58_{-4.70}^{+6.84}$ & $36.17_{-5.69}^{+4.50}$ & $0.20_{-0.16}^{+0.14}$ & $0.07_{-0.06}^{+0.07}$ & 20 & 4096 & -0.49 & SEOB\\
        
 $\mathrm{GW190828\_065509}$ & $<1\times10^{-5}$ & $17.36_{-0.72}^{+0.61}$ & $31.03_{-9.09}^{+10.08}$ & $13.22_{-2.82}^{+5.23}$ & $0.08_{-0.18}^{+0.18}$ & $0.05_{-0.05}^{+0.09}$ & 20 & 4096 & -0.69 & SEOB\\
        
 $\mathrm{GW190910\_112807}$ & $2.87\times10^{-3}$ & $43.24_{-4.49}^{+4.36}$ & $56.36_{-7.29}^{+8.61}$ & $44.45_{-9.59}^{+7.52}$ & $-0.02_{-0.21}^{+0.19}$ & $0.05_{-0.05}^{+0.08}$ & 20 & 4096 & -0.69 & SEOB\\
        
 $\mathrm{GW190915\_235702}$ & $<1\times10^{-5}$ & $32.25_{-2.21}^{+2.46}$ & $41.55_{-4.92}^{+7.62}$ & $33.31_{-5.72}^{+4.63}$ & $0.01_{-0.17}^{+0.16}$ & $0.09_{-0.08}^{+0.08}$ & 20 & 4096 & -0.35 & TEOB\\
        
 $\mathrm{GW190916\_200658}$ & 4.7 & $53.98_{-13.64}^{+16.13}$ & $82.05_{-23.30}^{+29.50}$ & $49.57_{-21.49}^{+22.86}$ & $0.31_{-0.36}^{+0.30}$ & $0.19_{-0.18}^{+0.24}$ & 5 & 4096 & -0.14 & SEOB\\
        
 $\mathrm{GW190917\_114630}$ & $6.56\times10^{-1}$ & $4.20_{-0.03}^{+0.04}$ & $9.60_{-3.75}^{+5.66}$ & $2.62_{-0.75}^{+1.37}$ & $-0.21_{-0.38}^{+0.38}$ & $0.03_{-0.03}^{+0.07}$ & 20 & 4096 & -0.89 & SEOB\\
        
 $\mathrm{GW190924\_021846}$ & $<1\times10^{-5}$ & $6.42_{-0.05}^{+0.04}$ & $10.12_{-2.50}^{+6.60}$ & $5.47_{-1.82}^{+1.64}$ & $0.02_{-0.10}^{+0.28}$ & $0.04_{-0.04}^{+0.05}$ & 20 & 8192 & -0.81 & SEOB\\
        
 $\mathrm{GW190925\_232845}$ & $7.2\times10^{-3}$ & $18.49_{-0.72}^{+0.74}$ & $24.69_{-3.23}^{+7.97}$ & $18.37_{-4.27}^{+2.76}$ & $0.09_{-0.15}^{+0.16}$ & $0.05_{-0.05}^{+0.09}$ & 20 & 4096 & -0.69 & SEOB\\
        
 $\mathrm{GW190926\_050336}$ & 1.1 & $47.13_{-13.56}^{+36.10}$ & $67.22_{-17.98}^{+61.70}$ & $45.32_{-20.91}^{+31.85}$ & $-0.003_{-0.34}^{+0.32}$ & $0.15_{-0.14}^{+0.25}$ & 10 & 4096 & -0.21 & SEOB\\
        
 $\mathrm{GW190929\_012149}$ & $1.55\times10^{-1}$ & $56.61_{-14.40}^{+14.81}$ & $99.29_{-19.52}^{+20.14}$ & $44.29_{-19.16}^{+27.77}$ & $-0.01_{-0.26}^{+0.23}$ & $0.10_{-0.09}^{+0.17}$ & 10 & 4096 & -0.41 & SEOB\\
        
 $\mathrm{GW190930\_133541}$ & $1.23\times10^{-2}$ & $9.78_{-0.24}^{+0.18}$ & $14.11_{-2.67}^{+13.65}$ & $8.97_{-3.73}^{+1.99}$ & $0.13_{-0.14}^{+0.31}$ & $0.07_{-0.06}^{+0.09}$ & 20 & 4096 & -0.56 & SEOB\\
        
 $\mathrm{GW191103\_012549}$ & $4.58\times10^{-1}$ & $9.89_{-0.26}^{+0.10}$ & $14.05_{-2.51}^{+7.39}$ & $9.21_{-2.78}^{+1.89}$ & $0.18_{-0.12}^{+0.16}$ & $0.08_{-0.07}^{+0.09}$ & 20 & 4096 & -0.43 & SEOB\\
        
 $\mathrm{GW191105\_143521}$ & $1.18\times10^{-2}$ & $9.55_{-0.16}^{+0.12}$ & $13.11_{-1.95}^{+5.43}$ & $9.22_{-2.41}^{+1.56}$ & $-0.03_{-0.09}^{+0.17}$ & $0.05_{-0.05}^{+0.08}$ & 20 & 8192 & -0.68 & SEOB\\
        
 $\mathrm{GW191109\_010717}$ & $1.8\times10^{-4}$ & $59.63_{-7.01}^{+6.07}$ & $77.72_{-8.87}^{+10.62}$ & $61.11_{-14.08}^{+10.98}$ & $-0.23_{-0.23}^{+0.26}$ & $0.10_{-0.09}^{+0.15}$ & 10 & 4096 & -0.40 & SEOB\\
        
 $\mathrm{GW191113\_071753}$ & $2.6\times10^{1}$ & $13.19_{-0.59}^{+1.88}$ & $34.51_{-15.48}^{+17.08}$ & $7.42_{-1.80}^{+6.90}$ & $-0.01_{-0.35}^{+0.41}$ & $0.10_{-0.09}^{+0.17}$ & 10 & 4096 & -0.40 & TEOB\\
        
 $\mathrm{GW191126\_115259}$ & 3.2 & $11.22_{-0.35}^{+0.27}$ & $15.98_{-2.83}^{+8.00}$ & $10.43_{-3.04}^{+2.17}$ & $0.21_{-0.15}^{+0.16}$ & $0.08_{-0.07}^{+0.11}$ & 20 & 4096 & -0.48 & SEOB\\
        
 $\mathrm{GW191127\_050227}$ & $2.49\times10^{-1}$ & $56.35_{-15.28}^{+18.75}$ & $85.80_{-23.82}^{+35.51}$ & $51.27_{-23.57}^{+26.01}$ & $0.22_{-0.35}^{+0.35}$ & $0.16_{-0.14}^{+0.26}$ & 10 & 4096 & -0.21 & SEOB\\
        
 $\mathrm{GW191129\_134029}$ & $<1\times10^{-5}$ & $8.48_{-0.07}^{+0.06}$ & $12.41_{-2.44}^{+5.68}$ & $7.73_{-2.09}^{+1.77}$ & $0.07_{-0.08}^{+0.18}$ & $0.04_{-0.04}^{+0.05}$ & 20 & 4096 & -0.81 & TEOB\\
        
 $\mathrm{GW191204\_110529}$ & 3.3 & $26.23_{-2.28}^{+2.45}$ & $36.47_{-6.13}^{+16.35}$ & $25.37_{-8.21}^{+5.36}$ & $0.04_{-0.26}^{+0.26}$ & $0.05_{-0.05}^{+0.09}$ & 20 & 4096 & -0.70 & SEOB\\
        
 $\mathrm{GW191204\_171526}$ & $<1\times10^{-5}$ & $9.65_{-0.05}^{+0.07}$ & $12.54_{-1.30}^{+2.81}$ & $9.83_{-1.65}^{+1.10}$ & $0.14_{-0.04}^{+0.06}$ & $0.07_{-0.05}^{+0.02}$ & 20 & 4096 & -0.07 & SEOB\\
        
 $\mathrm{GW191215\_223052}$ & $<1\times10^{-5}$ & $24.97_{-1.27}^{+1.36}$ & $33.33_{-4.42}^{+8.21}$ & $24.90_{-5.03}^{+3.79}$ & $-0.004_{-0.17}^{+0.17}$ & $0.06_{-0.05}^{+0.08}$ & 20 & 4096 & -0.63 & SEOB\\
        
 $\mathrm{GW191216\_213338}$ & $<1\times10^{-5}$ & $8.93_{-0.05}^{+0.06}$ & $13.60_{-2.97}^{+7.02}$ & $7.85_{-2.29}^{+2.04}$ & $0.12_{-0.07}^{+0.19}$ & $0.03_{-0.03}^{+0.04}$ & 20 & 4096 & -0.99 & SEOB\\
        
 $\mathrm{GW191219\_163120}$ & 4.0 & $4.74_{-0.07}^{+0.14}$ & $25.93_{-6.30}^{+12.40}$ & $1.56_{-0.32}^{+0.30}$ & $-0.36_{-0.44}^{+0.48}$ & $0.06_{-0.05}^{+0.11}$ & 10 & 8192 & -0.66 & SEOB\\
        
 $\mathrm{GW191222\_033537}$ & $<1\times10^{-5}$ & $51.26_{-6.42}^{+6.93}$ & $67.21_{-9.78}^{+13.66}$ & $52.62_{-13.76}^{+10.32}$ & $-0.04_{-0.24}^{+0.20}$ & $0.06_{-0.05}^{+0.10}$ & 20 & 4096 & -0.66 & TEOB\\
        
 $\mathrm{GW191230\_180458}$ & $5.02\times10^{-2}$ & $63.06_{-9.17}^{+8.51}$ & $82.71_{-12.33}^{+16.08}$ & $64.70_{-18.41}^{+12.84}$ & $0.0001_{-0.29}^{+0.26}$ & $0.12_{-0.11}^{+0.20}$ & 10 & 4096 & -0.33 & SEOB\\
        
        
 $\mathrm{GW200112\_155838}$ & $<1\times10^{-5}$ & $33.65_{-2.24}^{+2.56}$ & $44.18_{-5.51}^{+8.28}$ & $34.25_{-6.50}^{+5.14}$ & $0.05_{-0.16}^{+0.16}$ & $0.03_{-0.02}^{+0.05}$ & 20 & 4096 & -0.97 & SEOB\\
        
 $\mathrm{GW200115\_042309}$ & $<1\times10^{-5}$ & $2.57_{-0.01}^{+0.01}$ & $5.43_{-1.82}^{+2.34}$ & $1.72_{-0.41}^{+0.72}$ & $-0.28_{-0.30}^{+0.30}$ & $0.05_{-0.04}^{+0.06}$ & 10 & 8192 & -0.67 & TEOB\\
        
 $\mathrm{GW200128\_022011}$ & $4.29\times10^{-3}$ & $49.30_{-5.52}^{+6.49}$ & $64.28_{-9.27}^{+12.56}$ & $50.87_{-11.84}^{+9.44}$ & $0.13_{-0.21}^{+0.21}$ & $0.06_{-0.05}^{+0.09}$ & 20 & 4096 & -0.67 & TEOB\\
        
 $\mathrm{GW200129\_065458}$ & $<1\times10^{-5}$ & $31.06_{-0.96}^{+1.14}$ & $39.19_{-3.24}^{+4.34}$ & $32.63_{-3.54}^{+3.12}$ & $0.10_{-0.07}^{+0.07}$ & $0.11_{-0.04}^{+0.03}$ & 20 & 4096 & 3.59 & TEOB\\
        
 $\mathrm{GW200202\_154313}$ & $<1\times10^{-5}$ & $8.12_{-0.08}^{+0.06}$ & $11.47_{-1.59}^{+3.75}$ & $7.65_{-1.65}^{+1.11}$ & $0.03_{-0.08}^{+0.14}$ & $0.04_{-0.04}^{+0.06}$ & 20 & 4096 & -0.77 & SEOB\\
        
 $\mathrm{GW200208\_130117}$ & $3.11\times10^{-4}$ & $40.34_{-5.10}^{+5.31}$ & $53.78_{-8.47}^{+11.99}$ & $40.72_{-10.41}^{+8.38}$ & $-0.01_{-0.25}^{+0.21}$ & $0.06_{-0.06}^{+0.08}$ & 20 & 4096 & -0.59 & SEOB\\
        
 $\mathrm{GW200209\_085452}$ & $4.64\times10^{-2}$ & $42.57_{-7.27}^{+8.82}$ & $54.93_{-9.33}^{+15.84}$ & $43.92_{-11.44}^{+10.53}$ & $-0.04_{-0.28}^{+0.22}$ & $0.09_{-0.08}^{+0.15}$ & 20 & 4096 & -0.46 & TEOB\\
        
 $\mathrm{GW200210\_092254}$ & 1.2 & $7.71_{-0.16}^{+0.15}$ & $26.91_{-8.90}^{+12.01}$ & $3.49_{-0.76}^{+1.16}$ & $-0.06_{-0.45}^{+0.34}$ & $0.04_{-0.04}^{+0.07}$ & 20 & 8192 & -0.74 & TEOB\\
        
 $\mathrm{GW200216\_220804}$ & $3.5\times10^{-1}$ & $59.95_{-17.09}^{+14.30}$ & $83.60_{-17.51}^{+25.60}$ & $59.02_{-29.26}^{+19.47}$ & $0.04_{-0.37}^{+0.36}$ & $0.20_{-0.18}^{+0.25}$ & 10 & 4096 & -0.12 & TEOB\\
        
 $\mathrm{GW200219\_094415}$ & $9.94\times10^{-4}$ & $43.27_{-6.28}^{+6.34}$ & $58.18_{-9.31}^{+13.59}$ & $43.54_{-13.73}^{+9.79}$ & $-0.04_{-0.32}^{+0.23}$ & $0.08_{-0.07}^{+0.34}$ & 20 & 4096 & -0.54 & TEOB\\
        
 $\mathrm{GW200220\_124850}$ & $3.0\times10^{1}$ & $47.68_{-7.54}^{+8.13}$ & $64.14_{-11.01}^{+15.19}$ & $47.95_{-14.60}^{+11.63}$ & $-0.04_{-0.32}^{+0.28}$ & $0.15_{-0.14}^{+0.23}$ & 10 & 4096 & -0.22 & SEOB\\
        
 $\mathrm{GW200224\_222234}$ & $<1\times10^{-5}$ & $40.15_{-3.33}^{+3.10}$ & $51.20_{-5.50}^{+7.26}$ & $42.04_{-8.05}^{+5.45}$ & $0.08_{-0.16}^{+0.15}$ & $0.05_{-0.04}^{+0.06}$ & 20 & 4096 & -0.73 & SEOB\\
        
 $\mathrm{GW200225\_060421}$ & $<1\times10^{-5}$ & $17.93_{-1.68}^{+0.87}$ & $23.07_{-2.58}^{+4.18}$ & $18.36_{-3.39}^{+2.34}$ & $-0.05_{-0.24}^{+0.14}$ & $0.04_{-0.04}^{+0.08}$ & 20 & 4096 & -0.75 & TEOB\\
        
 $\mathrm{GW200302\_015811}$ & $1.12\times10^{-1}$ & $29.61_{-3.81}^{+6.59}$ & $47.78_{-10.19}^{+10.48}$ & $25.20_{-7.30}^{+12.09}$ & $0.02_{-0.25}^{+0.23}$ & $0.05_{-0.05}^{+0.09}$ & 20 & 4096 & -0.69 & SEOB\\
        
 $\mathrm{GW200306\_093714}$ & $2.4\times10^{1}$ & $25.64_{-5.76}^{+3.11}$ & $38.48_{-9.27}^{+22.65}$ & $22.44_{-10.07}^{+7.39}$ & $0.36_{-0.45}^{+0.25}$ & $0.07_{-0.07}^{+0.13}$ & 20 & 4096 & -0.55 & SEOB\\
        
 $\mathrm{GW200311\_115853}$ & $<1\times10^{-5}$ & $32.21_{-2.59}^{+2.34}$ & $40.95_{-4.47}^{+7.39}$ & $33.62_{-6.48}^{+4.33}$ & $-0.05_{-0.20}^{+0.16}$ & $0.03_{-0.03}^{+0.05}$ & 20 & 4096 & -0.95 & SEOB\\
        
 $\mathrm{GW200316\_215756}$ & $<1\times10^{-5}$ & $10.65_{-0.17}^{+0.15}$ & $17.48_{-4.83}^{+17.06}$ & $8.72_{-3.56}^{+3.05}$ & $0.15_{-0.14}^{+0.33}$ & $0.04_{-0.04}^{+0.07}$ & 20 & 4096 & -0.79 & TEOB\\
        
 $\mathrm{GW230518\_125908}$ & $<1\times10^{-5}$ & $2.94_{-0.01}^{+0.01}$ & $8.57_{-1.16}^{+1.24}$ & $1.52_{-0.14}^{+0.18}$ & $-0.02_{-0.13}^{+0.11}$ & $0.02_{-0.02}^{+0.03}$ & 13.33 & 8192 & -1.08 & TEOB\\
        
 $\mathrm{GW230529\_181500}$ & $2.2\times10^{-4}$ & $2.02_{-0.00}^{+0.00}$ & $3.23_{-0.82}^{+1.46}$ & $1.71_{-0.45}^{+0.54}$ & $-0.19_{-0.10}^{+0.23}$ & $0.02_{-0.02}^{+0.02}$ & 13.33 & 8192 & -1.17 & TEOB\\
        
 $\mathrm{GW230601\_224134}$ & $<1\times10^{-5}$ & $72.40_{-10.60}^{+8.76}$ & $101.44_{-15.89}^{+19.15}$ & $70.03_{-22.96}^{+17.02}$ & $0.004_{-0.30}^{+0.27}$ & $0.09_{-0.08}^{+0.16}$ & 13.33 & 4096 & -0.45 & SEOB\\
        
 $\mathrm{GW230605\_065343}$ & $<1\times10^{-5}$ & $14.40_{-0.29}^{+0.32}$ & $21.34_{-4.36}^{+10.06}$ & $12.97_{-3.55}^{+3.14}$ & $0.07_{-0.12}^{+0.20}$ & $0.06_{-0.06}^{+0.11}$ & 13.33 & 4096 & -0.61 & SEOB\\
        
 $\mathrm{GW230606\_004305}$ & $4.1\times10^{-4}$ & $38.79_{-5.63}^{+6.45}$ & $55.07_{-10.69}^{+16.68}$ & $37.03_{-11.71}^{+10.74}$ & $-0.13_{-0.31}^{+0.27}$ & $0.07_{-0.07}^{+0.13}$ & 13.33 & 4096 & -0.55 & SEOB\\
        
 $\mathrm{GW230608\_205047}$ & $1.2\times10^{-3}$ & $51.26_{-9.65}^{+8.71}$ & $74.46_{-13.49}^{+14.69}$ & $48.08_{-17.77}^{+16.04}$ & $0.05_{-0.25}^{+0.24}$ & $0.09_{-0.08}^{+0.16}$ & 13.33 & 4096 & -0.46 & SEOB\\
        
 $\mathrm{GW230609\_064958}$ & $1.4\times10^{-4}$ & $39.97_{-5.84}^{+5.71}$ & $53.93_{-8.48}^{+12.07}$ & $40.06_{-12.33}^{+9.09}$ & $-0.12_{-0.27}^{+0.22}$ & $0.07_{-0.06}^{+0.12}$ & 13.33 & 4096 & -0.57 & TEOB\\
        
 $\mathrm{GW230624\_113103}$ & $1.8\times10^{-4}$ & $24.27_{-1.94}^{+2.27}$ & $35.12_{-6.81}^{+14.45}$ & $22.52_{-6.10}^{+5.37}$ & $0.15_{-0.23}^{+0.24}$ & $0.07_{-0.07}^{+0.13}$ & 13.33 & 4096 & -0.54 & SEOB\\
        
 $\mathrm{GW230627\_015337}$ & $<1\times10^{-5}$ & $6.41_{-0.01}^{+0.01}$ & $8.87_{-1.35}^{+1.86}$ & $6.15_{-0.96}^{+1.05}$ & $0.02_{-0.03}^{+0.08}$ & $0.02_{-0.02}^{+0.03}$ & 13.33 & 4096 & -1.11 & SEOB\\
        
 $\mathrm{GW230628\_231200}$ & $<1\times10^{-5}$ & $35.42_{-2.91}^{+2.76}$ & $44.22_{-4.44}^{+6.39}$ & $37.68_{-6.32}^{+4.44}$ & $-0.03_{-0.18}^{+0.15}$ & $0.05_{-0.04}^{+0.08}$ & 13.33 & 4096 & -0.73 & SEOB\\
        
 $\mathrm{GW230630\_125806}$ & $1.6\times10^{-1}$ & $64.30_{-15.68}^{+12.48}$ & $90.96_{-18.53}^{+22.70}$ & $61.97_{-26.53}^{+19.28}$ & $0.22_{-0.31}^{+0.29}$ & $0.12_{-0.11}^{+0.30}$ & 13.33 & 4096 & -0.34 & TEOB\\
        
 $\mathrm{GW230630\_234532}$ & $4.2\times10^{-4}$ & $8.53_{-0.12}^{+0.07}$ & $11.87_{-1.89}^{+4.48}$ & $8.12_{-1.95}^{+1.48}$ & $-0.06_{-0.09}^{+0.17}$ & $0.07_{-0.06}^{+0.09}$ & 13.33 & 4096 & -0.56 & SEOB\\
        
 $\mathrm{GW230702\_185453}$ & $<1\times10^{-5}$ & $32.96_{-4.19}^{+5.88}$ & $61.79_{-21.30}^{+33.64}$ & $24.67_{-8.55}^{+12.49}$ & $0.08_{-0.31}^{+0.34}$ & $0.06_{-0.06}^{+0.11}$ & 13.33 & 4096 & -0.64 & SEOB\\
        
 $\mathrm{GW230704\_021211}$ & $2.1\times10^{-1}$ & $32.26_{-4.16}^{+3.61}$ & $48.17_{-9.85}^{+14.22}$ & $29.14_{-9.25}^{+8.63}$ & $-0.005_{-0.26}^{+0.23}$ & $0.07_{-0.06}^{+0.12}$ & 13.33 & 4096 & -0.57 & SEOB\\
        
 $\mathrm{GW230704\_212616}$ & $5.1\times10^{-1}$ & $116.33_{-38.66}^{+21.60}$ & $173.84_{-33.77}^{+49.29}$ & $106.30_{-61.16}^{+38.82}$ & $0.27_{-0.36}^{+0.32}$ & $0.11_{-0.10}^{+0.14}$ & 13.33 & 4096 & -0.34 & SEOB\\
        
 $\mathrm{GW230706\_104333}$ & $2.3\times10^{-1}$ & $15.83_{-0.52}^{+0.54}$ & $21.46_{-3.02}^{+6.17}$ & $15.53_{-3.26}^{+2.47}$ & $0.18_{-0.14}^{+0.13}$ & $0.08_{-0.07}^{+0.14}$ & 13.33 & 4096 & -0.51 & SEOB\\
        
 $\mathrm{GW230707\_124047}$ & $1.1\times10^{-3}$ & $59.63_{-7.30}^{+6.54}$ & $76.97_{-10.08}^{+13.14}$ & $61.98_{-15.52}^{+10.40}$ & $-0.03_{-0.27}^{+0.25}$ & $0.10_{-0.09}^{+0.17}$ & 13.33 & 4096 & -0.41 & SEOB\\
        
 $\mathrm{GW230708\_053705}$ & $2.2\times10^{-1}$ & $34.12_{-3.50}^{+4.11}$ & $44.65_{-6.58}^{+10.33}$ & $34.95_{-7.79}^{+6.18}$ & $0.07_{-0.24}^{+0.24}$ & $0.09_{-0.09}^{+0.16}$ & 13.33 & 4096 & -0.43 & SEOB\\
        
 $\mathrm{GW230708\_230935}$ & $3.7\times10^{-3}$ & $68.58_{-14.06}^{+13.91}$ & $100.62_{-20.03}^{+25.28}$ & $64.10_{-25.29}^{+22.23}$ & $0.04_{-0.30}^{+0.27}$ & $0.10_{-0.09}^{+0.30}$ & 13.33 & 4096 & -0.42 & TEOB\\
        
 $\mathrm{GW230709\_122727}$ & $1.1\times10^{-2}$ & $55.51_{-14.17}^{+9.79}$ & $76.07_{-14.56}^{+17.51}$ & $54.92_{-24.02}^{+15.16}$ & $0.08_{-0.32}^{+0.30}$ & $0.16_{-0.14}^{+0.28}$ & 13.33 & 4096 & -0.25 & TEOB\\
        
 $\mathrm{GW230712\_090405}$ & $1.8\times10^{-2}$ & $24.68_{-6.92}^{+17.91}$ & $44.17_{-15.73}^{+15.81}$ & $23.21_{-12.74}^{+22.13}$ & $-0.05_{-0.31}^{+0.29}$ & $0.10_{-0.09}^{+0.26}$ & 13.33 & 4096 & -0.42 & SEOB\\
        
 $\mathrm{GW230723\_101834}$ & $3.4\times10^{-3}$ & $14.82_{-0.45}^{+0.46}$ & $21.12_{-3.74}^{+6.72}$ & $13.87_{-3.12}^{+2.86}$ & $-0.19_{-0.15}^{+0.17}$ & $0.07_{-0.06}^{+0.11}$ & 13.33 & 4096 & -0.55 & SEOB\\
        
 $\mathrm{GW230726\_002940}$ & $<1\times10^{-5}$ & $37.38_{-3.57}^{+3.72}$ & $48.42_{-6.46}^{+10.01}$ & $38.33_{-7.63}^{+6.25}$ & $-0.01_{-0.24}^{+0.21}$ & $0.06_{-0.06}^{+0.11}$ & 13.33 & 4096 & -0.62 & SEOB\\
        
 $\mathrm{GW230729\_082317}$ & $1.8\times10^{-1}$ & $10.84_{-0.32}^{+0.22}$ & $16.18_{-3.46}^{+10.43}$ & $9.67_{-3.24}^{+2.46}$ & $0.12_{-0.15}^{+0.23}$ & $0.08_{-0.08}^{+0.15}$ & 13.33 & 4096 & -0.49 & SEOB\\
        
 $\mathrm{GW230731\_215307}$ & $<1\times10^{-5}$ & $9.45_{-0.08}^{+0.08}$ & $12.56_{-1.56}^{+3.57}$ & $9.41_{-1.88}^{+1.29}$ & $-0.04_{-0.06}^{+0.11}$ & $0.05_{-0.04}^{+0.08}$ & 13.33 & 4096 & -0.73 & SEOB\\
        
 $\mathrm{GW230803\_033412}$ & $3.1\times10^{-1}$ & $54.32_{-10.05}^{+9.10}$ & $75.55_{-14.54}^{+18.88}$ & $52.95_{-18.73}^{+14.22}$ & $0.05_{-0.31}^{+0.28}$ & $0.09_{-0.08}^{+0.16}$ & 13.33 & 4096 & -0.45 & SEOB\\
        
 $\mathrm{GW230805\_034249}$ & $3.7\times10^{-3}$ & $35.18_{-5.08}^{+10.15}$ & $49.15_{-10.28}^{+13.88}$ & $35.34_{-10.96}^{+12.59}$ & $0.05_{-0.29}^{+0.27}$ & $0.07_{-0.06}^{+0.13}$ & 13.33 & 4096 & -0.57 & SEOB\\
        
 $\mathrm{GW230806\_204041}$ & $3.7\times10^{-3}$ & $66.82_{-13.05}^{+11.51}$ & $92.34_{-17.27}^{+21.39}$ & $65.92_{-24.70}^{+17.94}$ & $0.08_{-0.28}^{+0.27}$ & $0.10_{-0.09}^{+0.19}$ & 13.33 & 4096 & -0.41 & SEOB\\
        
 $\mathrm{GW230811\_032116}$ & $<1\times10^{-5}$ & $33.51_{-2.70}^{+2.76}$ & $46.64_{-7.08}^{+8.54}$ & $32.26_{-6.88}^{+6.78}$ & $0.03_{-0.18}^{+0.18}$ & $0.05_{-0.04}^{+0.08}$ & 13.33 & 4096 & -0.76 & SEOB\\
        
 $\mathrm{GW230814\_061920}$ & $6.3\times10^{-4}$ & $75.99_{-18.00}^{+12.96}$ & $116.02_{-21.95}^{+21.72}$ & $67.27_{-27.66}^{+25.80}$ & $0.15_{-0.27}^{+0.26}$ & $0.10_{-0.09}^{+0.18}$ & 13.33 & 4096 & -0.43 & SEOB\\
        
 $\mathrm{GW230814\_230901}$ & $<1\times10^{-5}$ & $28.65_{-0.54}^{+0.57}$ & $36.69_{-3.07}^{+4.81}$ & $29.60_{-3.25}^{+2.72}$ & $0.01_{-0.06}^{+0.07}$ & $0.02_{-0.02}^{+0.03}$ & 13.33 & 4096 & -1.16 & SEOB\\
        
 $\mathrm{GW230819\_171910}$ & $1.1\times10^{-2}$ & $71.62_{-20.28}^{+14.50}$ & $110.85_{-23.77}^{+40.12}$ & $63.63_{-33.23}^{+25.56}$ & $-0.14_{-0.37}^{+0.30}$ & $0.28_{-0.25}^{+0.21}$ & 13.33 & 4096 & -0.00 & TEOB\\
        
 $\mathrm{GW230820\_212515}$ & $2.4\times10^{-1}$ & $66.34_{-27.18}^{+15.44}$ & $97.64_{-20.96}^{+28.68}$ & $62.37_{-40.41}^{+24.20}$ & $0.14_{-0.29}^{+0.32}$ & $0.13_{-0.12}^{+0.22}$ & 13.33 & 4096 & -0.34 & TEOB\\
        
 $\mathrm{GW230824\_033047}$ & $<1\times10^{-5}$ & $66.78_{-10.71}^{+8.00}$ & $89.95_{-12.80}^{+16.92}$ & $67.23_{-22.95}^{+13.95}$ & $0.01_{-0.25}^{+0.24}$ & $0.07_{-0.06}^{+0.12}$ & 13.33 & 4096 & -0.55 & TEOB\\
        
 $\mathrm{GW230825\_041334}$ & $1.0\times10^{-1}$ & $51.72_{-9.61}^{+8.38}$ & $73.78_{-14.26}^{+17.66}$ & $48.89_{-16.72}^{+14.38}$ & $0.30_{-0.32}^{+0.24}$ & $0.08_{-0.07}^{+0.15}$ & 13.33 & 4096 & -0.52 & SEOB\\
        
 $\mathrm{GW230831\_015414}$ & $2.9\times10^{-1}$ & $55.72_{-9.25}^{+28.91}$ & $75.28_{-13.01}^{+46.25}$ & $56.26_{-18.17}^{+24.79}$ & $0.05_{-0.29}^{+0.27}$ & $0.13_{-0.11}^{+0.27}$ & 13.33 & 4096 & -0.32 & TEOB\\
        
 $\mathrm{GW230904\_051013}$ & $<1\times10^{-5}$ & $9.03_{-0.11}^{+0.07}$ & $12.78_{-2.22}^{+5.76}$ & $8.46_{-2.29}^{+1.69}$ & $0.05_{-0.09}^{+0.18}$ & $0.06_{-0.05}^{+0.08}$ & 13.33 & 4096 & -0.65 & SEOB\\
        
 $\mathrm{GW230911\_195324}$ & $1.0\times10^{-3}$ & $30.14_{-2.86}^{+2.68}$ & $41.23_{-6.33}^{+8.30}$ & $29.42_{-6.30}^{+5.73}$ & $0.02_{-0.21}^{+0.19}$ & $0.07_{-0.06}^{+0.13}$ & 13.33 & 4096 & -0.58 & SEOB\\
        
 $\mathrm{GW230914\_111401}$ & $<1\times10^{-5}$ & $58.83_{-11.04}^{+7.79}$ & $87.25_{-12.57}^{+12.23}$ & $53.09_{-18.98}^{+18.86}$ & $0.13_{-0.19}^{+0.19}$ & $0.04_{-0.04}^{+0.07}$ & 13.33 & 4096 & -0.76 & SEOB\\
        
 $\mathrm{GW230919\_215712}$ & $<1\times10^{-5}$ & $26.23_{-1.14}^{+1.17}$ & $33.94_{-3.63}^{+6.39}$ & $26.94_{-4.64}^{+3.29}$ & $0.17_{-0.13}^{+0.12}$ & $0.04_{-0.04}^{+0.07}$ & 13.33 & 4096 & -0.78 & SEOB\\
        
 $\mathrm{GW230920\_071124}$ & $<1\times10^{-5}$ & $35.31_{-3.37}^{+3.62}$ & $47.14_{-6.82}^{+11.31}$ & $35.45_{-8.79}^{+6.63}$ & $-0.01_{-0.23}^{+0.21}$ & $0.07_{-0.06}^{+0.11}$ & 13.33 & 4096 & -0.55 & SEOB\\
        
 $\mathrm{GW230922\_020344}$ & $<1\times10^{-5}$ & $38.21_{-3.18}^{+3.54}$ & $50.57_{-7.10}^{+10.64}$ & $38.56_{-7.59}^{+6.35}$ & $0.06_{-0.20}^{+0.20}$ & $0.05_{-0.05}^{+0.09}$ & 13.33 & 4096 & -0.69 & SEOB\\
        
 $\mathrm{GW230922\_040658}$ & $<1\times10^{-5}$ & $105.06_{-19.02}^{+14.87}$ & $143.04_{-22.78}^{+35.02}$ & $105.05_{-42.16}^{+24.44}$ & $0.33_{-0.28}^{+0.25}$ & $0.12_{-0.11}^{+0.15}$ & 13.33 & 4096 & -0.35 & TEOB\\
        
 $\mathrm{GW230924\_124453}$ & $<1\times10^{-5}$ & $31.47_{-2.17}^{+2.16}$ & $39.94_{-4.24}^{+6.68}$ & $32.96_{-5.54}^{+3.98}$ & $0.02_{-0.19}^{+0.17}$ & $0.05_{-0.04}^{+0.08}$ & 13.33 & 4096 & -0.71 & SEOB\\
        
 $\mathrm{GW230927\_043729}$ & $<1\times10^{-5}$ & $40.66_{-4.01}^{+4.34}$ & $53.09_{-7.22}^{+10.73}$ & $41.69_{-9.27}^{+7.13}$ & $0.01_{-0.22}^{+0.21}$ & $0.06_{-0.05}^{+0.10}$ & 13.33 & 4096 & -0.68 & SEOB\\
        
 $\mathrm{GW230927\_153832}$ & $<1\times10^{-5}$ & $20.18_{-0.36}^{+0.40}$ & $26.11_{-2.63}^{+4.39}$ & $20.66_{-2.92}^{+2.31}$ & $0.04_{-0.07}^{+0.08}$ & $0.03_{-0.03}^{+0.06}$ & 13.33 & 4096 & -0.91 & SEOB\\
        
 $\mathrm{GW230928\_215827}$ & $<1\times10^{-5}$ & $60.75_{-13.45}^{+10.52}$ & $91.07_{-20.44}^{+26.05}$ & $54.57_{-20.91}^{+18.70}$ & $0.39_{-0.29}^{+0.20}$ & $0.08_{-0.08}^{+0.15}$ & 13.33 & 4096 & -0.50 & SEOB\\
        
 $\mathrm{GW230930\_110730}$ & $1.7\times10^{-1}$ & $44.01_{-6.16}^{+6.37}$ & $59.65_{-10.35}^{+17.54}$ & $43.66_{-13.22}^{+9.91}$ & $0.04_{-0.28}^{+0.25}$ & $0.08_{-0.07}^{+0.13}$ & 13.33 & 4096 & -0.53 & SEOB\\
        
 $\mathrm{GW231001\_140220}$ & $<1\times10^{-5}$ & $81.27_{-25.52}^{+18.50}$ & $126.49_{-27.66}^{+30.13}$ & $71.30_{-36.48}^{+30.54}$ & $0.04_{-0.31}^{+0.30}$ & $0.36_{-0.30}^{+0.11}$ & 13.33 & 4096 & 0.30 & TEOB\\
        
 $\mathrm{GW231004\_232346}$ & $1.6\times10^{-1}$ & $70.18_{-15.00}^{+12.30}$ & $107.04_{-22.27}^{+27.17}$ & $62.02_{-23.64}^{+23.36}$ & $-0.08_{-0.33}^{+0.27}$ & $0.19_{-0.18}^{+0.29}$ & 13.33 & 4096 & -0.18 & TEOB\\
        
 $\mathrm{GW231005\_021030}$ & $1.0\times10^{-2}$ & $108.98_{-18.57}^{+16.32}$ & $151.93_{-26.14}^{+36.33}$ & $105.23_{-34.38}^{+26.52}$ & $0.09_{-0.33}^{+0.34}$ & $0.24_{-0.22}^{+0.22}$ & 13.33 & 4096 & -0.14 & TEOB\\
        
 $\mathrm{GW231005\_091549}$ & $4.0\times10^{-2}$ & $34.32_{-3.44}^{+3.67}$ & $45.32_{-6.48}^{+12.68}$ & $34.77_{-8.88}^{+6.21}$ & $-0.02_{-0.26}^{+0.23}$ & $0.08_{-0.07}^{+0.13}$ & 13.33 & 4096 & -0.53 & SEOB\\
        
 $\mathrm{GW231008\_142521}$ & $1.6\times10^{-3}$ & $44.49_{-8.61}^{+9.48}$ & $67.93_{-14.72}^{+19.57}$ & $40.14_{-15.88}^{+15.44}$ & $-0.01_{-0.29}^{+0.25}$ & $0.06_{-0.06}^{+0.12}$ & 13.33 & 4096 & -0.62 & SEOB\\
        
 $\mathrm{GW231014\_040532}$ & $2.1\times10^{-1}$ & $21.53_{-2.10}^{+2.01}$ & $28.63_{-4.15}^{+7.79}$ & $21.47_{-4.95}^{+3.65}$ & $0.18_{-0.24}^{+0.22}$ & $0.09_{-0.08}^{+0.15}$ & 13.33 & 4096 & -0.46 & SEOB\\
        
 $\mathrm{GW231018\_233037}$ & $6.8\times10^{-1}$ & $10.14_{-0.26}^{+0.15}$ & $15.11_{-3.16}^{+7.30}$ & $9.06_{-2.61}^{+2.27}$ & $0.004_{-0.13}^{+0.20}$ & $0.08_{-0.07}^{+0.12}$ & 13.33 & 4096 & -0.48 & SEOB\\
        
 $\mathrm{GW231020\_142947}$ & $<1\times10^{-5}$ & $9.96_{-0.12}^{+0.11}$ & $14.81_{-3.15}^{+12.23}$ & $8.91_{-3.41}^{+2.23}$ & $0.14_{-0.12}^{+0.27}$ & $0.05_{-0.04}^{+0.08}$ & 13.33 & 4096 & -0.76 & TEOB\\
        
 $\mathrm{GW231028\_153006}$ & $<1\times10^{-5}$ & $109.63_{-8.16}^{+7.70}$ & $138.73_{-14.03}^{+26.18}$ & $115.92_{-26.34}^{+13.20}$ & $0.49_{-0.13}^{+0.15}$ & $0.06_{-0.05}^{+0.09}$ & 13.33 & 4096 & -0.64 & SEOB\\
        
 $\mathrm{GW231029\_111508}$ & $<1\times10^{-5}$ & $67.85_{-13.26}^{+10.50}$ & $96.21_{-15.42}^{+17.62}$ & $64.78_{-24.60}^{+19.54}$ & $0.14_{-0.22}^{+0.23}$ & $0.06_{-0.05}^{+0.11}$ & 13.33 & 4096 & -0.66 & SEOB\\
        
 $\mathrm{GW231102\_071736}$ & $<1\times10^{-5}$ & $72.05_{-9.25}^{+6.85}$ & $94.71_{-11.73}^{+14.91}$ & $73.89_{-21.81}^{+12.41}$ & $0.07_{-0.21}^{+0.20}$ & $0.08_{-0.07}^{+0.13}$ & 13.33 & 4096 & -0.52 & TEOB\\
        
 $\mathrm{GW231104\_133418}$ & $<1\times10^{-5}$ & $11.28_{-0.13}^{+0.12}$ & $15.57_{-2.39}^{+6.30}$ & $10.84_{-2.78}^{+1.90}$ & $0.14_{-0.08}^{+0.12}$ & $0.05_{-0.04}^{+0.09}$ & 13.33 & 4096 & -0.73 & SEOB\\
        
 $\mathrm{GW231108\_125142}$ & $<1\times10^{-5}$ & $23.78_{-1.04}^{+1.00}$ & $31.38_{-3.81}^{+6.26}$ & $23.93_{-4.10}^{+3.33}$ & $-0.07_{-0.14}^{+0.13}$ & $0.05_{-0.05}^{+0.09}$ & 13.33 & 4096 & -0.69 & SEOB\\
        
 $\mathrm{GW231110\_040320}$ & $<1\times10^{-5}$ & $18.05_{-0.52}^{+0.73}$ & $25.18_{-4.01}^{+7.89}$ & $17.27_{-3.89}^{+3.22}$ & $0.18_{-0.12}^{+0.13}$ & $0.05_{-0.05}^{+0.09}$ & 13.33 & 4096 & -0.71 & SEOB\\
        
 $\mathrm{GW231113\_122623}$ & $2.8\times10^{-1}$ & $44.20_{-8.44}^{+8.22}$ & $61.26_{-12.29}^{+24.20}$ & $42.52_{-14.81}^{+11.61}$ & $0.34_{-0.33}^{+0.25}$ & $0.10_{-0.09}^{+0.17}$ & 13.33 & 4096 & -0.42 & SEOB\\
        
 $\mathrm{GW231113\_200417}$ & $<1\times10^{-5}$ & $9.79_{-0.11}^{+0.09}$ & $13.36_{-1.95}^{+5.81}$ & $9.52_{-2.57}^{+1.54}$ & $0.12_{-0.08}^{+0.14}$ & $0.06_{-0.05}^{+0.09}$ & 13.33 & 4096 & -0.63 & TEOB\\
        
 $\mathrm{GW231114\_043211}$ & $1.3\times10^{-4}$ & $14.59_{-0.36}^{+0.35}$ & $31.65_{-9.48}^{+13.86}$ & $9.48_{-2.31}^{+3.50}$ & $0.14_{-0.21}^{+0.23}$ & $0.06_{-0.05}^{+0.11}$ & 13.33 & 4096 & -0.65 & SEOB\\
        
 $\mathrm{GW231118\_005626}$ & $<1\times10^{-5}$ & $17.52_{-0.57}^{+0.41}$ & $25.96_{-5.18}^{+6.64}$ & $15.74_{-2.96}^{+3.84}$ & $0.36_{-0.11}^{+0.09}$ & $0.07_{-0.07}^{+0.12}$ & 13.33 & 4096 & -0.53 & SEOB\\
        
 $\mathrm{GW231118\_071402}$ & $2.8\times10^{-3}$ & $51.60_{-9.83}^{+10.24}$ & $71.00_{-13.17}^{+19.98}$ & $51.15_{-19.09}^{+14.08}$ & $0.16_{-0.29}^{+0.29}$ & $0.11_{-0.10}^{+0.32}$ & 13.33 & 4096 & -0.40 & TEOB\\
        
 $\mathrm{GW231118\_090602}$ & $<1\times10^{-5}$ & $10.51_{-0.10}^{+0.11}$ & $15.35_{-3.03}^{+13.32}$ & $9.56_{-3.78}^{+2.21}$ & $0.07_{-0.08}^{+0.31}$ & $0.04_{-0.04}^{+0.07}$ & 13.33 & 4096 & -0.77 & SEOB\\
        
 $\mathrm{GW231119\_075248}$ & $1.9\times10^{-2}$ & $69.79_{-15.51}^{+20.60}$ & $97.86_{-22.53}^{+34.88}$ & $68.72_{-27.70}^{+25.81}$ & $0.003_{-0.30}^{+0.28}$ & $0.10_{-0.09}^{+0.25}$ & 13.33 & 4096 & -0.41 & TEOB\\
        
 $\mathrm{GW231123\_135430}$ & $<1\times10^{-5}$ & $145.72_{-10.85}^{+11.32}$ & $201.96_{-24.03}^{+30.19}$ & $139.24_{-19.98}^{+24.96}$ & $0.58_{-0.16}^{+0.16}$ & $0.46_{-0.04}^{+0.03}$ & 13.33 & 4096 & 6.05 & TEOB\\
        
 $\mathrm{GW231127\_165300}$ & $1.0\times10^{-2}$ & $53.37_{-13.33}^{+11.89}$ & $77.43_{-17.71}^{+22.42}$ & $50.11_{-21.04}^{+18.22}$ & $0.05_{-0.33}^{+0.31}$ & $0.10_{-0.09}^{+0.27}$ & 13.33 & 4096 & -0.39 & TEOB\\
        
 $\mathrm{GW231129\_081745}$ & $5.6\times10^{-2}$ & $45.24_{-9.52}^{+7.89}$ & $72.39_{-14.78}^{+15.72}$ & $38.14_{-14.12}^{+16.14}$ & $0.03_{-0.28}^{+0.26}$ & $0.10_{-0.09}^{+0.19}$ & 13.33 & 4096 & -0.41 & SEOB\\
        
 $\mathrm{GW231206\_233134}$ & $<1\times10^{-5}$ & $41.34_{-4.99}^{+4.33}$ & $53.29_{-6.83}^{+8.81}$ & $42.91_{-10.41}^{+7.07}$ & $-0.09_{-0.25}^{+0.22}$ & $0.08_{-0.07}^{+0.13}$ & 13.33 & 4096 & -0.55 & SEOB\\
        
 $\mathrm{GW231206\_233901}$ & $<1\times10^{-5}$ & $35.86_{-2.42}^{+2.43}$ & $48.02_{-6.26}^{+8.10}$ & $35.86_{-7.07}^{+6.05}$ & $-0.04_{-0.15}^{+0.14}$ & $0.04_{-0.03}^{+0.06}$ & 13.33 & 4096 & -0.85 & SEOB\\
        
 $\mathrm{GW231213\_111417}$ & $<1\times10^{-5}$ & $44.63_{-6.37}^{+7.50}$ & $58.39_{-9.33}^{+14.94}$ & $45.90_{-12.18}^{+9.97}$ & $0.08_{-0.24}^{+0.25}$ & $0.08_{-0.07}^{+0.14}$ & 13.33 & 4096 & -0.52 & SEOB\\
        
 $\mathrm{GW231221\_135041}$ & $5.4\times10^{-1}$ & $55.51_{-9.99}^{+8.33}$ & $74.81_{-12.50}^{+19.47}$ & $55.68_{-20.46}^{+12.35}$ & $-0.03_{-0.37}^{+0.35}$ & $0.21_{-0.19}^{+0.28}$ & 13.33 & 4096 & -0.12 & TEOB\\
        
 $\mathrm{GW231223\_032836}$ & $3.8\times10^{-4}$ & $57.06_{-10.24}^{+7.93}$ & $76.03_{-11.86}^{+16.36}$ & $57.81_{-20.52}^{+12.27}$ & $-0.15_{-0.32}^{+0.27}$ & $0.13_{-0.11}^{+0.22}$ & 13.33 & 4096 & -0.31 & TEOB\\
        
 $\mathrm{GW231223\_075055}$ & $5.5\times10^{-1}$ & $9.31_{-0.18}^{+0.41}$ & $13.90_{-2.91}^{+9.89}$ & $8.38_{-3.03}^{+2.09}$ & $0.09_{-0.14}^{+0.25}$ & $0.07_{-0.07}^{+0.13}$ & 13.33 & 4096 & -0.54 & SEOB\\
        
 $\mathrm{GW231223\_202619}$ & $2.0\times10^{-3}$ & $9.84_{-0.23}^{+0.22}$ & $13.14_{-1.70}^{+5.26}$ & $9.75_{-2.52}^{+1.41}$ & $0.11_{-0.11}^{+0.15}$ & $0.08_{-0.07}^{+0.14}$ & 13.33 & 4096 & -0.48 & SEOB\\
        
 $\mathrm{GW231224\_024321}$ & $<1\times10^{-5}$ & $8.47_{-0.07}^{+0.06}$ & $10.96_{-1.12}^{+2.41}$ & $8.66_{-1.43}^{+0.96}$ & $-0.003_{-0.05}^{+0.07}$ & $0.06_{-0.05}^{+0.09}$ & 13.33 & 4096 & -0.66 & SEOB\\
        
 $\mathrm{GW231226\_101520}$ & $<1\times10^{-5}$ & $39.74_{-1.46}^{+1.74}$ & $48.80_{-3.26}^{+4.88}$ & $43.01_{-4.76}^{+3.29}$ & $-0.06_{-0.09}^{+0.09}$ & $0.02_{-0.02}^{+0.04}$ & 13.33 & 4096 & -1.05 & TEOB\\
        
 $\mathrm{GW231230\_170116}$ & $4.2\times10^{-1}$ & $71.53_{-14.50}^{+13.69}$ & $96.79_{-17.15}^{+35.76}$ & $71.12_{-27.39}^{+17.88}$ & $-0.18_{-0.35}^{+0.28}$ & $0.14_{-0.12}^{+0.32}$ & 13.33 & 4096 & -0.32 & TEOB\\
        
 $\mathrm{GW231231\_154016}$ & $<1\times10^{-5}$ & $20.55_{-0.68}^{+0.68}$ & $26.86_{-3.05}^{+6.11}$ & $20.83_{-3.82}^{+2.65}$ & $-0.04_{-0.12}^{+0.12}$ & $0.05_{-0.04}^{+0.09}$ & 13.33 & 4096 & -0.72 & SEOB\\
        
 $\mathrm{GW240104\_164932}$ & $<1\times10^{-5}$ & $42.60_{-4.24}^{+4.42}$ & $55.73_{-7.12}^{+9.20}$ & $43.92_{-10.26}^{+7.39}$ & $0.12_{-0.19}^{+0.18}$ & $0.07_{-0.06}^{+0.12}$ & 13.33 & 4096 & -0.54 & SEOB\\
        
 $\mathrm{GW240107\_013215}$ & $2.8\times10^{-2}$ & $71.95_{-25.84}^{+15.45}$ & $103.32_{-23.68}^{+35.15}$ & $68.41_{-39.68}^{+24.46}$ & $0.28_{-0.36}^{+0.32}$ & $0.16_{-0.14}^{+0.23}$ & 13.33 & 4096 & -0.23 & SEOB\\
        
 $\mathrm{GW240109\_050431}$ & $2.3\times10^{-4}$ & $25.17_{-1.90}^{+2.22}$ & $36.50_{-6.86}^{+8.74}$ & $23.36_{-5.09}^{+5.64}$ & $-0.07_{-0.23}^{+0.20}$ & $0.06_{-0.06}^{+0.12}$ & 13.33 & 4096 & -0.61 & SEOB\\
    \end{longtable*}
    \setlength\tabcolsep{6pt}

%% file: special_event_table.tex
    \begin{table}[h!]
    \centering
    \begin{tabular}[t]{l c c}
    \hline \hline
    Name & $\log_{10} \mathcal{B}_\text{SEOB}$ & $\log_{10} \mathcal{B}_\text{TEOB}$ \\
    \hline

 $\mathrm{GW200129\_065458}$ & $3.12$ & $3.59$ \\

 $\mathrm{GW231001\_140220}$ & $-0.12$ & $0.30$ \\

 $\mathrm{GW231123\_135430}$ & $0.04$ & $6.05$ \\

 \hline
 
 $\mathrm{GW190521\_074359}$ & $-0.67$ & $-0.83$ \\
        
 $\mathrm{GW191204\_171526}$ & $-0.07$ & $-0.31$ \\

 \SkipForEarlyCirculation{
 $\mathrm{GW200105\_162426}$ & $-0.85$ & $-0.27$ \\
        }
 \hline
    \end{tabular}
    \caption{The log-10 Bayes factors ($\log_{10}\mathcal{B}$) from both the SEOBNRv5EHM and TEOBResumS-Dali analyses for events with substantial eccentricity (top) and events containing a nonzero eccentricity peak in at least one analysis (below).}
    \label{tab:spec_events}
    \end{table}